\RequirePackage{fix-cm}
\documentclass[smallextended]{svjour3wide} 
\smartqed  
\usepackage{amsmath,amssymb,bm}
\usepackage{mathrsfs,amsbsy,amssymb,latexsym,amsfonts,amsmath,fancyhdr,lastpage}
\usepackage[dvipdfmx]{graphicx}
\usepackage{epsfig,amsfonts,amsbsy}
\usepackage{fancyhdr}
\usepackage{color}
\usepackage{url}

\newcommand{\pderf}[3]{\left({\frac{\partial #1}{\partial  #2}}\right)_{#3}}
\newcommand{\der}[2]{\frac{d #1}{d  #2}}

\DeclareMathOperator*{\argmin}{arg\,min} 
 
\newcommand{\intx}{\int_{0}^{L_x} dx~}
\newcommand{\intL}{\int_{0}^{X} dx~}
\newcommand{\intG}{\int_{X}^{L_x} dx~}
\newcommand{\intr}{\int_{V} d^3{\bm r}~}
\newcommand{\intrd}{\int_{{\cal D}'} d^3{\bm r'}~}
\newcommand{\ep}{\varepsilon}
\newcommand{\oep}{O(\varepsilon^3)}
\newcommand{\oet}{O(\varepsilon^2)}

\newcommand{\eq}{\mathrm{eq}}

\newcommand{\LE}{\mathrm{LE}}

\newcommand{\vdW}{\mathrm{vdW}}
\newcommand{\subC}{\mathrm{c}}
\newcommand{\subL}{\mathrm{L}}
\newcommand{\subG}{\mathrm{G}}
\newcommand{\subLG}{\mathrm{L/G}}

\newcommand{\subR}{\mathrm{R}}
\newcommand{\Tc}{T_\mathrm{c}}
\newcommand{\mT}{T_\mathrm{m}}
\newcommand{\Tint}{\theta}
\newcommand{\xc}{X_\mathrm{c}}
\newcommand{\xint}{X}
\newcommand{\pex}{p^\mathrm{ex}}
\newcommand{\Ps}{p_\mathrm{s}}
\newcommand{\Vs}{V_\mathrm{s}}

\newcommand{\bT}{{\tilde T}}
\newcommand{\bmu}{\tilde\mu}
\newcommand{\bP}{{p}}
\newcommand{\bG}{{G}}
\newcommand{\bF}{{F}}
\newcommand{\bS}{{S}}
\newcommand{\bH}{{H}}
\newcommand{\bU}{{U}}
\newcommand{\bC}{{C}}
\newcommand{\bA}{{A}}

\newcommand{\bTG}{\tilde T^\mathrm{G}}
\newcommand{\bTL}{\tilde T^\mathrm{L}}
\newcommand{\NL}{N^\mathrm{L}}
\newcommand{\Ng}{N^\mathrm{G}}

\renewcommand{\arraystretch}{1.5}

\makeatletter
\@addtoreset{equation}{section}
\makeatother

\journalname{Journal of Statistical Physics}
\begin{document}

\title{Global Thermodynamics for Heat Conduction Systems}

\author{Naoko Nakagawa \and Shin-ichi Sasa}

\institute{N. Nakagawa \at
              Department of Physics, Ibaraki University, Mito 310-8512, Japan\\             
              \email{naoko.nakagawa.phys@vc.ibaraki.ac.jp}
                            \and
              S. Sasa \at
              Department of Physics, Kyoto University, Kyoto 606-8502, Japan \\
              \email{sasa@scphys.kyoto-u.ac.jp}
}

\date{\today}

\maketitle

\begin{abstract}
  We propose the concept of global temperature for spatially non-uniform
  heat conduction systems. With this novel quantity, we present
  an extended framework of thermodynamics for the whole
  system such that the fundamental relation of thermodynamics holds, which
  we call ``global thermodynamics'' for heat conduction systems. 
  Associated with this global thermodynamics, we formulate a
  variational principle for determining thermodynamic properties
  of the liquid-gas phase coexistence in heat conduction, which corresponds 
  to the natural extension of the Maxwell construction for equilibrium
  systems. We quantitatively predict that the  temperature   of the
  liquid-gas interface deviates from the equilibrium   transition
  temperature. This result indicates that a super-cooled gas stably
  appears near the interface. 
\end{abstract}


\setcounter{tocdepth}{3}
\tableofcontents



\section{Introduction}

\begin{figure}[tb]
\centering
\includegraphics[scale=0.6]{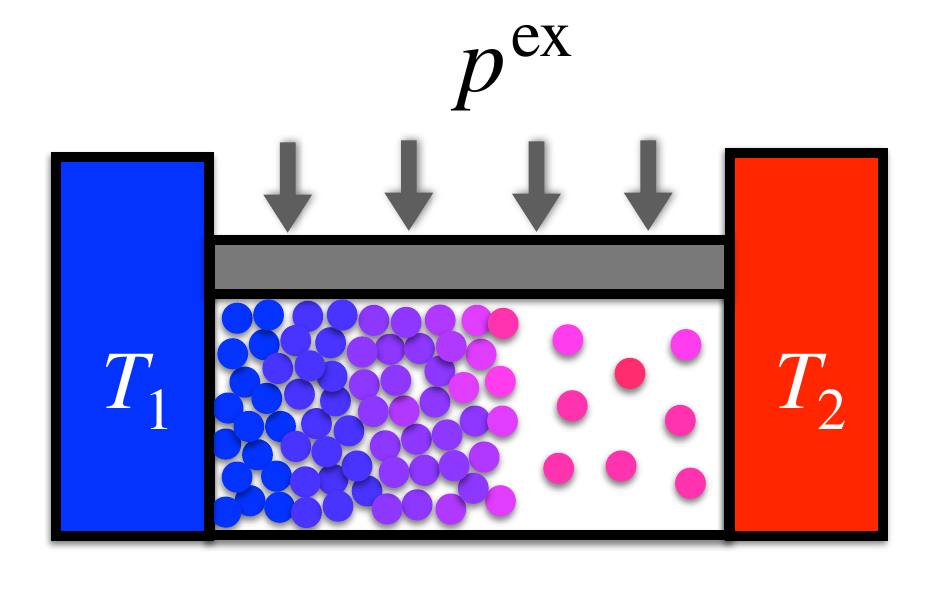}
\caption{Example of the system under study.}
\label{fig:Fig-intro}
\end{figure}

The behavior of  liquids and gases close to equilibrium
have been extensively studied for a long time. As a macroscopic universal
theory describing these phases of matter, hydrodynamics is believed to be
well-established \cite{Landau-Lifshitz-Fluid},
and the connection of the hydrodynamics with the classical and quantum
mechanics of atoms have been discussed for over a century \cite{Zubarev,Mclennan}.  
Nevertheless, in this paper, we  construct a new universal
theory for thermodynamic properties in the linear response regime.
There are two main messages. First, a new concept, {\it global temperature},
is found, with which a novel framework of {\it global  thermodynamics} is
constructed to describe the whole of non-uniform non-equilibrium systems with
local equilibrium thermodynamics. This outcome provides a fresh viewpoint for
the description of systems out of equilibrium. Second, this formulation
provides non-trivial quantitative predictions. As an example, let us
consider pure water under a pressure of $\pex=1.013\times 10^5$ Pa,
where two heat baths of temperature $T_1=368.0$ K and $T_2=378.0$ K are in contact with
the sides of the system. See Fig.~\ref{fig:Fig-intro} for
an illustration. Recall that the liquid-gas transition temperature is
$373.1$ K. Based on  global thermodynamics, 
in this paper, we predict that  the interface temperature of the liquid-gas
coexistence is $368.3$ K. This result means that {\it super-cooled
gas stably appears in heat conduction}, which may be tested
in experiments.


In the remaining part of this introduction, we first present
a brief summary of development in non-equilibrium statistical
mechanics, confirming that the phenomenon described above
has never been discussed by established theories. We then provide
a review of extended frameworks of thermodynamics so that readers
can understand  how the global thermodynamics proposed in this
  paper is different from previous frameworks. 
At the end of the introduction, 
we summarize the achievements of this paper. 

\subsection{Non-equilibrium statistical mechanics}


In the universal theory of macroscopic irreversible dynamics built
by Onsager,  a steady-state current is assumed to be linear
in the thermodynamic forces \cite{Onsager1931}.
The linear coefficients form a non-negative symmetric matrix,
which is referred to as the Onsager matrix. This  theory
with a variational principle for determining a current and/or
a  thermodynamic force is called  {\it irreversible thermodynamics},
 which is valid near equilibrium  \cite{Groot-Mazur}. 
Onsager theory can be interpreted as
a universal theory for the dynamical fluctuations of thermodynamic variables, 
where the properties of the Onsager matrix are connected to 
the stability of the equilibrium state and the time-reversal symmetry in microscopic systems. 
In short, this fluctuation theory is represented by a simple stochastic process with
a symmetry property \cite{Onsager1931,Schmitz}. In accordance with this theory,
statistical mechanics of trajectories
  has been  studied based on microscopic dynamics,
  which leads to the expression
of probability densities, linear and non-linear response formulas,
and macroscopic deterministic dynamics 
\cite{Zubarev,Mclennan,Kubo,Nakano,Zwanzig,Mori,Kawasaki-Gunton}.


After these developments, in the last two decades, our understanding of phenomena related
to thermodynamics has progressed greatly because of the following
two reasons. First, the development of experimental techniques for
the measurement and manipulation of biological molecular machines
\cite{Ashkin,Bustamante,Svoboda,Chu,Noji}
naturally leads to the extension of thermodynamics to mesoscopic
scales, which is now called stochastic thermodynamics
\cite{SeifertRPP,Sekimoto-book,Kalges-Just-Jarzynski}. 
Second, the newly discovered relations, such  as 
the fluctuation theorem \cite{Evans-Cohen-Morriss}
and the Jarzynski equality \cite{JarzynskiPRL}, have emerged as
universal \cite{Gallavotti,Kurchan,LS,Maes,Crooks,JarzynskiDFT},
providing a new starting point for the re-organization of previous
theories with greatly simplified derivations of the formulas
\cite{CrooksNRL,KN,KNST-rep,Maes-rep,SasaFluid}. These two developed
directions, stochastic thermodynamics and new universal
relations, are related to each other and have given rise to a new
connection with large deviation theory 
\cite{Bodineau-Derrida,BertiniETAL,Maes-LD,Nemoto,Bertini-rev},
and information theory \cite{Information}.


These results are quite useful when we specify a phenomenon under study.
Indeed, the phase coexistence in heat conduction can be studied
through a tough calculation based on the non-equilibrium statistical
mechanics \cite{SNIN}. However, we could not conjecture the non-trivial
nature of this phenomenon immediately from principles of non-equilibrium
statistical mechanics. For example, although one expects a variational
formulation for the phase coexistence in heat conduction by recalling the
minimum entropy production principle \cite{min-ent}, we cannot obtain
thermodynamic properties directly from the principle because
the variational principle basically determines the statistical ensemble
in the linear response regime as the minimizer of the entropy production
\cite{Klein}. In order to develop a variational principle for
thermodynamic  quantities, we have to start with an extended framework
of thermodynamics.

\subsection{Extended frameworks of thermodynamics}


Equilibrium thermodynamics provides  a unified description
of thermodynamic properties 
of materials at equilibrium. It also formalizes the second law,
which leads to a variational principle for determining equilibrium
states \cite{Callen,Prigogine-Kondepudi}.
The variational principle naturally suggests the law
of fluctuation of thermodynamic variables \cite{Einstein},
which is formulated as a large deviation theory. 
From this viewpoint, a framework
of statistical mechanics may be  constructed in a consistent manner
with the fluctuation theory of thermodynamic variables \cite{Oono}.


Therefore, in order to establish a universal theory for thermodynamic
properties out of equilibrium, it is natural to consider an
extended framework of thermodynamics. A naive attempt is to
extend  the equilibrium fundamental relation of thermodynamics
\begin{equation}
  dF=-SdT-p dV
\end{equation}
for the case of simple fluids, where $F$ is the Helmholtz free energy
and $S$ is the entropy. Examples of such attempts can be seen in Refs.
\cite{Keizer,Eu,Jou-book}.
The heart of the problem  for the extension is  to confirm the two conditions:
First, the theory is  self-consistent
and self-contained; second, new predictions specific to the extension
are presented. 
However, the two conditions are not confirmed in many studies.
The derivation of the extended thermodynamics from the microscopic theory, regardless of its importance,  does not make sense unless the extended framework
satisfies the above two conditions.  Here, we do not give a complete
review of previous studies on the extended framework of thermodynamics,
but study heat conduction systems from the viewpoint of the extended
framework of thermodynamics.


Heat conduction is described as a spatially extended system
in which thermodynamic variables slowly vary in space. The local
subsystems, which are small  but  macroscopic,  are regarded
as a local equilibrium state
\cite{Landau-Lifshitz-Fluid,Groot-Mazur,Bedeauz86}.
If we followed this standard description,
we could not  have a strong motivation to seek a thermodynamic
framework because all thermodynamic properties can be calculated
by the heat conduction equation and the local equilibrium thermodynamics.
Nevertheless, we still have two possibilities for considering an extended framework
of thermodynamics.


The first attempt is to go beyond the local equilibrium thermodynamics.
In this approach, which may be adapted in Refs. \cite{Jou,Sasa-Tasaki}, 
the equation of state for the local subsystems is modified to contain the
influence of the heat flux. Since such a contribution is
quite small in the linear response regime, the theory is not useful
even if it is correct. Although this approach may still be effective
for spatially homogeneous driven systems, in which intensive
parameters may describe the balance of extensive variables
\cite{Sasa-Tasaki,Hayashi-Sasa,Bertin,Seifert-contact,Dickman}, we
do not  deal with such systems in this paper. 


In the second approach, we retain the standard description 
for a spatially extended system with
local equilibrium thermodynamics. On this basis, we then seek a thermodynamic framework
for the whole system. As an example, we first consider the extension
of the second law, by which a state variable ``entropy'' is defined,
and we then derive the thermodynamic relation which corresponds to an
extension of the fundamental relation of thermodynamics. A concrete
procedure of the first step was proposed by introducing the concept
of {\it excess heat} \cite{Jou-book,Landauer,Oono-Paniconi}. This idea
was also studied from  semi-macroscopic and microscopic theories
\cite{Hatano-Sasa,Ruelle,KNST,NN,Jona-thermo,Maes-thermo,Spinney-Ford}.
However, the fundamental relation of thermodynamics
has not been derived from the extended entropy. One reason may be 
that an interesting phenomenon associated with the extended
entropy was not addressed. Nevertheless, one can continue to seek 
a possible framework without considering certain phenomena. Indeed, for a
specific model of heat conduction, the extended entropy was numerically
estimated \cite{Chiba-Nakagawa}, which suggests that the extended
entropy is close to the spatial integration of the local equilibrium
entropy density. Since the extended entropy can be obtained in experiments,
a natural question is  ``What is temperature?''  If
the fundamental relation of thermodynamics is formulated, there should be
a temperature satisfying it.  We consider this question seriously,
putting aside specific phenomena.

\subsection{Summary of results}


We first focus on  single-phase systems 
(either liquid or gas)  in heat conduction.
By assuming local equilibrium
thermodynamics, we  define the entropy $S$ and the Helmholtz free energy
$F$ in heat conduction by the spatial integration of the local density
fields corresponding to these variables. The pressure
field $p$ is homogeneous in space. Then, the problem is to find a
temperature $\bT$  satisfying the fundamental relation
\begin{equation}
  dF=-S d \bT-pdV
\label{f-relation}
\end{equation}
in the linear response regime. 
In \S\ref{s:single phase},  we solve this problem by 
defining $\tilde T$ as the kinetic temperature averaged over
particles in the system. We call this temperature {\it global
temperature}. It should be noted that the local temperature $T(\bm r)$,
which  depends on the position $\bm r$, satisfies the thermodynamic
relations for each point $\bm r$. In contrast to the local relations,
(\ref{f-relation}) is a global relation applied for the whole
system as if the system is at equilibrium. We call such a 
thermodynamic framework   {\it global thermodynamics}.
This formulation is also interpreted as a  mapping of
each heat conduction system to an equilibrium system 
through the novel quantity of the global temperature,  $\tilde T$.


The formulation for a single-phase system is indeed derived from
 fluid dynamics with local equilibrium thermodynamics.
That is, this formulation  is interpreted as a different formulation for
describing thermodynamic quantities in the linear response
regime. The prediction by using global thermodynamics
can also be predicted by hydrodynamics with local equilibrium
thermodynamics, in principle. Here, we go one step further. 
We consider a thermodynamic phenomenon that cannot be described by the
standard hydrodynamics with local equilibrium thermodynamics.
This is the phenomenon of  phase coexistence in heat conduction. In \S\ref{s:LG-SS},
we study this phenomenon and we show how existing theories are not
appropriate for determining the thermodynamic properties.
The essential point is that the condition for the connection
of the two phases is outside of the local equilibrium thermodynamics.


We study the phase coexistence in heat conduction with the
framework of global thermodynamics. We first
formulate a variational principle
for determining the local temperature of the liquid-gas
interface. The idea is quite simple. Fixing the global
temperature of the whole system, we naturally extend
the variational principle for  equilibrium systems 
to that for heat conduction systems. 
This idea was proposed in our previous paper \cite{NS}.
Remarkably, by using the solution of the variational equation,
in \S\ref{s:variational principle}, we derive
a universal relation among the interface temperature $\Tint$,
the equilibrium transition temperature $\Tc$, the global temperature
$\bT$, and the mean temperature of the two heat baths $\mT$,
which is
\begin{align}
  \Tint-\Tc=\bT-\mT.
\end{align}  
We call this relation  the {\it temperature relation}.  From the
temperature relation, we find that the temperature of
the liquid-gas interface deviates from the equilibrium
transition temperature. That is,  super-cooled gas
stably appears near the interface in heat conduction.
This is a qualitatively new phenomenon that has never
been considered in previous studies. 


Since the steady state as determined by the variational principle
is expressed as a function of the global temperature,
the prediction of measurable quantities for  given
conditions is indirect. Thus, in \S\ref{s:steady LG}, we re-express
all quantities
in the steady state in terms of the two temperatures of
the heat baths. We present several formulas of the interface
temperature by directly using measurable quantities. Furthermore,
we illustrate  examples of quantitative results for
a van der Waals fluid and pure water. 


For the steady state determined by the variational principle,
we further develop thermodynamics for heat conduction
systems with phase coexistence. 
First, in \S\ref{s:liquid-gas phase constant P},
we derive the fundamental relation associated with the
Gibbs free energy $G$.
At first sight, the result does not seem to contain non-equilibrium extensions.
This is because $G$ is not differentiable
at the transition point at equilibrium. By performing a careful
analysis near equilibrium, we find that the fundamental relation
holds in an appropriate equilibrium limit. The heat capacity and
the compressibility, which are singular at equilibrium, are also
obtained as a regularized form while breaking the additivity.
In \S\ref{s:liquid-gas phase constant V}, we derive the fundamental
relation associated
with the Helmholtz free energy $F$. Since the free energy $F$
is defined for the coexistence phase in equilibrium cases, its
non-equilibrium extension can be written as a perturbation
from the equilibrium form. We derive this expression explicitly.  


It should be noted that the quantitative prediction is made based on a
fundamental assumption of the variational principle. As is often observed
in universal theories, one may replace the fundamental assumption
by another one. In \S\ref{s:equivalence}, we formulate the theory
starting from assumptions other than the variational
principle. For example, when we assume the fundamental relation
of thermodynamics for the whole system by using the global
temperature, we can derive the results of the variational
principle. As another example, one may focus on how the volume
change near the transition temperature  at constant pressure
exhibits the singularity in the equilibrium limit. Supposing
the simplest form of a singularity, we can derive the results
of the variational principle. These findings indicate that the theory
itself possesses an elegant structure.


In \S\ref{s:general configuration}, we extend  the theory
in \S\ref{s:single phase}
to cover steady states with an arbitrarily shaped container
beyond the linear response regime,  while restricting our
attention to single-phase systems. Our theory quantitatively
predicts a new relation among the global quantities in this setup.

\section{Preliminaries}  

\subsection{Equilibrium thermodynamics}\label{s:Eq}


We consider a macroscopic material at equilibrium.
As the simplest example, we focus on a simple fluid
whose thermodynamic state is characterized by 
temperature $T$, the volume $V$ of a container,
and the amount of material $N$. For a system in contact
with a heat bath of temperature $T$ which may be controlled
externally, there exists a state variable $S(T,V,N)$, called
entropy, which  satisfies the Clausius equality 
\begin{equation}
  dS=\frac{d'Q}{T}
\label{clausius}
\end{equation}
for infinitesimal quasi-static heat $d'Q$ from the heat bath.
The infinitesimal change of internal energy $U(T,V,N)$ of the material is determined by
\begin{equation}
  dU=d'Q+d'W,
\label{1st}  
\end{equation}
which  is referred to as the first law of thermodynamics.   
$d'W$ is the infinitesimal quasi-static work required in the 
infinitesimal change of the volume $dV$, which is given by
\begin{equation}
  d'W=-pdV.
\label{work}  
\end{equation}
The substitution of  (\ref{clausius}) and (\ref{work}) into
(\ref{1st}) leads to the fundamental relation of thermodynamics:
\begin{align}
dU=TdS-pdV.
\end{align}
From this expression, we find that it is useful to consider
$U$ as a function of $(S,V,N)$. Indeed, a single function
$U(S,V,N)$ leads to all thermodynamic properties such as
 equation of state
\begin{align}
  p=p(T,V,N)
\label{eq-state}
\end{align}
and heat capacity
\begin{align}
  C_V=C_V(T,V,N).
\end{align}
We then define chemical
potential $\mu$ as
\begin{equation}
\mu \equiv  \pderf{U}{N}{S,V}.
\end{equation}


Various thermodynamic functions equivalent to $U(S,V,N)$
can be defined by the Legendre transformation of $U(S,V,N)$: 
\begin{eqnarray}
  F(T,V,N) &\equiv&  \min_{S}[U(S,V,N)-TS], \\
  H(S,p,N) &\equiv&  \min_{V}[U(S,V,N)+pV]. \\
  G(T,p,N) &\equiv&  \min_{S}[H(S,p,N)-TS].
\end{eqnarray}
The fundamental relations associated with these functions are
\begin{align}
&dF = -S dT -pdV+\mu dN ,\\
&dH = T dS +Vdp+\mu dN , \\
&dG = -S dT +Vdp+\mu dN  \label{fr:G}.
\end{align}
Substituting $G(T,p,N)=NG(T,p,1)$ into (\ref{fr:G}),
we find 
\begin{align}
G=\mu N.
\label{e:Extensivity Eq}
\end{align}
This form together with \eqref{fr:G} leads to the Gibbs-Duhem relation
\begin{align}
-SdT+Vdp-Nd\mu=0.
\label{e:Gibbs-Duhem Eq}
\end{align}


The extensivity of thermodynamic quantity leads to the concept
of density defined as the quantity per unit volume, similarly to
particle density $\rho=N/V$. 
For instance, entropy density is defined as 
\begin{align}
s(T,\rho) \equiv \frac{S(T, V, N)}{V}.
\end{align}
By using $\rho=\rho(T,p)$, which is obtained from (\ref{eq-state}),
one may consider the entropy density as a function of $(T,p)$, which is expressed as 
\begin{align}
s(T,p) = s(T,\rho(T,p)),
\end{align}
following the convention in thermodynamics. 
Similarly to the entropy density, for any extensive quantity
$A$, its density is defined as
\begin{align}
&a(T,\rho)\equiv \frac{A(T, V, N)}{V},\\
&a(T,p)=a(T,\rho(T,p)).\label{e:TP}
\end{align}
We here consider free energy density $g=G/V$. Substituting
$G=gV$ into (\ref{e:Extensivity Eq}) and (\ref{e:Gibbs-Duhem Eq}),
we obtain
\begin{align}
&g=\mu\rho, \label{e:Extensivity density}\\
  &s d T +\rho d\mu=d p,
  \label{e:Gibbs-Duhem density}
\end{align}
where $u=U/V$. From these, we have 
\begin{align}
d g=-s d T +\mu d\rho+d p ,
\label{e:Gibbs density}
\end{align}
where $dT$, $d\rho$ and $dp$ are not independent, because
$T$, $\rho$ and $p$ are connected by the equation of state
(\ref{eq-state}) such that
\begin{align}
d \rho = \left(\frac{\partial \rho}{\partial T}\right)_{p}dT
+\left(\frac{\partial \rho}{\partial p}\right)_{T}d p.
\label{e:EqOfState}
\end{align}


Furthermore, we  define extensive
quantities per one particle, as
\begin{align}
\hat a(T, p)\equiv \frac{a(T,p)}{\rho(T,p)}=\frac{A(T,p,N)}{N}.
\label{e:A/N}
\end{align}
Note that $\hat g$ is equivalent to the chemical potential $\mu$.
We then have 
\begin{align}
&\mu=\hat u-T\hat s+\hat\phi p,
\label{e: LeGendre density 1mol}\\
&d\mu=-\hat s d T+\hat\phi d p,
\label{e:Gibbs density 1mol}
\end{align}
where $\hat \phi$ is specific volume $\hat\phi=1/\rho$.

\subsection{Setup of heat conduction systems}

\begin{figure}[bt]
\centering
\includegraphics[scale=0.42]{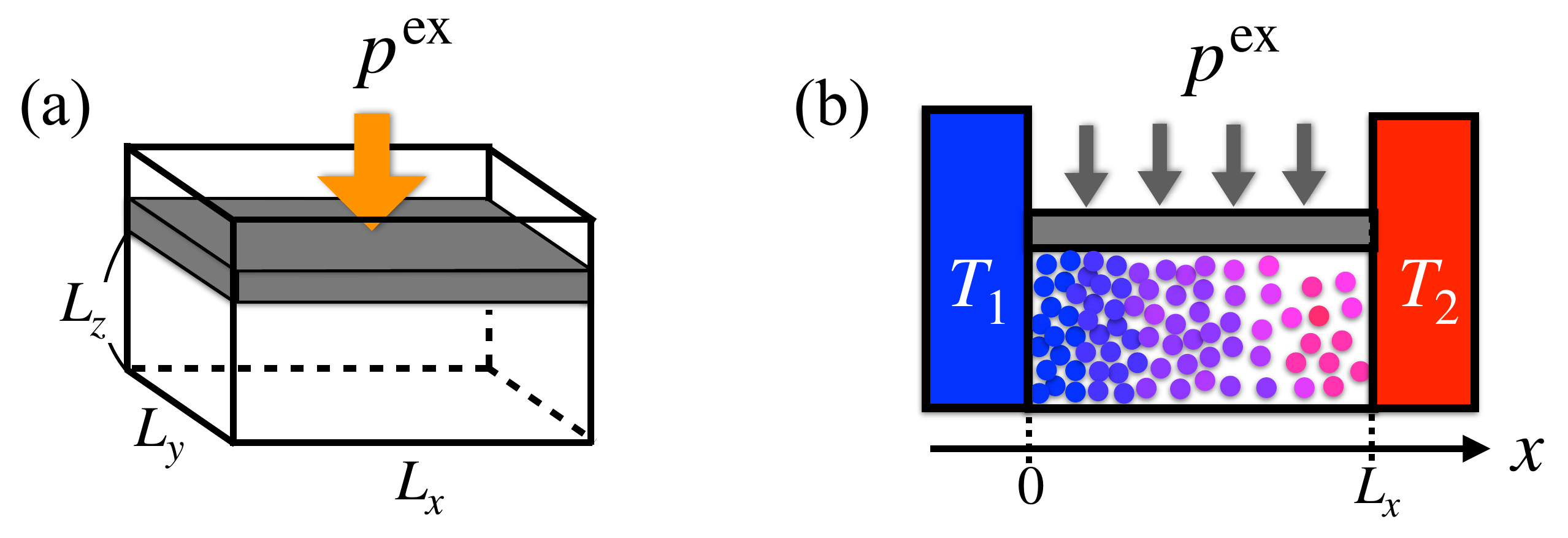}
\caption{Schematic figures of heat conduction systems at constant pressure.
(a) Shape of a container with a movable top plate. (b)
Layout of a system with two heat baths.}
\label{fig:Fig-setup}
\end{figure}

Throughout this paper, we consider a system of $N$-particles, which are
packed in a rectangular container with lengths of side $L_x$, $L_y$ and
$L_z$, as shown in Fig.~\ref{fig:Fig-setup}(a).
We ignore the effect of gravity.
$L_x$ and $L_y$ are fixed throughout this paper.
$L_z$ is fixed for a constant volume system,
or  not fixed at constant pressure. 
We study heat conduction states driven by the
temperature difference between two heat baths.
As schematically described in Fig. ~\ref{fig:Fig-setup}(b),
a heat bath of temperature $T_1$ is attached to the left end ($x=0$)
and another heat bath of temperature $T_2$ to the right end ($x=L_x$).
Other four boundaries are thermally insulating.
We take $T_1\le T_2$ without loss of generality,
and $(T_2-T_1)/T_2$ is assumed to be so  small that
the system reaches a unique nonequilibrium steady state in the linear response regime.
In such an idealized steady state without any convection, 
the system is regarded as a one-dimensional system.
Local states are homogeneous inside any section perpendicular
to the $x$ axis, and therefore, local thermodynamic  quantities are
considered as  functions of $x$.

We introduce  a dimensionless parameter $\ep$
that indicates the degree of non-equilibrium as
\begin{align}
\ep \equiv \frac{\Xi}{\mT},
\end{align}
where $\Xi$ is the temperature difference and 
$\mT$ is the mean temperature of the heat baths, which
are defined by 
\begin{align}
&\Xi \equiv T_2-T_1,\\
&\mT\equiv \frac{T_1+T_2}{2}.
\end{align}
When we focus on the linear response regime around an
equilibrium state, we ignore the contribution of $O(\ep^2)$.

\subsection{Local equilibrium thermodynamics}

For such macroscopic non-equilibrium systems, the hypothesis
of local equilibrium thermodynamics works well. That is,
local thermodynamic quantities are assumed to satisfy
local thermodynamic relations at each space and time
\cite{Landau-Lifshitz-Fluid,Groot-Mazur}.
Now, suppose that the temperature profile $T(x)$, 
the density profile $\rho(x)$ and the pressure $p(x)$
are determined by experimental observation.
Any local thermodynamic quantities, such as $a(x)$
and $\hat a(x)$, are expressed  as 
\begin{align}
a(x)&=a(T(x), \rho(x)),\\
\hat a(x)&=\hat a(T(x), p(x)),
\end{align}
where  functions $a(T,\rho)$ and $\hat a(T,p)$ are determined
in thermodynamics, as described in the previous subsection.
According to \eqref{e:TP},
$a(x)$ may be also written as $a(x)=a(T(x),p(x))$
via $\rho(x)=\rho(T(x),p(x))$.
Then, local equilibrium thermodynamics means 
\begin{align}
& g(x)=\mu(x)\rho(x),\label{e:localExtensivity}\\
&s(x) d T(x) +\rho(x)d\mu(x)=d p(x),\label{e:localGibbs-Duhem(x)}\\
&d g(x)=-s(x)d T(x) +\mu(x)d\rho(x)+d p(x),\label{e:dg(x)}
\end{align}
which correspond to the relations \eqref{e:Extensivity density},
\eqref{e:Gibbs-Duhem density} and \eqref{e:Gibbs density}, respectively.
For the quantities per one particle, we also have 
\begin{align}
  &\mu(x)=\hat u(x)-T(x)\hat s(x)+\hat\phi(x)p(x),
  \label{e:local-LeGendre-1mol}\\
  &d\mu(x)=-\hat s(x)d T(x)+\hat\phi(x)d p(x),
  \label{e:localGibbs(x)-1mol}
\end{align}
which correspond to the local version of the thermodynamic
relations \eqref{e: LeGendre density 1mol} and \eqref{e:Gibbs density 1mol}.

\subsection{Global conditions for steady states} \label{s:constraints}

For steady state  heat conduction, the local pressure $p(x)$
satisfies 
\begin{align}
p(x)=p(x')
\label{e:Pconst}
\end{align}
for any $x$ and $x'$.
Especially, for the system at the constant pressure $\pex$, 
\begin{align}
p(x)=\pex
\label{e:Pexconst}
\end{align}
holds for any $x$. These equalities for the local pressure may be regarded
as  global relations because they are not obtained in the local
thermodynamics.
Furthermore, heat flux $J(x)$ is uniform in $x$
as expressed in the form
\begin{align}
J(x)=J(x')
\label{e:Jconst}
\end{align}
for any $x$ and $x'$, which may work as another global
relation for the local quantities. In order to connect the heat flux
with the local thermodynamic quantities, we assume a heat conduction equation
\begin{align}
J(x)=-\kappa(T(x), \rho(x))\frac{\partial T(x)}{\partial x},
\label{e:eqnHeat}
\end{align}
in which $\kappa(T,\rho)$ is heat conductivity as a function of
$(T,\rho)$. 
We assume that there is no temperature gap at the boundaries, i.e.,
\begin{align}
\lim_{x\rightarrow +0}T(x)=T_1, \quad \lim_{x\rightarrow L_x-0}T(x)=T_2.
\label{e:boundary}
\end{align}
Finally, the conservation of particle number is written as
\begin{align}
L_y L_z\intx \rho(x) =N.
\label{e:Nconst}
\end{align}
These global relations, \eqref{e:Pconst}, \eqref{e:Pexconst},
\eqref{e:Jconst}, \eqref{e:Nconst}, together with the equation
of state \eqref{eq-state}, the heat conduction equation \eqref{e:eqnHeat},
and the boundary condition \eqref{e:boundary}, are sufficient to determine the
profiles of the local temperature $T(x)$ and the local density $\rho(x)$
provided that the systems consist of a single phase.


As we will see in \S\ref{s:LG-SS},
there is a case where liquid and gas coexist in the container.
For this special situation, the above global relations are not sufficient
to determine the local states. This problem will be seriously
studied in  later sections.

\subsection{Global thermodynamic quantities} \label{s:global quantities}

Since thermodynamic quantities in heat conduction are not uniform
in the space, they are basically described as local fields.
Nevertheless, from the fact that the local states are governed
by the global relations as  explained in \S\ref{s:constraints},
we expect that some properties of the heat conduction systems are
explained from a global point of view. Toward  the characterization
of  global properties, we here define global thermodynamic quantities
for heat conduction systems.

For extensive variables $A$ originally defined
for equilibrium states, we define the following global quantities
as an extension to those for heat conduction states:
\begin{align}
\bA=L_yL_z\intx a(T(x),p),
\label{e:globalA}
\end{align}
where we have used the same notation $A$ as that for equilibrium states.
When we interpret $A$ as a state function of heat conduction states, 
we explicitly write  $A(T_1, T_2, p, N)$ which is in contrast to  $A(T,p,N)$
for equilibrium states. 
For instance, global entropy and global Gibbs free
energy are defined as
\begin{align}
&\bS=L_yL_z\intx s(T(x),p), \label{e:globalS} \\
&\bG=L_yL_z\intx g(T(x),p). \label{e:globalG}
\end{align}

Here, we investigate the reference state dependence of these global
thermodynamic quantities. 
In equilibrium thermodynamics, entropy density and  internal energy
density are defined up to an additive constant which depends on the choice
of their reference state. Thus, the entropy density $s(x)$ and
the internal energy density $u(x)$ possess this property. Concretely, 
let $\hat s_0$ and $\hat u_0$ be the shift of the additive
arbitrary constants  of  entropy and energy  per one particle,
respectively, for the change of the reference state. We express
this transformation by 
\begin{align}
&s(x)\rightarrow s(x)+\hat s_0 \rho(x),\\
&u(x)\rightarrow u(x)+\hat u_0 \rho(x).
\end{align}
The shift of other thermodynamic quantities are induced as
\begin{align}
&g(x)\rightarrow g(x)-\hat s_0\rho(x)T(x)+\hat u_0 \rho(x),\\
&\mu(x) \rightarrow \mu(x)-\hat s_0T(x)+\hat u_0.
\end{align}
Note that local thermodynamic relations are invariant under the
transformation. 

Now, we consider the transformation of the global thermodynamic
quantities, which  are
defined by the spatial integral of the local quantities.
It is obvious that 
\begin{align}
\bS &\rightarrow \bS +\hat s_0 L_yL_z \intx~ \rho(x)= \bS+\hat s_0 N,\\
\bU &\rightarrow \bU +\hat u_0 L_yL_z\intx~  \rho(x)= \bU+\hat u_0 N,
\end{align}
while for the global Gibbs free energy $\bG$, we have 
\begin{align}
  \bG &\rightarrow \bG
  - \hat s_0 L_yL_z\intx~ \rho(x)T(x) +\hat u_0 L_yL_z \intx~ \rho(x).
\end{align}
The $\hat s_0$ dependence of $\bG$ is far from trivial.
Here, we introduce the {\it global temperature} $\bT$ such that
the transformation is written as
\begin{align}
\bG &\rightarrow \bG- \hat s_0 N\bT+\hat u_0 N.
\label{G-trans}
\end{align}
Explicitly, $\bT$ is given as
\begin{align}
\bT&=\frac{\intx~ \rho(x)T(x)}{\intx~ \rho(x)},
\label{e:globalT}
\end{align}
which means that global temperature $\bT$
corresponds to the kinetic temperature averaged over particles.
The transformation (\ref{G-trans})  suggests the
consistency among $\bG$, $\bS$ and $\bT$.
Indeed, we will show the global thermodynamic relations for these quantities.
Similarly, we also define the {\it global chemical potential} as
\begin{align}
\bmu&=\frac{\intx~ \rho(x)\mu(x)}{\intx~ \rho(x)}.
\label{e:globalMu}
\end{align}
Since  $\bmu=\bG/N$, the global chemical potential corresponds
to the Gibbs free energy per one particle.

\section{Global thermodynamics for single-phase systems
  in the linear response regime} \label{s:single phase}

In this section, we restrict ourselves to a single phase
where the heat conduction system  is occupied
by either liquid or gas at the constant pressure $\pex$.
When the environmental parameters $T_1$, $T_2$, and $\pex$ are fixed,
the steady state is uniquely determined from the equation
of state (\ref{eq-state}), the heat conduction equation \eqref{e:eqnHeat},
and the conservation law \eqref{e:Nconst}.
That is, the values of $(T(x), \rho(x), L_z)$ are determined
by 
\begin{align}
&p(T(x),\rho(x))=\pex, \label{68}\\
&\frac{\partial}{\partial x}\left[\kappa(T(x), \rho(x))\frac{\partial T(x)}{\partial x}\right]=0, \label{69} \\
&L_z=\frac{N}{L_y\intx \rho(x)} \label{70}.
\end{align}
Since the local thermodynamic quantities obey equilibrium thermodynamics, 
the profile $(T(x),\rho(x))$ leads to any local thermodynamic quantities
such as $s(x)$ and $\mu(x)$. Then, the global quantities such as
$\bT$, $\bmu$, and $\bS$ are  calculated for the steady state.

\begin{figure}[bt]
\centering
\includegraphics[scale=0.65]{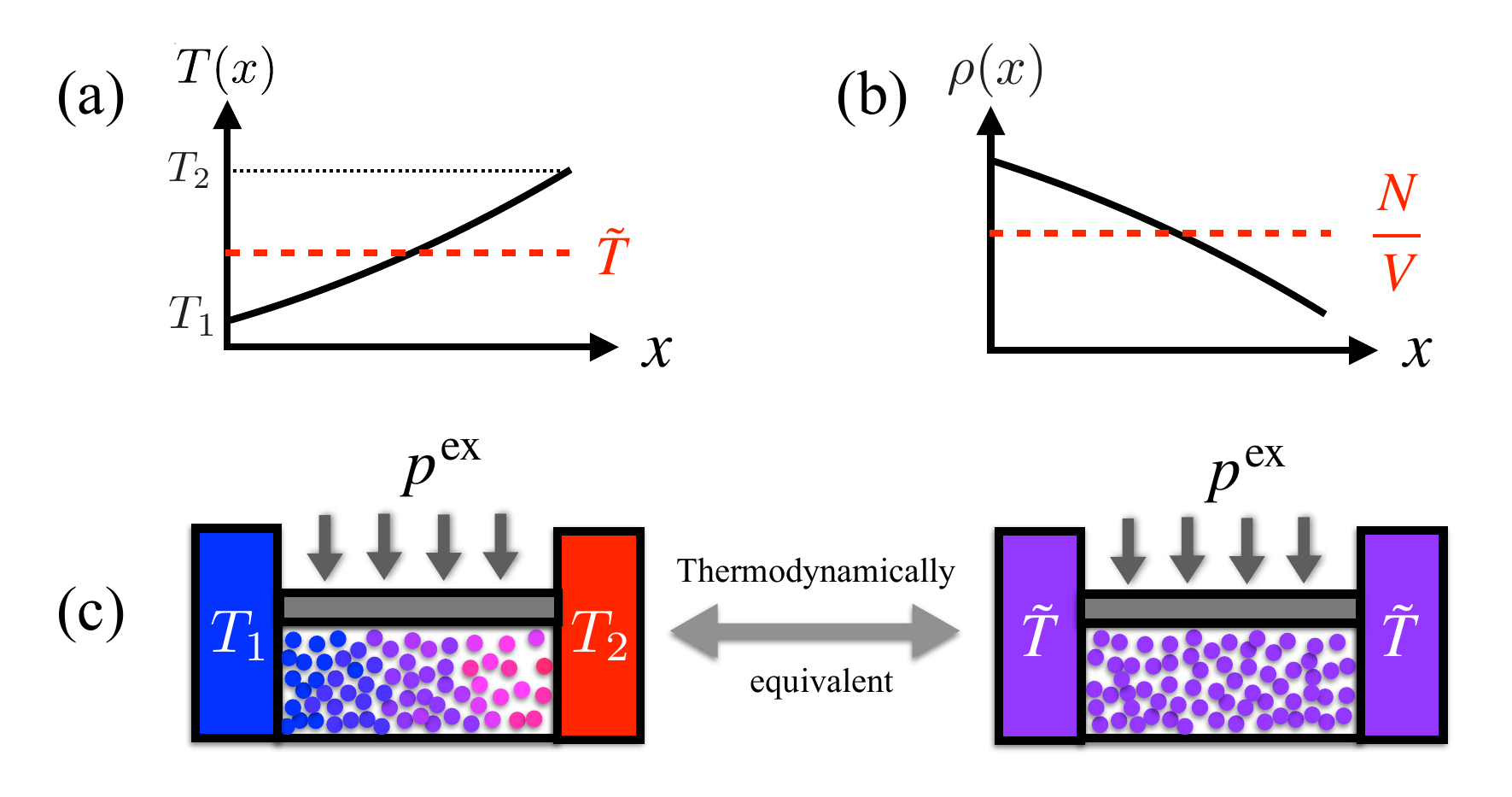}
\caption{(a) Typical temperature profile $T(x)$.
  The global temperature $\bT$ is shown in the red dotted line, which satisfies $\bT=(T_1+T_2)/2+O(\ep^2)$. See \eqref{e:mT-bT}. 
  (b) Typical density profile $\rho(x)$, where its spatial average $N/V$
  is displayed  by the red dotted line. The global version of the equation
  of state is given as $\pex=p(\bT, N/V)+O(\ep^2)$ in \eqref {e:global-eos}.
  (c) Correspondence of the heat conduction system to the
  equilibrium system of $(\bT,\pex,N)$.}
\label{fig:Fig-equiv}
\end{figure}

Now, we show that the global thermodynamic functions satisfy 
\begin{align}
&  \bG(T_1,T_2,p,N)=G(\bT,p,N)+O(\ep^2), \label{3-main-1} \\
&  \bS(T_1,T_2,p,N)=S(\bT,p,N)+O(\ep^2). \label{3-main-2}
\end{align}
This means that the global free energy and entropy, which are functions
of $(T_1, T_2, p, N)$, are expressed as the equilibrium free energy
and the equilibrium entropy when we use the global temperature $\bT$.
This result leads to relations among the global quantities as
\begin{align}
&G=U-\bT S+pV +O(\ep^2), \\
&d \bG=-\bS d\bT+V dp+\bmu dN+ O(\ep^2),
\end{align}
which represents the fundamental relation of thermodynamics
extended to heat conduction states in the linear
response regime. We call such a framework the {\it global
thermodynamics} for heat conduction systems.
These results indicate that, as schematically shown in Fig.~\ref{fig:Fig-equiv}, 
the global nature of the heat conduction system is equivalent to that of the equilibrium system.
Any global quantity $A$ defined by \eqref{e:globalA}
is connected to a corresponding equilibrium quantity $A(\bT,p,N)$ as
\begin{align}
A(T_1,T_2,p,N)=A(\bT,p,N)+O(\ep^2).
\label{e:globalA-TP}
\end{align}

\subsection{Proof of (\ref{3-main-1}) and (\ref{3-main-2})}
\label{s:globalGibbs-linear}

When the environmental parameters $(T_1, T_2, \pex)$ are slightly
changed to $(T_1+\delta T_1, T_2+\delta T_2, \pex+\delta\pex)$, 
the solution of the equations (\ref{68}), (\ref{69}), and (\ref{70})
is modified slightly.  We express the corresponding change as
\begin{align}
(T(x), \rho(x)) &\rightarrow (T(x)+\delta T(x), \rho(x)+\delta \rho(x)),\\
L_z& \rightarrow L_z+\delta L_z,
\end{align}
which leads to the change of any local thermodynamic quantity. For instance,
the change of the local Gibbs free energy density is given by
\begin{align}
\delta g(x)=g(T(x)+\delta T(x), \pex+\delta \pex)-g(T(x), \pex).
\end{align}
The change of the local quantities brings the change of the global
quantities as 
\begin{align}
\bT\rightarrow \bT+\delta\bT,\\
\bG\rightarrow \bG+\delta \bG,
\end{align}
where 
\begin{align}
  \delta \bG
  &=L_y\intx~[ (L_z+\delta L_z) g(T(x)+\delta T(x), \pex+\delta \pex)
- L_z g(T(x),  \pex)],\nonumber\\
&=L_y\intx ~ (L_z \delta g(x) +g(x) \delta L_z ).
\label{e:deltaG-org}
\end{align} 
Here, since \eqref{e:dg(x)} holds for the local densities,
the variation of each density satisfies a thermodynamic relation such as
\begin{align}
  \delta g(x)=-s(x)\delta T(x) +\mu(x)\delta \rho(x)
  +\delta \pex. \label{e:localGibbs}
\end{align}
By substituting the local relations \eqref{e:localExtensivity}
and \eqref{e:localGibbs} into \eqref{e:deltaG-org}, we have
\begin{align}
\delta \bG 
&=
-\bS \delta \bT +
V\delta \pex-\intx~(\hat s(x)n(x)\delta\eta(x)-\mu(x)\delta n(x)),
\label{e:deltaG1}
\end{align}
where
$\hat s(x) = s(x)/\rho(x)$ and we have defined 
\begin{align}
\eta(x)\equiv T(x)-\bT,
\quad
n(x)\equiv L_yL_z \rho(x).
\end{align}
Note that $n(x)dx$ is the particle number in $[x,x+dx]$
and that $n(x)$ is  estimated as $n(x)=N/L_x +O(\ep)$.
$\eta(x)$ is the deviation of the local temperature $T(x)$ from
the global temperature $\bT$, and is estimated as
$\eta(x)=O(\ep)$.

Since any local thermodynamic quantity $a(x)$ is regarded as
$a(T(x),\pex)=a(\bT+\eta(x),\pex)$, 
we may estimate $a(x)=a(\bT,\pex)+O(\ep)$. Then, the integrant
in \eqref{e:deltaG1} is estimated  as
\begin{align}
\hat s(x)n(x)\delta\eta(x)-\mu(x)\delta n(x)
&=
\hat s(x)\delta(n(x)\eta(x))-(\hat s(x)\eta(x)+\mu(x))\delta n(x)
\nonumber\\
&=
\hat s(\bT, p)\delta(n(x)\eta(x))-\mu(\bT,\pex)\delta  n(x) +\oet,
\end{align}
where $\delta n(x)=O(\ep)$ for $N$ and $L_x$ fixed.
We find that 
\begin{align}
\intx~n(x)\eta(x)=0
\label{e:globalT-eta}
\end{align}
holds from \eqref{e:globalT}. Obviously, the conservation of
the particle number leads to 
\begin{align}
\intx~ \delta n(x)=0.
\end{align}
Thus, the integral in \eqref{e:deltaG1} is estimated as
\begin{align}
\intx~(\hat s(x)n(x)\delta\eta(x)-\mu(x)\delta n(x))=\oet,
\end{align}
and then \eqref{e:deltaG1} becomes
\begin{align}
\delta \bG 
&=
-\bS \delta \bT +V\delta \pex +O(\ep^2).
\label{e:globalGibbs}
\end{align}
We emphasize that the definition of $\bT$ in \eqref{e:globalT}
is essential for this relation.

Here, let us recall that the value of $\bG$ is determined for
a given $(T_1, T_2, \pex, N)$. Since there is 
 one-to-one correspondence between $(\bT, \ep)$ and $(T_1, T_2)$, 
 $\bG$ is 
given as a function of $(\bT, \pex, N, \ep)$.
The relation (\ref{e:globalGibbs}) implies that $\bG$
is independent of $\ep$ in the linear response regime,
and thus given as a function of $(\bT, \pex, N)$.
Therefore, we conclude (\ref{3-main-1}). 
Next, we show \eqref{3-main-2}.
From the relation \eqref{e:globalGibbs}, the left-hand side
of \eqref{3-main-2} is expressed as
\begin{align}
  &\bS(T_1,T_2,\pex,N)=
  -\left(\frac{\partial \bG}{\partial \bT}\right)_{\pex,N}
  +\oet,
\label{e:globalS-differential}
\end{align}
while the equilibrium fundamental relation results in 
\begin{align}
  & S(\bT, \pex,N)=
  -\left(\frac{\partial \bG}{\partial \bT}\right)_{\pex,N}
 \label{e:globalS-differential-2}.
\end{align}
Combining these two equalities, we obtain (\ref{3-main-2}).


The definition \eqref{e:globalMu} for the global
chemical potential $\bmu$ indicates $\bmu=G/N$.
Then,  (\ref{3-main-1}) leads to
\begin{align}
\bmu(T_1,T_2,p) =\mu(\bT,p)+O(\ep^2).
\end{align}
We thus find 
\begin{align}
d\bmu&=-\hat s d\bT+\hat\phi d p+O(\ep^2),
\label{e:globalGibbs-1mol}
\end{align}
where 
\begin{align}
\hat s\equiv\frac{S}{N}, \quad\hat \phi\equiv\frac{V}{N},
\end{align}
are entropy per one particle and specific volume
of the heat conduction system.

\subsection{Various global thermodynamic functions} \label{s:GtoF}

From a local relation $f(x)=g(x)-p$, we have 
\begin{align}
\bF=\bG-pV.
\label{e:GtoF}
\end{align}
By using (\ref{3-main-1}), we obtain
\begin{align}
\bF(T_1,T_2,V,N)=F(\bT, V,N)+O(\ep^2).
\label{e:GtoF-2}
\end{align}
We thus have the fundamental relation for $\bF$
\begin{align}
d \bF=-\bS d\bT-pdV+\bmu dN+ O(\ep^2)
\end{align}
in the linear response regime.

Next, we consider the global internal energy $\bU$ 
and the global enthalpy $\bH$. 
From local thermodynamic relations $u(x)=f(x)+T(x)s(x)$ and $h(x)=g(x)+T(x)s(x)$,
we write
\begin{align}
\bU= \bF+\bT \bS +\intx~\hat s(x) \eta(x)n(x),\\
\bH=\bG+\bT \bS +\intx~\hat s(x) \eta(x)n(x).
\label{e:LeGendre0}
\end{align}
Remembering that $\eta(x)=O(\ep)$,  the integral is estimated as
\begin{align}
\intx~\hat s(x) \eta(x)n(x)&=\hat s(\bT, P) \intx~n(x)\eta(x)+\oet\nonumber\\
&=\oet,
\end{align}
where we have used \eqref{e:globalT-eta}. 
Thus, we obtain
 \begin{align}
   \bU &=  \bF+\bT \bS+\oet  \label{e:FtoU}\\
  \bH  &= \bG+\bT \bS+\oet \label{e:GtoH}
\end{align}
Substituting \eqref{3-main-1},\eqref{3-main-2},\eqref{e:GtoF-2} into \eqref{e:FtoU} and \eqref{e:GtoH}, we have
\begin{align}
   \bU(T_1,T_2,V,N) &=    U(\bS, V,N) +\oet,\\
  \bH(T_1,T_2,p,N)  &= H(\bS,p,N) +\oet.   
\end{align}

\subsection{Clausius equality}\label{Clausius}

Since the thermodynamic relations are extended with keeping
the same forms as those in equilibrium thermodynamics, we may define
 quasi-static heat $d'Q$ in an infinitely small quasi-static
process $(T_1,T_2, \pex) \to (T_1+\delta T_1,T_2+\delta T_2,
\pex+\delta \pex)$ as 
\begin{align}
{d' Q}=\bT \delta S .
\label{e:globalClausius}
\end{align}
This $d'Q$ corresponds to the absorbed heat by the system, 
which is the sum of the heat from the left and right heat baths
during the infinitesimal change. 
There, by summing the two heats, the net heat flow is canceled.
If the heat absorbed from the right
heat bath is exactly the same as  the heat released from the
left heat bath at every moment during the process, the net heat vanishes for
the heat conduction. Such systems are thought to be 
{\it  adiabatic} in the sense of  global thermodynamics,
and
\begin{align}
dS=0
\end{align}
in such a quasi-static adiabatic process in the linear response regime.

Then, constant-pressure heat capacity  is defined as	
\begin{align}
  C_p \equiv \left(\frac{d' Q}{d \bT}\right)_p.
\label{defCp}
\end{align}
Applying the thermodynamic relation, we obtain
\begin{align}
  C_p(T_1,T_2,p,N) &= \left(\frac{\partial H}{\partial \bT}\right)_p+\oet \\
      &=\bT \left(\frac{\partial S}{\partial \bT}\right)_p+\oet.
  \label{e:globalCp}
\end{align}
Note that $C_p$ is defined as the response to the change
of the global temperature $\bT$. By changing the temperatures
of heat baths, $\bT$ may change in accordance with absorbing heat
corresponding to $C_p \delta\bT$.

\subsection{Correspondence of global thermodynamic quantities to
  equilibrium quantities}\label{s:trapezoidal}

In the previous subsections,
we have obtained thermodynamic relations in the  linear response regime 
and found that they are equivalent to the equilibrium ones.
The key concept for extending thermodynamics is the
global temperature $\bT$ and the global chemical potential $\bmu$,
by which the heat conduction system of $(\bT, \bmu)$
is mapped to the equilibrium system of $(\bT,\bmu)$.
The connection between the two thermodynamic frameworks
becomes clearer in the argument below by considering the estimation
method of global thermodynamic quantities.

In a single phase system, local temperature
$T(x)$ is a continuous monotonic function of $x$.
Any extensive quantity  $A$, which  is defined as a spatial
integral of local density $a(x)$,  can be transformed
into an integral over temperature:
\begin{align}
\bA &=L_y L_z\intx~ a(T(x), p) \nonumber\\
& =\frac{V}{T_2-T_1}\int_{T_1}^{T_2} dT ~\psi(T, p), 
\label{e:Tintegral}
\end{align}
where
\begin{align}
&\psi(T,p)\equiv\frac{a(T,p)}{{\cal J}(T)}, \\
&{\cal J}(T)\equiv\frac{L_x}{T_2-T_1} \left(\frac{dx(T)}{dT}\right)^{-1}.
\end{align}
We expand $\psi(T,P)$ around $T=\mT$ in the form
\begin{align}
  \psi(T,p)=\psi(\mT,p)+\left.
  \left(\frac{\partial \psi}{\partial T}\right)_{p}\right|_{\mT}(T-\mT)+\oet.
\end{align}
The second term is canceled when the integral \eqref{e:Tintegral}
is performed. We thus have an estimation
\begin{align}
\bA&=V\psi(\mT,p)+\oet\nonumber\\
&=V\frac{a(\mT,p)}{{\cal J}(\mT)}+\oet.
\label{e:trapezoidal0}
\end{align}
For $A=V$, we have $a(x)=1$ for all $x$. By combining this 
with  \eqref{e:trapezoidal0}, we find 
\begin{align}
{{\cal J}(\mT)}=1+\oet.
\label{e:Jaccobian}
\end{align}
We then have arrived at a universal estimation
of the global quantity $A$ as
\begin{align}
\bA(T_1,T_2,p,N)=A(\mT,p,N)+\oet,
\label{e:trapezoidal}
\end{align}
where $A(\mT, p,N)=V a(\mT,p)$. The result 
\eqref{e:trapezoidal} directly connects
the global quantity in the heat conduction state
to the corresponding equilibrium quantity  at $\mT$.
 
We may apply the similar method to the global temperature.
The global temperature is also written as 
\begin{align}
\bT&=\frac{L_y L_z}{N}\intx~\rho(x)T(x) \\
   &=\frac{V}{N}\rho(\mT,p)\mT+\oet, \label{tochu}
\end{align}
in which
\begin{align}
  N &=L_y L_Z \intx~\rho(x) \\
    &=V \rho(\mT,p).
    \label{e:rho-TP}
\end{align}
By substituting it into (\ref{tochu}), we obtain
\begin{align}
\bT=\mT+\oet.
\label{e:mT-bT}
\end{align}
Thus, in the linear response regime, the global
temperature $\bT$ in  the heat conduction system is equal
to the mean temperature $\mT$ of the heat baths.
The relation \eqref{e:trapezoidal}, together with \eqref{e:mT-bT},
concludes \eqref{e:globalA-TP}.
We also note that \eqref{e:rho-TP},
which is rewritten as $\rho=\rho(\bT, p)+O(\ep^2)$ with $\rho=N/V$,
corresponds to the global version of the equation of state
\begin{align}
p=p(\bT,N/V)+O(\ep^2).
\label{e:global-eos}
\end{align}

We here remark that the relation $\bT=\mT$ in the
linear response regime is a specific feature of 
systems in a rectangular container.
We will show  in \S\ref{s:arbitrary shaped} 
that the global temperature deviates
from the mean temperature when  the shape of the
container is not a rectangle, whereas the relation \eqref{e:globalA-TP}
still holds.

\section{Liquid-gas coexistence in heat conduction}\label{s:LG-SS}

From now on, we study liquid-gas coexistence in heat
conduction. In \S\ref{s:LG-EQ}, we review  thermodynamics under
equilibrium conditions while paying attention to the  description of 
metastable states. In \S\ref{s:LG-HC}, we describe heat
conduction states based on local equilibrium thermodynamics.
We explain how the temperature profile and  the density profile
are determined for a given position of the liquid-gas interface.
In \S\ref{s:global-X}, we present global thermodynamic
quantities as a function  of the interface position. 
In \S\ref{s:problem-LG},
we address the main problem of thermodynamics for
the heat conduction systems with the liquid-gas coexistence.
Hereafter,  superscripts $\subL$ and $\subG$
are attached to quantities related to liquid and gas, respectively.

\subsection{Thermodynamics under equilibrium conditions}\label{s:LG-EQ}

\begin{figure}[tb]
\centering
\includegraphics[scale=0.7]{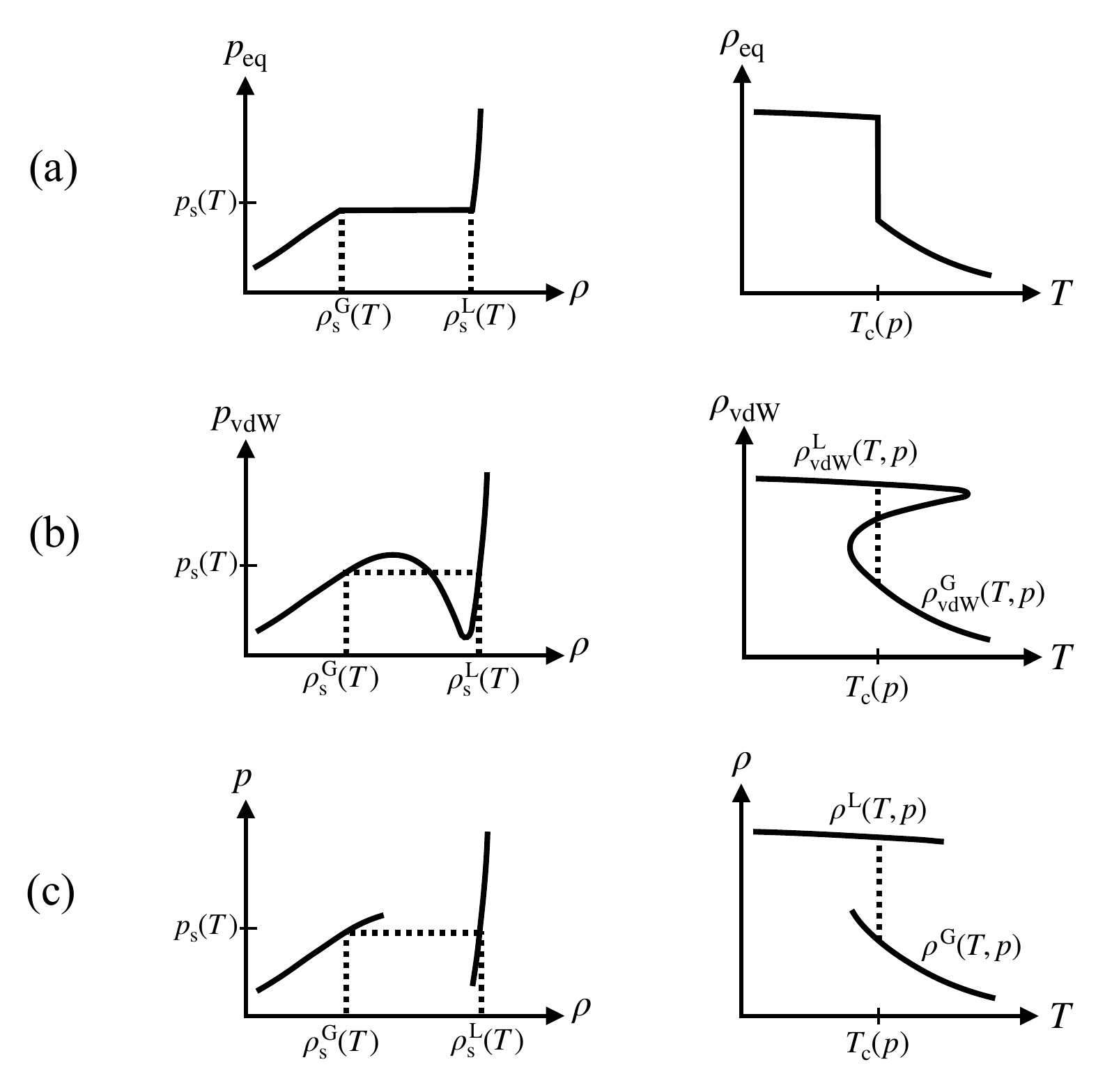}
\caption{Three kinds of equation of state: (a) Equilibrium pressure,
(b) van der Waals pressure, and (c) experimentally measured pressure.
$p_{\rm s}(T)$ is saturated pressure in liquid-gas
transition for a given $T$ below the critical temperature. 
$\Tc(p)$ is transition temperature for a given $p$.
$\rho^\subLG_{\rm s}$ denotes saturated density, which is defined by
$\rho^\subLG_{\rm s}=\rho^\subLG(T,p_{\rm s}(T))$.}
\label{fig:eos}
\end{figure}

We first consider equilibrium thermodynamics. For a given $(T,\rho)$,
the pressure $p$ is uniquely determined even when phase coexistence
is observed. This equation of state is expressed as
\begin{equation}
  p=p_{\rm eq}(T,\rho)
\label{e:eqnState-LG0}
\end{equation}
so as to emphasize that the pressure is the equilibrium value.
See Fig.~\ref{fig:eos}(a).  A remarkable fact is that there is a plateau
region $[\rho^\subG_{\rm s}(T),\rho^\subL_{\rm s}(T)]$ for a given
temperature $T$ below 
the critical temperature.  From this equation of state, we find
that the density $\rho$ exhibits the discontinuous change at $T=\Tc(p)$
when the temperature is changed with the pressure $p$ fixed, as shown
in Fig.~\ref{fig:eos}(a). That is, the whole system is occupied by
liquid when $T<\Tc(p)$, while by gas when $T>\Tc(p)$. 


In experiments, metastable states, which also may be called
quasi-equilibrium states, are often observed. 
The most typical approach to the
description of such meta-stable states is to employ the van der Waals
equation of state: 
\begin{align}
  p=&  p_\vdW(T,\rho) \nonumber \\
   =&  \frac{\rho k_{\rm B} T}{1-\rho b} -a \rho^2,
\label{v-dw}
\end{align}
where $k_\mathrm{B}$ is the Boltzmann constant. The constants $a$ and $b$ are van der Waals parameters.
When $T$ is fixed as that below the critical temperature,  the
equation of state contains the unstable states
$(\partial \rho/\partial p)_T <0$. See Fig.~\ref{fig:eos}(b). 
Following the standard procedure of the Maxwell construction, 
we obtain the equilibrium equation of state from (\ref{v-dw}).
Here, we extract the two stable branches satisfying
$(\partial \rho/\partial p)_T \ge 0$ from the curve defined by
(\ref{v-dw}), each of which is connected to the equilibrium liquid
phase  or to the equilibrium gas phase. By solving (\ref{v-dw})
in $\rho$ for each region, we obtain
\begin{align}
&\rho=\rho_\vdW^\subL(T,p)\quad\mbox{for liquid},\label{meta-vdW1}\\
&\rho=\rho_\vdW^\subG(T,p)\quad\mbox{for gas}.
\label{meta-vdW}
\end{align}
When $\rho_\vdW^\subG(T,p) > \rho^\subG_{\rm s}(T)$, the spatially homogeneous
phase of $(T,p)$ corresponds to a meta-stable gas. 
Similarly, when $\rho_\vdW^\subL(T,p) < \rho^\subL_{\rm s}(T)$, the homogeneous
phase of $(T,p)$ corresponds to a meta-stable liquid.
By plotting $\rho$ as a function of $T$ with $p$
fixed as shown in Fig.~\ref{fig:eos}(b), 
one finds that the meta-stable gas and the meta-stable liquid
are  super-cooled gas and  super-heated liquid, respectively.

More generally, without using the van der Waals equation of state,
we may obtain 
\begin{align}
&\rho=\rho^\subL(T,P)\quad\mbox{for liquid},\label{meta1}\\
&\rho=\rho^\subG(T,P)\quad\mbox{for gas},
\label{meta}
\end{align}
by experimental measurements. See Fig.~\ref{fig:eos}(c).
Below we assume (\ref{meta1}) and (\ref{meta}), whereas one may
interpret them  as (\ref{meta-vdW1}) and (\ref{meta-vdW}).
We then define densities for the liquid phase and the gas phase as
\begin{align}
a^\subL(T,p)\equiv a(T, \rho^\subL(T,p)),
\qquad
a^\subG(T,p)\equiv a(T, \rho^\subG(T,p)),
\end{align}
and define quantities per one particle for the liquid phase
and the gas phase as
\begin{align}
\hat a^\subL(T,p)\equiv \frac{a(T, \rho^\subL(T,p))}{\rho^\subL(T,p)},
\qquad
\hat a^\subG(T,p)\equiv \frac{a(T, \rho^\subG(T,p))}{\rho^\subG(T,p)}.
\end{align}

\subsection{Heat conduction states}\label{s:LG-HC}

\begin{figure}[tb]
\centering
\includegraphics[scale=0.65]{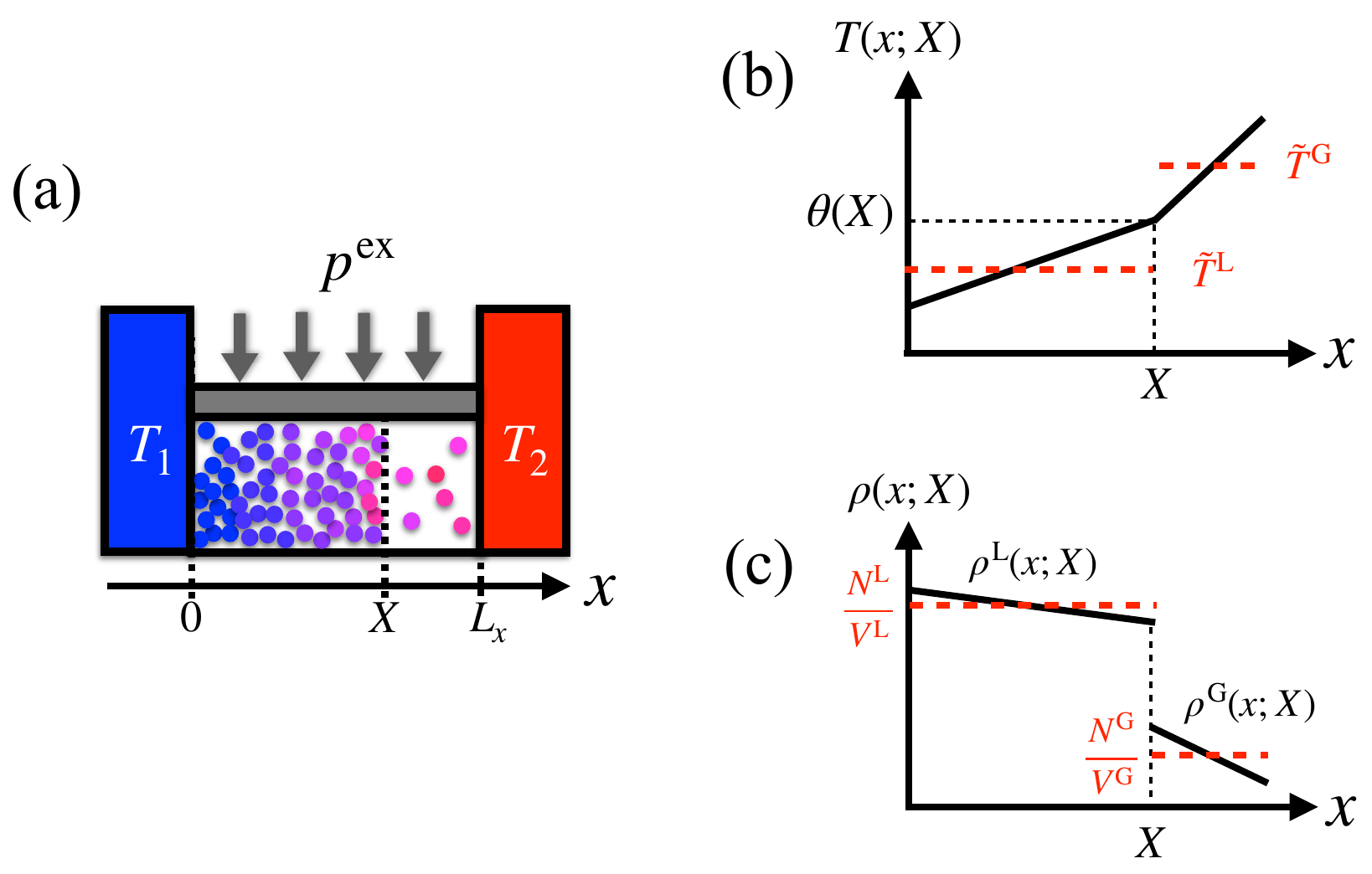}
\caption{(a) Schematic figure of  the liquid-gas coexistence
  with the interface position $X$.
  (b) Temperature profile $T(x;X)$ is continuous but not smooth at $x=X$,
  where the interface temperature satisfies $\theta(X)=T(X;X)$. The global temperatures in the liquid
  region and the gas region, $\bT^\subL$ and $\bT^\subG$, are displayed
  by the red dotted lines, respectively.
  See \eqref{e:bTLG}.
  (c) Density profile $\rho(x;X)$ shows a jump at $x=X$, where
  $\rho^\subL(x;X)\equiv\rho^\subL(T(x;X),p)$ and $\rho^\subG(x;X)\equiv\rho^\subL(T(x;X),p)$.
Since global thermodynamics holds for each region, the equation of state \eqref{e:global-eos} leads to   $\pex=p(\bT^\subL,N^\subL/V^\subL)+O(\ep^2)=p(\bT^\subG,N^\subG/V^\subG)+O(\ep^2)$.
}
\label{fig:FigLG}
\end{figure}

When $T_1<\Tc(\pex)<T_2$, we may observe liquid-gas coexistence
at the constant pressure $\pex$, which is in contrast with
equilibrium cases. Concretely, the liquid 
occupies a lower temperature region, while the gas does a higher
temperature region,  as shown in Fig.~\ref{fig:FigLG}(a). 
We assume that both the interface width and the temperature gap
at the interface are negligible. See Fig.~\ref{fig:FigLG}(b). 
Then, let $X$  denote the position of the interface. 
Although the value of $X$ should be uniquely determined under given conditions, 
we shall formally treat $X$ as if it were an independent variable in what follows. 
The density profile for a given $X$ is then expressed as
\begin{align}
\rho(x;X)=
\left\{
\begin{array}{lll}
\rho^\subL(T(x;X),\pex) \quad &~~\mbox{for}  & ~~0\le x<X,\\
\rho^\subG(T(x;X),\pex) \quad &~~\mbox{for}  & ~~X<x\le L_x.
\end{array}
\right.
\label{e:eqnState-LG2}
\end{align}
See Fig.~\ref{fig:FigLG}(b).
The temperature profile $T(x;X)$ is determined
by the heat conduction equation
\renewcommand{\arraystretch}{2.}
\begin{align}
J(X)=
\left\{
\begin{array}{lll}
  -\kappa^\subL(T(x;X), \pex)
  {\displaystyle\frac{\partial T(x;X)}{\partial x}}
  \quad &~~\mbox{for} & ~~0<x<X, \\
  -\kappa^\subG(T(x;X), \pex)
  {\displaystyle\frac{\partial T(x;X)}{\partial x}}
  \quad &~~\mbox{for} &~~ X<x<L_x,
\end{array}
\right.
\label{e:J-LG0}
\end{align}
where $J(X)$ is the steady heat current which
is constant in $x$ due to the conservation of energy.
The boundary conditions for the equation (\ref{e:J-LG0})
are 
\begin{align}
\lim_{x\rightarrow +0}T(x;X)=T_1,\quad \lim_{x\rightarrow L_x-0}T(x;X)=T_2. 
\end{align}
\renewcommand{\arraystretch}{1.5}

Concretely, $T(x;X)$ is obtained as follows.
We first express the interface temperature  as $\theta(X)$:
\begin{equation}
  \Tint(X)=T(X;X).
\end{equation}
By setting 
\begin{align}
\kappa^\subLG_\subC\equiv \kappa^\subLG(\Tc(\pex),\pex),
\end{align}
we find that the local heat conductivity is estimated as
\begin{align}
\kappa^\subLG(T(x), \pex)=\kappa^\subLG_\subC+O(\ep) .
\end{align}
Then,  \eqref{e:J-LG0} leads to 
\begin{align}
  J(X)=-\kappa^\subL_\subC\frac{\Tint(X)-T_1}{X}
  +O(\ep^2)=-\kappa^\subG_\subC\frac{T_2-\Tint(X)}{L_x-X}+\oet.
\label{e:J-LG} 
\end{align}
Solving the second equality of \eqref{e:J-LG} in $\Tint(X)$,
we obtain 
\begin{align}
  \Tint(X)=\mT
  +\frac{\Xi}{2}\left(\frac{X}{\kappa^\subL_\subC}
  -\frac{L_x-X}{\kappa^\subG_\subC}\right)\left(\frac{X}{\kappa^\subL_\subC}
  +\frac{L_x-X}{\kappa^\subG_\subC}\right)^{-1}+\oet.
\label{e:Tint-X}
\end{align}
Defining 
the scaled position $\zeta$ of the interface and an effective heat conductivity $\kappa_{\rm eff}(\zeta)$ as
\begin{align}
\zeta\equiv\frac{X}{L_x}, \qquad
  \kappa_{\rm eff}(\zeta)  \equiv\left(\frac{\zeta}{\kappa^\subL_\subC}
  +\frac{1-\zeta}{\kappa^\subG_\subC}\right)^{-1},
\end{align}
we have
\begin{align}
J(X)=-\kappa_{\rm eff}(\zeta)\frac{\Xi}{L_x}+\oet,
\label{e:J-X}
\end{align}
which is obtained by substituting \eqref{e:Tint-X} into \eqref{e:J-LG}.

By using $\Tint(X)$ given by \eqref{e:Tint-X}, the temperature
profile is written as 
\renewcommand{\arraystretch}{2.}
\begin{align}
T(x;X)=
\left\{
\begin{array}{ll}
T_1+{\displaystyle \frac{x}{X}}(\Tint(X)-T_1)+\oet \quad & \mbox{for} ~~0<x<X,\\
T_2-{\displaystyle \frac{L_x-x}{L_x-X}}(\Tint(X)-T_2)+\oet \quad & \mbox{for} ~~X<x<L_x.
\end{array}
\right.
\end{align}
\renewcommand{\arraystretch}{1.5}
Substituting this profile into the equation of
state \eqref{e:eqnState-LG2},
we have $\rho(x;X)$ explicitly as exemplified in Fig.~\ref{fig:FigLG}(c).

\subsection{Global quantities as a function of $X$
  } \label{s:global-X}

The system with the phase coexistence is interpreted as
a composite system of a liquid region and a gas region.
For each subsystem, we may define global thermodynamic
quantities introduced in \S\ref{s:single phase}.
In this section, we express global thermodynamic quantities
as a function of $X$.

First, the global temperatures in the liquid region and in the
gas region are defined by
\begin{align}
\bT^\subL=\frac{\intL \rho(x)T(x)}{\intL\rho(x)},\qquad
\bT^\subG=\frac{\intG \rho(x)T(x)}{\intG\rho(x)}.
\label{e:bTLG0}
\end{align}
In each region, the temperature profile and the density profile are smooth.
We describe each subsystem as a single phase system discussed
in \S\ref{s:single phase}. The boundary
temperatures of the liquid are $T_1$ and $\Tint$, and 
those of  the gas are $\Tint$ and $T_2$.  Thus, applying
the trapezoidal rule explained in \S\ref{s:trapezoidal} to
\eqref{e:bTLG0}, the global temperature in each region is
expressed as
\begin{align}
\bT^\subL(X)=\frac{T_1+\Tint(X)}{2}+\oet, \qquad
\bT^\subG(X)=\frac{\Tint(X)+T_2}{2}+\oet,
\label{e:bTLG}
\end{align}
where we explicitly write the dependence on the interface position $X$.
See Fig.~\ref{fig:FigLG}(b).

Next, the volume of the whole system $V(X)$ is expressed as
\begin{align}
V(X)&=L_x\frac{N}{\intx \rho(x;X)}\nonumber\\
&=\frac{N}{\zeta \rho^\subL(\bT^\subL(X),\pex)
  +(1-\zeta)\rho^\subG(\bT^\subG(X),\pex)}+\oet,
\label{e:V-X}
\end{align}
and thus the volume of the liquid region
$V^\subL(X)$ and the volume of the gas region 
$V^\subG(X)$ are written as 
\begin{align}
V^\subL(X)=\zeta V(X), \qquad V^\subG(X)=(1-\zeta) V(X).
\end{align}
Then, the particle number in the liquid region, $N^\subL$,
is obtained as
\begin{align}
N^\subL(X)&=V^\subL(X) \rho^\subL(\bT^\subL(X),\pex)\nonumber\\
&=N\frac{\zeta \rho^\subL(\bT^\subL(X),\pex)}{\zeta \rho^\subL(\bT^\subL(X),\pex)+(1-\zeta)\rho^\subG(\bT^\subG(X),\pex)}+\oet.
\label{e:NL-X}
\end{align}
The particle number in the gas region, $N^\subG$,
is immediately given by 
\begin{align}
N^\subG(X)=N-N^\subL(X).
\end{align}

By using $\bT^\subL(X)$, $\bT^\subG(X)$, and $N^\subL(X)$,
we can express all global thermodynamic quantities
in the  liquid region and the gas region. Explicitly, the extensive
quantities are defined as
\begin{align}
&\bA^\subL\equiv L_yL_z\intL a^\subL(T(x),\pex),\\
&\bA^\subG\equiv L_yL_z\intG a^\subG(T(x),\pex).
\end{align}
Noting \eqref{e:globalA-TP} in  \S\ref{s:single phase} with \eqref{e:A/N}, 
we conclude that the global extensive quantities are written as
\begin{align}
&\bA^\subL(X)=N^\subL(X) \hat a^\subL(\bT^\subL(X),\pex)+\oet,\\
&\bA^\subG(X)=N^\subG(X) \hat a^\subG(\bT^\subG(X),\pex)+\oet.
\end{align}
Finally, we define  the global temperature for the whole system
by the formula \eqref{e:globalT}, which is rewritten as 
\begin{align}
\bT(X)=\frac{N^\subL(X)}{N}\bT^\subL(X)+\frac{N^\subG(X)}{N}\bT^\subG(X).
\label{e:globalT-X}
\end{align}

\subsection{Problem}\label{s:problem-LG}


We have shown that all thermodynamic quantities are determined
for a given position $X$ of the liquid-gas interface. Since 
the liquid-gas interface in heat conduction may be at rest
at constant pressure, the position $X$ is uniquely
determined for $(T_1, T_2, \pex)$.


For equilibrium cases, the phase coexistence is observed in
a container with $V$ and $T$ fixed. When we start with the van der
Waals equation of state, we can determine the interface position 
of the two phases by the Maxwell construction, which is equivalent
to the continuity of the chemical potential at the interface.
Recalling this theory, one may impose the  continuity of the
local chemical potential $\mu(x)$ at the interface
even in heat conduction \cite{Bedeauz86}. 
Since $\mu^\subL(\Tc(p))=\mu^\subG(\Tc(p))$, it leads to $\theta=\Tc(\pex)$.
That is, under this assumption,
the interface temperature is equal to the equilibrium
transition temperature.


Here, the important observation is that  there is no justification
of this condition for heat conduction systems. To our best knowledge,
there are no experimental measurements on this issue, no numerical
simulations of sufficiently 
large systems, and no reliable theory for supporting  this condition.
To be more precise, although the continuity of the chemical potential
was reported in numerical simulations \cite{Bedeaux00,Ogushi},
the system size was
too small to draw a definite conclusion. One may also recall 
from a viewpoint of non-equilibrium statistical mechanics that 
the local equilibrium distribution may be the leading contribution
\cite{Bertini-rev}.
However, it
should be noted that  in the standard approach, fluctuations
are described as those around the most probable profile, while the most
probable profile itself is undetermined in the current problem. 
Therefore, the previous studies starting with the local equilibrium
distribution do not result in the continuity of the chemical potential
without imposing an additional assumption. 

Furthermore, as another
theoretical approach, one may study  the stationary solution of the
deterministic equation for the generalized Navier-Stokes equation
with the interface thermodynamics \cite{Bedeaux03,Onuki}.
We can estimate the discontinuous jump of the chemical
potential at the interface as $\ep (w/L_x) \Tc$ based
on reasonable assumptions \cite{SNIN}, where $w$ represents
the interface width. This result indicates the validity of
the continuity of the chemical potential at the interface,
because we may assume $w / L_x \ll 1$. However, as carefully
argued in the paper \cite{SNIN},  fluctuation effects
should be taken into account for the generalized Navier-Stokes
equation, so as to quantitatively
describe the thermodynamic behavior.


Now, based on these facts, we reconsider the continuity
of the chemical potential at the liquid-gas interface within the
framework of equilibrium thermodynamics. We then find that the
continuity is equivalent to the thermodynamic variational principle. 
If the principle were applied to a local sub-system including the interface,
the continuity of the chemical potential would be concluded. However,
the constraint of  a variational principle is generally applied to a whole system
but not to the local subsystem,
and therefore no variational principle is expected for the local subsystem near the interface.
Thus  we cannot justify the continuity of the chemical potential. This suggests an
alternative approach to determine the interface position $X$. Namely,
we formulate a variational principle for the  whole heat conduction
system  as a natural extension of the equilibrium variational principle.

\section{Variational principle for determining the
  liquid-gas interface position}\label{s:variational principle}

All our discussion on global thermodynamics so far has been firmly based on well-established local equilibrium thermodynamics.
Now, we formulate the variational principle for
determining the interface position $X$ and solve
the variational problem. In \S\ref{s:var}, we start
with the simplest example of the variational principle
for equilibrium systems and naturally extend it to 
that for heat conduction systems. In \S\ref{s:solution-var}
we derive the steady-state value by solving the minimization
of the variational function. The solution is then reformulated
as the {\it temperature relation} in \S\ref{s:Temp relation}.
By using these results, in \S\ref{s:T-Xi}, we express the global
quantities in terms of $(\tilde T, \pex, \Xi)$. 

\subsection{Variational principle}\label{s:var}

\subsubsection{Equilibrium systems}

We start with an example of the variational
principle for equilibrium systems. We consider a system
at the constant pressure $\pex$ in contact with a heat bath
of the temperature $T$. Since the volume $V$ of the system
is not fixed, we determine $V$ for a given $(T,\pex)$. When
the equation of state is assumed, it is
determined from the pressure balance equation
\begin{equation}
  \pex=p(T,V/N).
\label{balance}
\end{equation}

We attempt to derive this condition from a  variational principle. 
We take a variational function ${\cal G}({\cal V}; T,\pex,N)$ as
\begin{equation}
  {\cal G}({\cal V}; T,\pex,N)= F(T,{\cal V},N)+\pex {\cal V}.
\label{Gmin}
\end{equation}
This corresponds to the Helmholtz free energy
minimum principle for the composite
system of the system and the volume reservoir whose
Helmholtz free energy is $\pex V+{\rm const}$.
We then find that the minimization of ${\cal G}$
with respect to $\cal V$ for $(T,\pex,N)$ fixed leads to
(\ref{balance}). 
The volume in the equilibrium state satisfies
\begin{align}
  V=\argmin_{\cal V}  {\cal G}({\cal V}; T,\pex,N),
\label{eq-val}
\end{align}
with which the Gibbs free energy is determined as
\begin{align}
G(T,\pex,N) &= \min_{\cal V} {\cal G}({\cal V}; T,\pex,N)\\
&=F(T,V,N)+\pex V .
\end{align}
It should be noted that this principle may be applied to
the case that $F(T,V,N)$ is given by the van der Waals
free energy:
\begin{align}
  F(T,V,N) = \frac{3}{2}N k_{\rm B}T-a\frac{N}{V}
  - N k_{\rm B} T \log\frac{T^{3/2}(V-Nb)}{N^{5/2}}.
\end{align}
The variational principle for this case is equivalent
to the Maxwell construction.

\subsubsection{Extension to heat conduction systems}

\begin{figure}[tb]
\centering
\includegraphics[scale=0.5]{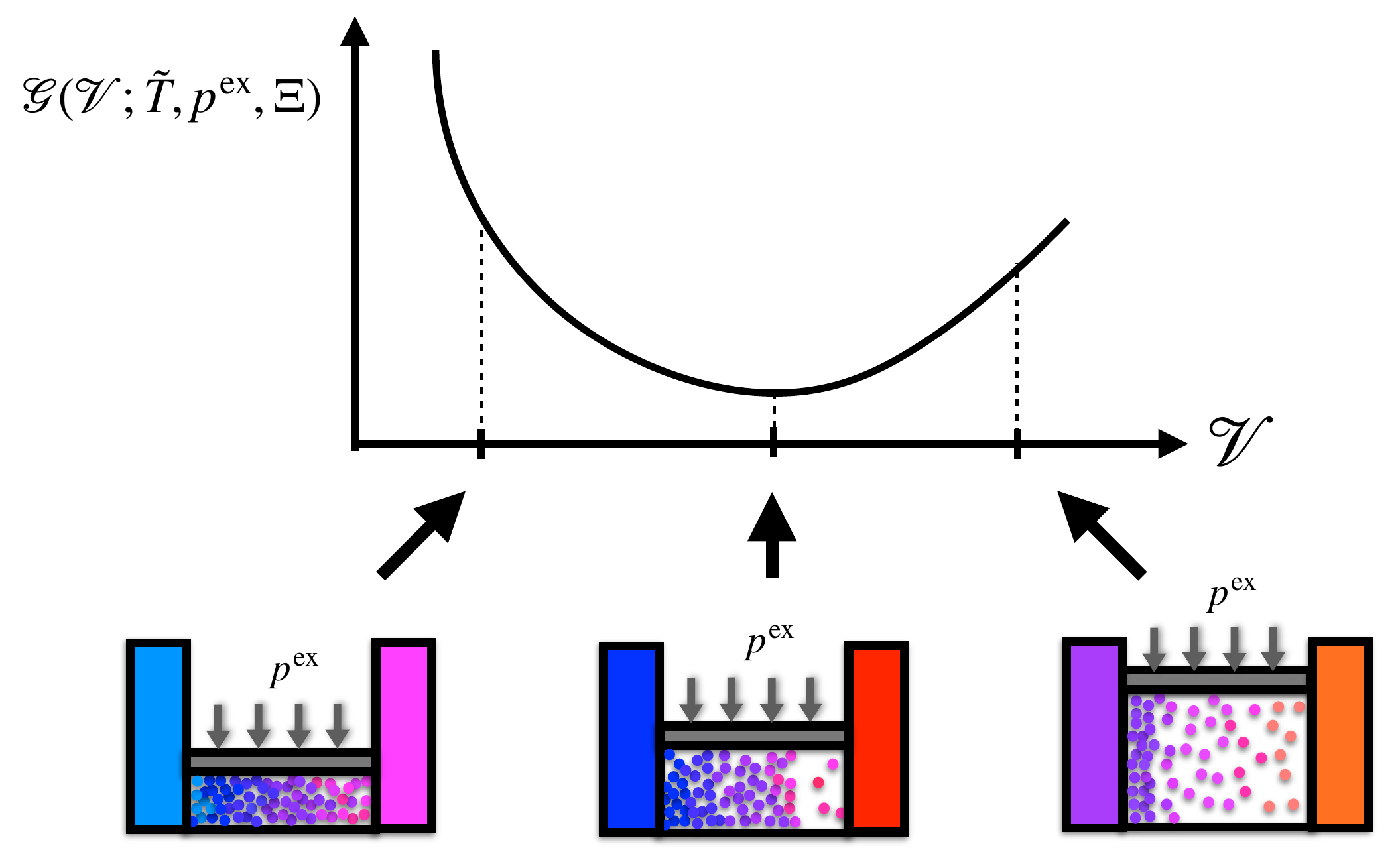}
\caption{The variational function ${\cal G}({\cal V})$
  with  $(\bT,\pex,\Xi)$ fixed for the heat conduction system. 
$T_1$ and $T_2$ may depend on ${\cal V}$ in order to fix $\tilde T$
  and $\Xi$. The configuration corresponding to the minimum
  of $\cal G$, the mid of the three configurations in the bottom
  of the figure, is selected as the steady state for $(\tilde T,\pex,\Xi)$.}
\label{fig:Fig-var}
\end{figure}

We study heat conduction systems at the constant pressure $\pex$. 
First, we focus on single-phase systems studied in \S\ref{s:single phase}. 
From the equivalence between a heat conduction system
and the corresponding equilibrium system, the volume $V$ of this
system is determined by the minimization of 
\begin{equation}
  {\cal G}({\cal V}; \bT,\pex,N)= \bF(\bT,{\cal V},N)+\pex {\cal V} ,
\label{Gmin-hc}
\end{equation}
where $\bT$ is the global temperature of the system. 
Since $\bF=F(\bT,V,N)+O(\ep^2)$, we rewrite (\ref{Gmin-hc}) as
\begin{equation}
  {\cal G}({\cal V}; \bT,\pex,N)=L_y L_z \int_0^{L_x}  dx [f(T(x),\rho(x))
    +\pex] +O(\ep^2),
\label{Gmin-hc-2}
\end{equation}
where $L_z={\cal V}/(L_xL_y)$ depends on $\cal V$ and 
$\bT$ should be fixed in the variation of $\cal V$. 
Then, the variational principle (\ref{eq-val}) holds
for single-phase systems. 
Since this formula relies on the equivalence to the equilibrium system, 
we do not need to fix the degree of nonequilibrium in the variation.

Next, we consider  a system with a liquid-gas interface. In this case,
$V$ is not determined when we use the equation of state for each
region, because the interface position $X$ is not determined yet
as described in the previous section. We then {\it assume} 
that the minimization 
of (\ref{Gmin-hc-2}) is also valid for such cases, where 
$\Xi=T_2-T_1$ is also fixed in the variation $V$.
The last property is necessary to determine
the value $V$ uniquely.  That is, we define a variational function
\begin{equation}
  {\cal G}({\cal V}; \bT,\pex,\Xi,N)=L_y L_z \int_0^{L_x}  dx [f(T(x),\rho(x))
    +\pex] 
\label{Gmin-hc-3}
\end{equation}
and propose that the volume of the system $V$ is determined as
\begin{align}
  &\left. \frac{\partial {\cal G}({\cal V}; \bT,\pex,\Xi,N)}
  {\partial {\cal V}}\right|_{{\cal V}=V}=0,
\label{e:variational principle V}
\end{align}
with
\begin{align}
  &\left. \frac{\partial^2 {\cal G}({\cal V}; \bT,\pex,\Xi,N)}
  {\partial {\cal V}^2}\right|_{{\cal V}=V}>0.
 \label{e:variational principle V2}
\end{align}
This variational principle was first proposed in Ref. \cite{NS}.

\subsubsection{Remarks}

Since the volume $V$ is uniquely determined by the interface position
$X$ with $(\bT,\pex,\Xi,N)$ fixed, as explicitly shown in  \eqref{e:V-X},
the variational principle \eqref{e:variational principle V}
with \eqref{e:variational principle V2} is  expressed as
\begin{align}
  \left. \frac{\partial {\cal G}({\cal X}; \bT,\pex,\Xi,N)}
       {\partial {\cal X}}\right|_{{\cal X}=X}=0,
\qquad
\left. \frac{\partial^2 {\cal G}({\cal X}; \bT,\pex,\Xi)}
     {\partial {\cal X}^2}\right|_{{\cal X}=X}>0.
\label{e:variational principle X}
\end{align}
Similarly,  since the particle number $N^\subL$ in the
liquid region is uniquely determined by $X$, we also have
another variational principle
\begin{align}
  \left. \frac{\partial {\cal G}({\cal N}^\subL; \bT,\pex,\Xi)}
       {\partial {\cal N}^\subL}\right|_{{\cal N}^\subL=N^\subL}=0,
\qquad
\left. \frac{\partial^2 {\cal G}({\cal N}^\subL; \bT,\pex,\Xi)}{\partial ({\cal N}^\subL)^2}\right|_{{\cal N}^\subL=N^\subL}>0.
\label{e:variational principle N}
\end{align}

As the second remark, we provide a physical interpretation of 
\eqref{e:variational principle X}. Let us consider a fluctuation
of  the interface position, $X\rightarrow X+\delta X$. We assume that
the motion of the interface is slowest and that the interface motion
can be observed in a hypothetical system where $(T_1, T_2)$ is
controlled such that $\bT$ and $\Xi$ are fixed. See Fig.~\ref{fig:Fig-var}.
It is then expected that
fluctuations in this hypothetical system may be described by equilibrium
statistical mechanics for the system with $(\bT, \pex)$. Thus, the
most probable position is given by the left equation of
\eqref{e:variational principle X} and the stability
of the interface position is expressed as the right inequality in
\eqref{e:variational principle X}.

\begin{figure}[bt]
\centering
\includegraphics[scale=0.5]{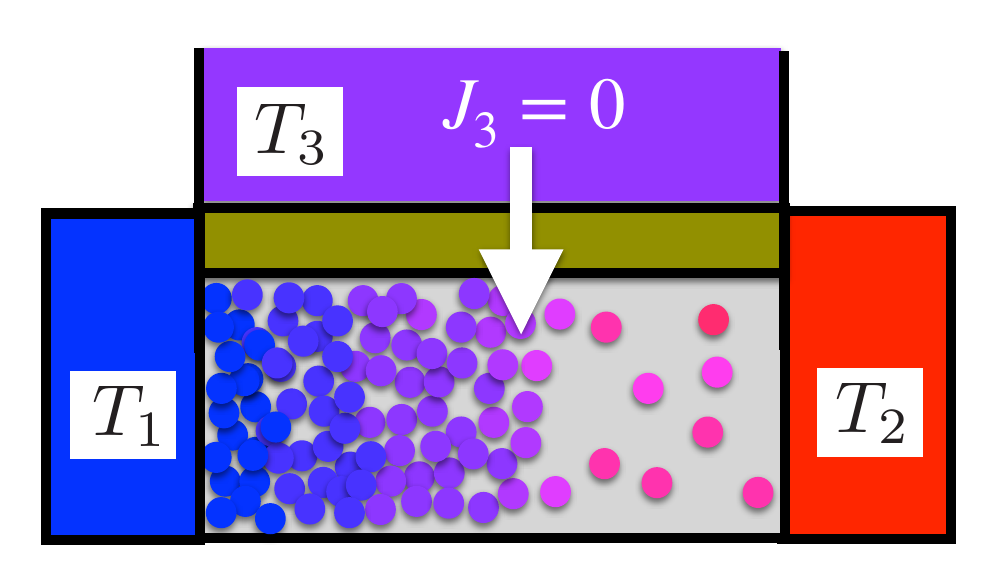}
\caption{Operational interpretation of the global temperature $\tilde T$.
A heat bath of $T_3$ contacts to the movable top plate.
When $T_3=\tilde T$, the heat current $J_3$ from this heat bath may vanish.
Then, the system behavior may be equivalent to the behavior of
the original system in Fig.~\ref{fig:Fig-setup}.}
\label{fig:Fig-T3}
\end{figure}

Moreover, we interpret $\bT$ from an operational viewpoint.
We virtually attach another heat bath  of the temperature $T_3$
to the rigid top plate as  shown in Fig.~\ref{fig:Fig-T3}.
Using the idealized assumption that
the motion of the top plate is sufficiently slower than the
other dynamical degrees of freedom, we control $T_1$ and $T_2$
with $\Xi=T_2-T_1$ fixed 
such that the heat flux $J_3$ to the heat bath of $T_3$ is zero.
Here, noting that the total
kinetic energy in the bulk is given by $N_\mathrm{deg} k_\mathrm{B} \bT/2$ 
with the total degrees $N_\mathrm{deg}$ of freedom in the system, 
$J_3$ is proportional to $\bT-T_3$ in the linear response
regime. That is, $J_3=0$ is equivalent to $\bT=T_3$.
When we focus on the position of the top plate, the motion
would be described as if it were in the equilibrium state
at the temperature $T_3$. From this picture for the 
specific setting, it is expected that 
\eqref{e:variational principle X} is equivalent to the equilibrium variational principle with $
\bT$.

As the last remark, we comment on the condition that $\Xi$ is fixed.
Since there are two temperatures $T_1$ and $T_2$, two variables
associated with $(T_1,T_2)$ should be fixed in the variation
of $\cal X$. Although we adopt $\Xi$ as a fixed variable in
addition to the global temperature $\tilde T$, one may conjecture
that $J$ would be a more plausible fixed-variable than $\Xi$.
As is formulated in \eqref{e:J-X}, 
$J$ is proportional to $\Xi$ but the proportional constant depends on the interface position.
Thus, the variational principle with $J$
fixed results in a different steady state from  the solution
of \eqref{e:variational principle V} with \eqref{e:variational principle V2}.
There are two reasons for our choice. The first is that
the final result becomes simplest among our trials.
Second, when $J$ is fixed in the variation, the enthalpy is
conserved in the variation. In this case, the corresponding
variational principle may be an extension of the equilibrium
variational principle for adiabatic (thermally isolated) systems.
From these two aspects, we assume that $\Xi$ is fixed in the
variation.

\subsection{Steady state determined from the variational principle}
\label{s:solution-var}

We solve the variational equation \eqref{e:variational principle N}.
Hereafter, we consider the variation of ${\cal N}^\subL$
with $(\bT,\pex,\Xi)$  fixed.  We abbreviate
$A({\cal N}^\subL;\bT,\pex,\Xi)$ as $A({\cal N}^\subL)$
or $A$ for the notational simplicity.

We express the variational function $\cal G$ as 
\begin{align}
  {\cal G}({\cal N}^\subL)
  \equiv \bF^\subL({\cal N}^\subL)+\bF^\subG({\cal N}^\subL)+\pex V({\cal N}^\subL),
\label{e:defG}
\end{align}
where the free energy of the liquid $F^\subL({\cal N}^\subL)$,
the free energy of the gas $F^\subG({\cal N}^\subL)$, and  the volume
of the system $ V({\cal N}^\subL)$ are given as functions of ${\cal N}^\subL$
with $(\bT,\pex,\Xi)$ fixed. 
Here, for any global quantity $A$ in each region,
such as $\bT^\subLG$ and $V^\subLG$, we define
\begin{align}
\delta \bA \equiv \bA({\cal N}^\subL+\delta {\cal N}^\subL)-\bA({\cal N}^\subL).
\label{e:definition variation}
\end{align}
We then have
\begin{align}
\delta {\cal G}=\delta F^\subL+\delta F^\subG +\pex\delta V.
\label{e:delta calG}
\end{align}
From the argument in \S \ref{s:single phase},  we obtain
\begin{align}
\bF^\subLG=F(\bT^\subLG, V^\subLG, N^\subLG).
\end{align}
Since every global thermodynamic relation obtained
in \S\ref{s:single phase} holds in each region, we have
\begin{align}
\bF^\subLG=\bmu^\subLG N^\subLG-\pex V^\subLG
\label{e:globalLeGendre-X}
\end{align}
with
\begin{align}
&\bmu^\subLG=\mu^\subLG(\bT^\subLG, \pex)+\oet.
\end{align}

Since the variation of the free energy in each region obeys
the fundamental relation of thermodynamics for $F$, we have
\begin{align}
  \delta F^\subLG=-S^\subLG\delta \bT^\subLG
  -\pex\delta V^\subLG+\bmu^\subLG\delta {\cal N}^\subLG,
\label{e:globalGibbs-X}
\end{align}
where
\begin{align}
&S^\subLG={\cal N}^\subLG\hat s^\subLG(\bT^\subLG, \pex)+\oet.
\end{align}
By using \eqref{e:globalGibbs-X}, we rewrite the variation of
${\cal G}$ in \eqref{e:defG}  as 
\begin{align}
  \delta {\cal G}=-S^\subL\delta\bT^\subL
  -S^\subG\delta\bT^\subG+(\bmu^\subL-\bmu^\subG)\delta {\cal N}^\subL,
\label{e:deltaG-X0}
\end{align}
where we used ${\cal N}^\subL+{\cal N}^\subG=N$ and $V^\subL+V^\subG=V$.

The global temperature $\bTL$ in the liquid region and $\bTG$ in
the gas region satisfy the relation
\begin{align}
\bTL=\bT-\frac{\Xi}{2}\frac{{\cal N}^\subG}{N}+\oet,\qquad
\bTG=\bT+\frac{\Xi}{2}\frac{{\cal N}^\subL}{N}+\oet,
\label{e:bT-bTL-bTR}
\end{align}
which are derived from the relations
\begin{align}
\bT=\frac{{\cal N}^\subL}{N}\bT^\subL+\frac{{\cal N}^\subG}{N}\bT^\subG,
\label{e:bT-LG}
\end{align}
and 
\begin{align}
\bT^\subG-\bT^\subL=\frac{\Xi}{2}+\oet.
\label{e:bTR-bTL}
\end{align}
Furthermore, (\ref{e:bT-bTL-bTR})  leads to
\begin{align}
  \delta\bT^\subL
  =\delta\bT+\frac{\Xi}{2}
  \delta\left(\frac{{\cal N}^\subL}{N}\right)
  -\frac{{\cal N}^\subG}{2N}\delta\Xi,
\label{e:bTL-bT}\\
\delta\bT^\subG=\delta\bT
+\frac{\Xi}{2}\delta\left(\frac{{\cal N}^\subL}{N}\right)
+\frac{{\cal N}^\subL}{2N}\delta\Xi.
\label{e:bTR-bT}
\end{align}
We have used ${\cal N}^\subL+{\cal N}^\subR=N$ in the first line.
Especially, the formulas \eqref{e:bTL-bT} and \eqref{e:bTR-bT},
with $\bT$, $\Xi$, and $N$ fixed, bring
\begin{align}
\delta \bTL=\delta \bTG=\frac{\Xi}{2}\frac{\delta {\cal N}^\subL}{N},
\label{e:dTLTG}
\end{align}
which simplifies \eqref{e:deltaG-X0}  as
\begin{align}
\delta {\cal G}=
\left(\bmu^\subL-\frac{\Xi}{2N}S^\subL-\bmu^\subG-\frac{\Xi}{2N}S^\subG
\right)\delta {\cal N}^\subL.
\label{e:deltaG-X1}
\end{align}
Here,  we confirm 
\begin{align}
\bmu^\subL-\frac{\Xi}{2N}S^\subL
&=
\mu^\subL(\bT^\subL,\pex)+\frac{\Xi}{2}\frac{{\cal N}^\subL}{N}\left(\frac{\partial \mu^\subL}{\partial T}\right)_{\pex}+\oet
\nonumber\\
&=\mu^\subL\left(\bT^\subL+\frac{\Xi}{2}\frac{{\cal N}^\subL}{N},\pex\right)+\oet,
\end{align}
and
\begin{align}
\bmu^\subG+\frac{\Xi}{2N}S^\subG
&=\mu^\subG\left(\bT^\subG-\frac{\Xi}{2}\frac{{\cal N}^\subG}{N},\pex\right)+\oet.
\end{align}

Let us define a specific temperature $T^\mathrm{s}$ as
\begin{align}
  T^\mathrm{s}({\cal N}^\subL) &
  \equiv\bT+\Xi\left(\frac{{\cal N}^\subL}{N}-\frac{1}{2}\right).
\label{e:Ts}
\end{align}
By using \eqref{e:bT-bTL-bTR}, we rewrite \eqref{e:Ts} as 
\begin{align}
T^\mathrm{s}({\cal N}^\subL) 
&=\bT^\subL({\cal N}^\subL)+\frac{\Xi}{2}\frac{{\cal N}^\subL}{N}=\bT^\subG({\cal N}^\subL)-\frac{\Xi}{2}\frac{{\cal N}^\subG}{N}.
\label{e:Ts-bTLG}
\end{align}
Thus, the variation \eqref{e:deltaG-X1} is further simplified as
\begin{align}
\delta {\cal G}=\left(\mu^\subL(T^\mathrm{s},\pex)-\mu^\subG(T^\mathrm{s},\pex)+\oet
\right)\delta {\cal N}^\subL.
\end{align}
This is equivalent to 
\begin{align}
\frac{\partial {\cal G}({\cal N}^\subL; \bT,\pex,\Xi)}{\partial {\cal N}^\subL}
=
\mu^\subL\left(T^\mathrm{s},\pex\right)
-
\mu^\subG\left(T^\mathrm{s},\pex\right)
+\oet.
\label{e:dGdN}
\end{align}
Since the functional form of $\mu^\subG$ is different from that
of $\mu^\subL$ due to the crucial difference between the liquid and the gas,
$\mu^\subL\left(T^\mathrm{s},\pex\right)$ is not identically equal
to $\mu^\subG\left(T^\mathrm{s},\pex\right)$. The equality
\begin{align}
\left. \mu^\subL(T^\mathrm{s},\pex)\right|_{{\cal N}^\subL=N^\subL}
=\left. \mu^\subG(T^\mathrm{s},\pex)\right|_{{\cal N}^\subL=N^\subL}+O(\ep^2)
\label{e:sol1}
\end{align}
holds only when
\begin{align}
\left. T^\mathrm{s}\right|_{{\cal N}^\subL=N^\subL}=\Tc (\pex)+O(\ep^2).
\label{e:Ts-Tc}
\end{align}
Now, \eqref{e:Ts} and  \eqref{e:Ts-Tc} yield the  unique value $N^\subL$
as 
\begin{align}
\frac{N^\subL(\bT,\pex,\Xi)}{N}=
\frac{1}{2}+\frac{\Tc(\pex)-\bT}{\Xi}+O(\ep).
\label{e:NL*}
\end{align}
Thus we conclude that 
the variational principle \eqref{e:variational principle N}
results in the unique steady value $N^\subL$
formulated by \eqref{e:NL*}.

Next, we consider the second derivative of $\cal G$.
By using \eqref{e:dGdN} with \eqref{e:Ts}, 
the second derivative of $\cal G$ is obtained  as
\begin{align}
  \frac{\partial^2 {\cal G}({\cal N}^\subL; \bT,\pex,\Xi)}
       {\partial ({\cal N}^{\subL})^2}
&=
\left[
\left(\frac{\partial \mu^\subL}{\partial T^{\rm s}}\right)_{\pex}
-
\left(\frac{\partial \mu^\subG}{\partial T^{\rm s}}\right)_{\pex}
\right]
\frac{dT^\mathrm{s}}{d N^\subL}
+\oet
\nonumber\\
&=
(\hat s^\subG(T^\mathrm{s},\pex)-\hat s^\subL(T^\mathrm{s},\pex))\frac{\Xi}{N}+\oet .
\end{align}
At the steady state satisfying \eqref{e:Ts-Tc},
the above entropy difference is connected to the latent heat $\hat q$
per one particle 
\begin{align}
  \hat q(\pex)= \Tc(\pex)\left[\hat s^\subG(\Tc(\pex),\pex)
    -\hat s^\subL(\Tc(\pex),\pex)\right].
\label{e:latent-heat}
\end{align}
Therefore, we estimate the second derivative as
\begin{align}
  \left.\frac{\partial^2 {\cal G}({\cal N}^\subL; \bT,\pex,\Xi)}
       {\partial ({\cal N}^{\subL})^2}
\right|_{{\cal N}^\subL=N^\subL}
&=
\frac{\hat q(\pex)}{N}\frac{\Xi}{\Tc(\pex)}+\oet,
\label{e:dGdN2}
\end{align}
Since $\Xi >0$ and $\hat q(\pex)>0$,
we conclude that
\begin{align}
  \left. \frac{\partial^2 {\cal G}({\cal N}^\subL; \bT,\pex,\Xi)}
       {\partial ({\cal N}^{\subL})^2}\right|_{{\cal N}^\subL=N^\subL}> 0
\label{e:stabilty}
\end{align}
in the linear response regime.

\subsubsection{Careful analysis of $\ep \to 0$} \label{s:eptozero-var}

\begin{figure}[bt]
\centering
\includegraphics[scale=0.5]{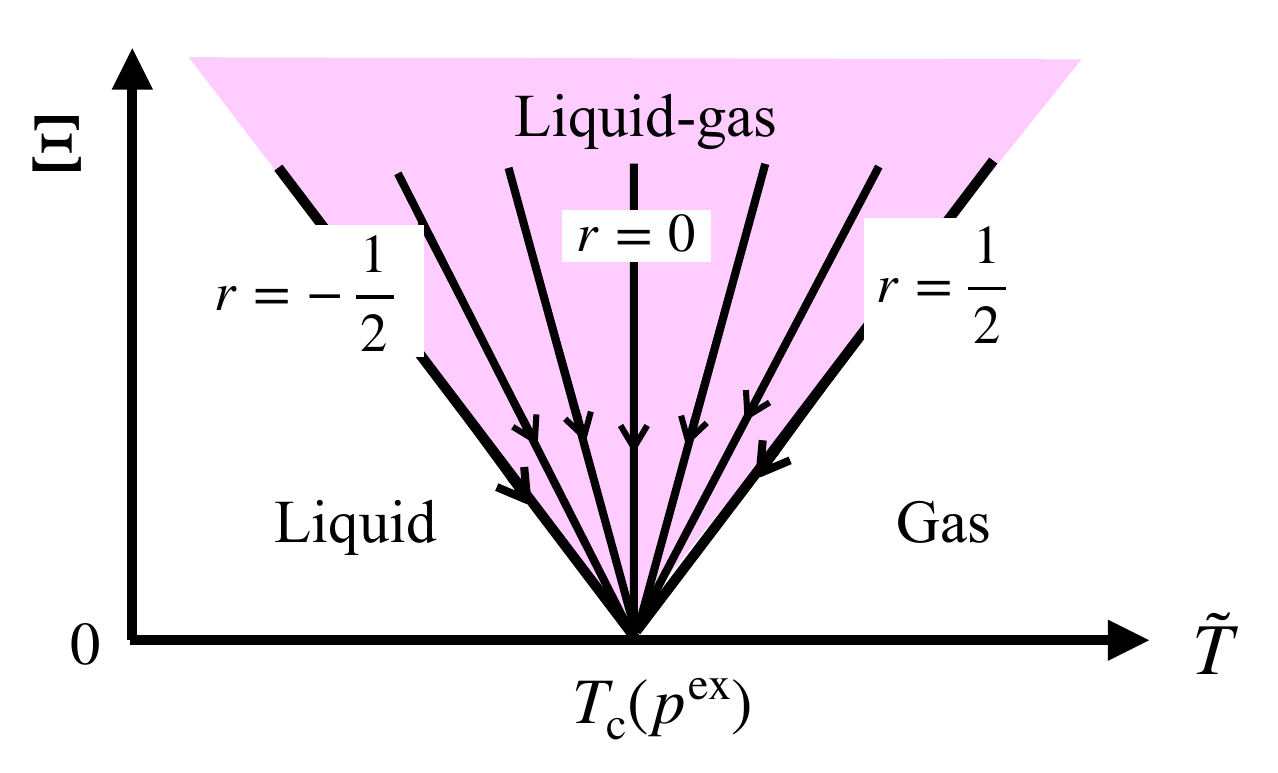}
\caption{Phase diagram for a given $\pex$.
  For equilibrium cases ($\Xi=0$), the system shows the first-order
  transition at $\bT=\Tc(\pex)$. The system is occupied by the liquid 
  when $\bT>\Tc(\pex)-{\Xi}/{2}$ and by the gas when
  $\bT<\Tc(\pex)+{\Xi}/{2}$. The liquid and the gas coexist when
  $\Xi>0$ and  $\Tc(\pex)-{\Xi}/{2}<\bT<\Tc(\pex)+{\Xi}/{2}$
  painted by pink. The ratio of the liquid and the gas 
  is kept along the line corresponding to each value of $r$,
  which goes to $T=\Tc(\pex)$ at $\Xi=0$. }
\label{fig:Fig-r}
\end{figure}

The solution \eqref{e:NL*} of the variational principle
includes an undetermined term of $O(\ep)$. Nevertheless,
\eqref{e:NL*} provides additional information to equilibrium
behavior. To clarify the situation, 
we consider the limit $\ep \to 0$ with keeping the phase coexistence.
This may be formalized by fixing a parameter $r$ defined as
\begin{align}
r\equiv \frac{\bT-\Tc(\pex)}{\Xi},
 \label{e:r}
\end{align}
whereas  \eqref{e:NL*}  indicates 
\begin{align}
\frac{N^\subL}{N}=\frac{1}{2}-r.
\label{N-r}
\end{align}
As $0< N^\subL<N$, $r$ satisfies
\begin{align}
-\frac{1}{2}<r <\frac{1}{2}.
\end{align}
In Fig.~\ref{fig:Fig-r}, we draw  straight lines connecting
from $(\bT,\Xi)=(\Tc+r\Xi, \Xi)$ to $(\Tc, 0)$ for several values of $r$
in the parameter region where the liquid-gas coexistence is observed.
The convergence of these lines at $(\Tc,0)$ indicates that the equilibrium
state $(\Tc,0)$ behaves as a singular point. By specifying the value of
$r$, we identify the corresponding line which undoes the degeneration
at equilibrium. $N^\subL$ is determined by considering the equilibrium
limit $\ep\rightarrow 0$, while it is not uniquely determined at equilibrium.

\subsection{Temperature relation}\label{s:Temp relation}

\begin{figure}[b]
\centering
\includegraphics[scale=0.6]{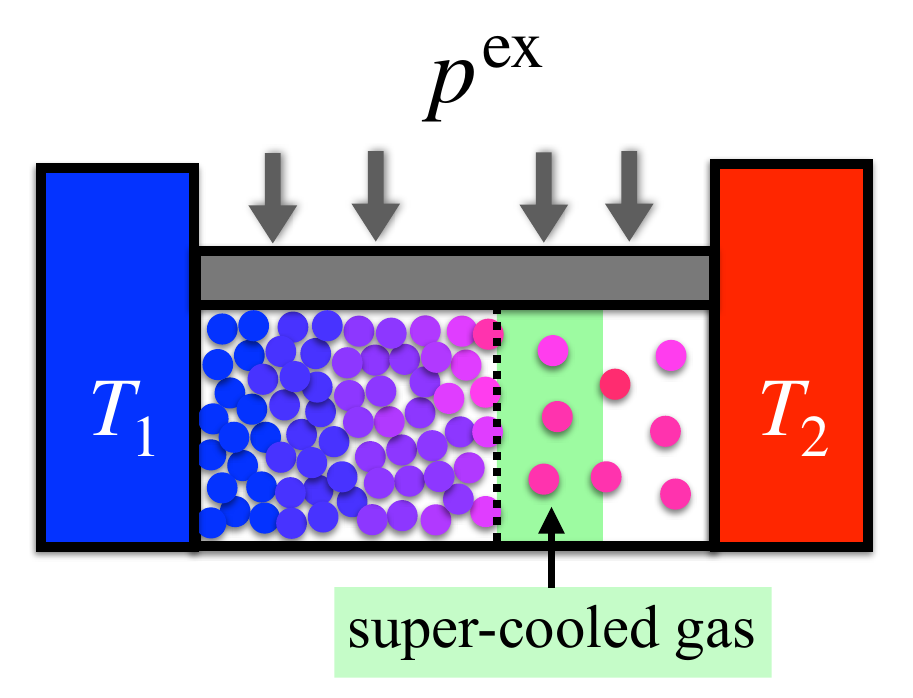}
\caption{Super-cooled gas stabilized by the steady  heat current.}
\label{fig:Fig-meta}
\end{figure}

We obtained the steady state value of $N^\subL$ as \eqref{e:NL*}
by solving the variational principle \eqref{e:variational principle N}. 
Substituting this solution into \eqref{e:Ts-bTLG} with \eqref{e:Ts-Tc},
we have the global temperature for each region as
\begin{align}
\bT^\subLG(\bT,\pex,\Xi)=\frac{\Tc(\pex)+\bT}{2}\mp\frac{\Xi}{4}+\oet.
\label{e:bTLG*}
\end{align}
We sum up the two relations in \eqref{e:bTLG*} and
substitute \eqref{e:bTLG} into $\bT^\subLG$.
We then obtain
\begin{align}
\bT-{\mT}=\Tint-\Tc(\pex)+\oet,
\label{e:tempRelation}
\end{align}
which we call {\it the temperature relation}. 
This non-trivial relation, which  results from the variational principle,
is a simple  condition that the steady states satisfy.
Generally, $\bT$ is not equal to ${\mT}$, because the particle density
of liquid is larger than that of gas, $\rho^\subL>\rho^\subG$. 
Therefore, the relation \eqref{e:tempRelation} implies that the liquid-gas coexistence temperature   $\Tint$ in a heat conduction system must be strictly lower than the equilibrium coexistence temperature $\Tc(\pex)$.
This means that metastable states stably appear near the liquid-gas
interface as schematically shown in Fig.~\ref{fig:Fig-meta}.

Now, suppose that \eqref{e:tempRelation} holds.
Summing up  the two equalities in \eqref{e:bT-bTL-bTR},
we obtain
\begin{align}
\mT+\Tint=2 \bT +\frac{\Xi}{2}\left( \frac{2N^\subL}{N}-1 \right).
\end{align}
Here, it should be noted that we do not use the variational
principle for the derivation of \eqref{e:bT-bTL-bTR}. 
By substituting the temperature relation \eqref{e:tempRelation} into
this result, we derive \eqref{e:NL*} without the variational
principle. In this sense, we may say that the temperature relation
\eqref{e:tempRelation} is equivalent to the variational principle.

\subsection{Global quantities as functions of $(\bT,\pex,\Xi)$}\label{s:T-Xi}

We determine the steady state values of several quantities.
Let $\Vs^\subLG(\pex)$ denote  the equilibrium saturated volume defined
as
\begin{align}
\Vs^\subLG(\pex)=\frac{N}{\rho^\subLG_\subC(\pex)}
\end{align}
with
\begin{align}
  \rho^\subLG_\subC(\pex)=\rho^\subLG(\Tc(\pex),\pex).
\end{align}  
The volume of the liquid (gas) region is given by 
\begin{align}
V^\subLG(\bT,\pex,\Xi)&=\frac{N^\subLG}{N}\Vs^\subLG(\pex)+O(\ep).
\label{e:VLG*}
\end{align}
Since $N^\subL$ is determined  with an error of $O(\ep)$
in \eqref{e:NL*},  \eqref{e:VLG*} is the  best estimate of $V^\subLG$
in $\ep \rightarrow 0$. Substituting \eqref{e:NL*} into \eqref{e:VLG*},
we obtain
\begin{align}
&V(\bT,\pex,\Xi)\nonumber\\
&=
\left(\frac{1}{2}-\frac{\bT-\Tc(\pex)}{\Xi}\right) \Vs^\subL(\pex)
+\left(\frac{1}{2}+\frac{\bT-\Tc(\pex)}{\Xi}\right) \Vs^\subG(\pex)+O(\ep).
\label{e:V*}
\end{align}
which corresponds to the solution of the  variational principle
\eqref{e:variational principle V}.
Then, the steady position $X$ of the liquid-gas interface,
\begin{align}
\frac{X}{L_x}=\frac{V^\subL}{V},
\end{align}
is determined as
\begin{align}
\frac{X(\bT,\pex,\Xi)}{L_x} 
=\frac{\displaystyle\frac{1}{2}-\frac{\bT-\Tc(\pex)}{\Xi}}
{\displaystyle\frac{1}{2}\left(\frac{\rho^\subL_\subC}{\rho^\subG_\subC}+1\right)
+\frac{\bT-\Tc(\pex)}{\Xi}\left(\frac{\rho^\subL_\subC}{\rho^\subG_\subC}-1\right)}
+O(\ep),
\label{e:X*}
\end{align}
which is the solution of the
variational principle \eqref{e:variational principle X}.

Then,  the interface temperature is obtained from
the elimination of $\mT$ from \eqref{e:tempRelation} and \eqref{e:Tint-X} as
\begin{align}
\Tint(\bT,\pex,\Xi)=\frac{\Tc(\pex)+\bT}{2}+\frac{\Xi}{4}\left(\frac{X}{\kappa^\subL_\subC}-\frac{L_x-X}{\kappa^\subG_\subC}\right)\left(\frac{X}{\kappa^\subL_\subC}+\frac{L_x-X}{\kappa^\subG_\subC}\right)^{-1}+\oet.
\label{e:Tint-X*-bT}
\end{align}
The deviation of $\theta$ from $\Tc$ and the relation between
$X$ and $\theta$ will be demonstrated in \S \ref{s:steady LG} for explicit  examples.

\section{Properties of the steady states with a
  liquid-gas interface}\label{s:steady LG}

In \S\ref{s:T-Xi}, we derived the steady state values 
for given $(\bT, \Xi)$ such as  $N^\subL=N^\subL(\bT,\pex,\Xi)$
in \eqref{e:NL*}, $V=V(\bT,\pex,\Xi)$ in \eqref{e:V*}
and $X=X(\bT,\pex,\Xi)$ in \eqref{e:X*}.
However, the global temperature $\bT$ is not easily controlled
in experiments, differently from the heat bath temperatures $T_1$ and $T_2$.
In this section, we express any quantity in the
steady state as a function of $(T_1, T_2, \pex)$.
In \S\ref{s:T1-T2}, we derive $(\tilde T, X)$ as a function
of $(T_1, T_2,\pex)$. Since we already have 
\begin{align}
A=A(X;\bT,\pex,\Xi),
\end{align}
the relations 
\begin{align}
\bT&=\bT(T_1,T_2,\pex), \label{bT-res}  \\
X&=X(T_1,T_2,\pex), \label{zeta-res}
\end{align}
lead to  the expression
\begin{align}
A=A(T_1, T_2,\pex,\Xi).
\end{align}
Particularly,  we consider the interface temperature $\theta$ and the jump of the chemical potential $\Delta \mu$ at the interface as a function
of $(T_1, T_2,\pex)$ in \S\ref{s:Tint in T1-T2} as \eqref{e:Tint-zeta*}
and \S\ref{s:dmu in T1-T2} as \eqref{e:dMuint},
respectively.  In \S\ref{s:local meta-stable}, we discuss how
the super-cooled gas appears near the interface quantitatively.
The last subsection \S\ref{s:vdW} is devoted to the demonstration
of examples.  
In what follows, we  use the notations
\begin{align}
u\equiv\frac{\rho^\subL_\subC(\pex)}{\rho^\subG_\subC(\pex)}, \qquad 
v\equiv\frac{\kappa^\subL_\subC(\pex)}{\kappa^\subG_\subC(\pex)} ,\qquad
\zeta\equiv\frac{X}{L_x}
\end{align}
for simplicity. 

\subsection{$\tilde T$ and $\zeta$ as functions of $(T_1,T_2,\pex)$}
\label{s:T1-T2}

Substituting \eqref{e:Tint-X*-bT} into \eqref{e:tempRelation}, we obtain
\begin{align}
\bT=2\mT-\Tc(\pex)+\frac{\Xi}{2}
\frac{\zeta-v(1-\zeta)}{\zeta+v(1-\zeta)}
+\oet,
\label{e:Tint-X*-T1T2}
\end{align}
where the dimensionless position $\zeta$ of the steady interface
is
\begin{align}
\zeta=
\frac{\displaystyle\frac{1}{2}-\frac{\bT-\Tc(\pex)}{\Xi}}
{\displaystyle\frac{1}{2}\left(1+u\right)
-\frac{\bT-\Tc(\pex)}{\Xi}\left(1-u\right)}
+O(\ep),
\label{e:X*-T1T2}
\end{align}
as is derived in \eqref{e:X*}.
Eliminating $\bT$ from \eqref{e:Tint-X*-T1T2} and \eqref{e:X*-T1T2},
we have
\begin{align}
  \mT-\Tc(\pex)=-\frac{\Xi}{2}
  \frac{u \zeta^2-v(1-\zeta)^2}{[u\zeta+(1-\zeta)][\zeta+v(1-\zeta)]}+\oet.
\label{e:implicit-zeta*}
\end{align}
Solving this equation in $\zeta$, we derive
\begin{align}
  \zeta(T_1,T_2,\pex)=\frac{tu+(1-t)v+t(u-1)(v-1)
    -\sqrt{(uv-1)^2t^2+uv}}{v-u+2t(u-1)(v-1)}+O(\ep),
\label{e:zeta*}
\end{align}
with
\begin{align}
t\equiv \frac{\mT-\Tc(\pex)}{\Xi}.
\end{align}
\eqref{e:zeta*} provides a concrete functional form of (\ref{zeta-res}).

Substituting \eqref{e:implicit-zeta*} into \eqref{e:Tint-X*-T1T2}, 
we obtain a  concrete form of (\ref{bT-res}),
\begin{align}
  \bT(T_1,T_2,\pex)
=\mT -\frac{\Xi}{2}\frac{(uv-1)\zeta(1-\zeta)}{[u\zeta+(1-\zeta)][\zeta+v(1-\zeta)]}+\oet,
\label{e:bT*-T1T2}
\end{align}
with the form of $\zeta$ in (\ref{e:zeta*}).

\subsection{Interface temperature as a function of $(T_1,T_2,\pex)$}\label{s:Tint in T1-T2}

Applying the temperature relation \eqref{e:tempRelation} to \eqref{e:bT*-T1T2}, 
we find that the interface temperature is written as
\begin{align}
\Tint(T_1,T_2,\pex)=\Tc(\pex)-\frac{\Xi}{2}\frac{(uv-1)\zeta(1-\zeta)}{[u\zeta+(1-\zeta)][\zeta+v(1-\zeta)]}+\oet,
\label{e:Tint-zeta*}
\end{align}
where $\zeta$ is given by \eqref{e:zeta*}.
This formula clarifies that the interface temperature $\Tint(T_1,T_2,\pex)$
generally deviates from $\Tc(\pex)$ when the temperature gradient
is imposed. Since  the particle density satisfies $\rho^\subL > \rho^\subG$,
i.e., $u > 1$,  and the heat conductivity is expected to be
$\kappa^\subL\neq \kappa^\subG$,  the interface temperature $\Tint$ is deviated
from the equilibrium transition temperature $\Tc(\pex)$
in the order of $\ep$, which is not negligible.

We have another formula for the interface temperature $\Tint$,
which is expressed solely by experimentally accessible quantities:
\begin{align}
\Tint=\Tc(\pex)-\frac{\zeta(1-\zeta)}{2}
\left[
-JL_x\left(\frac{1}{\kappa^\subG_\subC}-\frac{1}{\kappa^\subL_\subC}\right)
+
\Xi\bar\phi\left(\rho^\subL_\subC-{\rho^\subG_\subC}\right)
\right]
+\oet,
\label{e:Tint}
\end{align}
where $\bar\phi=V/N$. The derivation of this formula is shown in \S\ref{s:derive Tint}.

Up to here, we have studied the systems with $T_1<T_2$,
where $\Xi> 0$ and $J<0$. More generally, from
the left-right symmetry of the system in Fig.~\ref{fig:Fig-setup},
we notice that the interface temperature is invariant for the transformation $(\Xi,\zeta)$ to $(-\Xi,1-\zeta)$.
That is, 
\eqref{e:Tint} is expressed as
\begin{align}
&\Tint=\Tc(\pex)-\frac{\zeta(1-\zeta)}{2}
\left[
|J| L_x\left(\frac{1}{\kappa^\subG_\subC}-\frac{1}{\kappa^\subL_\subC}\right)
+
|\Xi|\bar\phi\left(\rho^\subL_\subC-{\rho^\subG_\subC}\right)
\right]
+\oet
\end{align}
for any $\Xi$.

One might guess that the coexistence temperature $\theta$ is uniquely determined by local quantities that characterize the phase boundary, namely, the pressure $\pex$ and the heat current per unit area.
Rather interestingly, this is not the case. 
Reflecting the global nature of our variational principle, the coexistence temperature $\theta$ explicitly depends on global conditions of the system, namely, the temperatures of the two heat baths.

\subsubsection{Derivetion of \eqref{e:Tint}} \label{s:derive Tint}

We first recall \eqref{e:J-LG} and \eqref{e:Tint-X}. From these,
\begin{align}
&\Xi =-J L_x\left(\frac{\zeta}{\kappa^\subL_\subC}+\frac{1-\zeta}{\kappa^\subG_\subC}\right)+\oet,\label{e:J-Xi-LG}\\
&\mT-\Tint=\frac{J L_x}{2}\left(\frac{\zeta}{\kappa^\subL_\subC}-\frac{1-\zeta}{\kappa^\subG_\subC}\right)+\oet.
\end{align}
Combining these two relations, we have another form of $\mT$ as
\begin{align}
\mT-\Tint
=\frac{\Xi}{2}+\zeta\frac{J L_x}{\kappa^\subL_\subC}
=-\frac{\Xi}{2}-(1-\zeta)\frac{J L_x}{\kappa^\subG_\subC}
\label{e:mT-Tint}
\end{align}
with an error of $\oet$.

Second, the result of the variational principle \eqref{e:NL*}
is written as 
\begin{align}
\bT-\Tc(\pex)
=\frac{\Xi}{2}-\Xi\frac{N^\subL}{N}+\oet
=-\frac{\Xi}{2}+\Xi\frac{N^\subG}{N}+\oet.
\label{e:bT-Tc0}
\end{align}
As $N^\subL=\zeta V \rho^\subL(\bTL,\pex)$
and $N^\subG=(1-\zeta) V\rho^\subG(\bTG,\pex)$
with $\rho^\subL(\bTL,\pex)=\rho^\subL_\subC+O(\ep)$ and 
$\rho^\subG(\bTG,P)=\rho^\subG_\subC+O(\ep)$,
the relation \eqref{e:bT-Tc0} is further expressed as
\begin{align}
\bT-\Tc(\pex)
=\frac{\Xi}{2}-\zeta\Xi\bar\phi\rho^\subL_\subC
=-\frac{\Xi}{2}+(1-\zeta)\Xi\bar\phi\rho^\subG_\subC
\label{e:bT-Tc-estimate}
\end{align}
with an error of $\oet$.
By subtracting \eqref{e:mT-Tint} from  \eqref{e:bT-Tc-estimate} and using
the temperature relation \eqref{e:tempRelation}, we have two forms
of the interface temperature:
\begin{align}
  \Tint-\Tc(\pex)&=-\frac{\zeta}{2}\left(\frac{J L_x}
           {\kappa^\subL_\subC}+\Xi\bar\phi \rho^\subL_\subC\right), \label{213}\\
           \Tint-\Tc(\pex)&=\frac{1-\zeta}{2}\left(\frac{J L_x}
                    {\kappa^\subG_\subC}+\Xi\bar\phi \rho^\subG_\subC\right).
                    \label{214}
\end{align}
By substituting \eqref{213} and \eqref{214} into the first term and the
second term of the right-hand side of the trivial identity
\begin{align}
\Tint-\Tc(\pex)=(1-\zeta) (\Tint-\Tc(\pex))+\zeta (\Tint-\Tc(\pex)),
\end{align}
we obtain \eqref{e:Tint}.

\subsection{$\Delta \mu$ as a function of $(T_1,T_2,\pex)$}\label{s:dmu in T1-T2}

Next, we discuss the chemical potential jump 
\begin{align}
\Delta\mu\equiv\mu^\subG(\Tint,\pex)-\mu^\subL(\Tint,\pex)
\label{e:Muint-def}
\end{align}
at the interface. Note that 
\begin{align}
\mu^\subG(\Tc(\pex),\pex)=\mu^\subL(\Tc(\pex),\pex),
\end{align}
which gives the definition of the transition temperature
$\Tc$. Thus, $\Tint \not = \Tc$ means  the imbalance
of the chemical potential at the interface, which is
quantitatively expressed as
\begin{align}
\Delta\mu
=-[\hat s^\subG(\Tc(\pex),\pex) -\hat s^\subL(\Tc(\pex),\pex)]
(\Tint-\Tc(\pex))+\oet.
\label{e:Muint}
\end{align}
Since the entropy difference is connected to the
latent heat $\hat q$ per one particle as
\begin{align}
\hat s^\subG(\Tc(\pex),\pex) -\hat s^\subL(\Tc(\pex),\pex)=
\frac{\hat q}{\Tc(\pex) },
\end{align}
the jump of the chemical potential is proportional to $\hat q$.
Now, substituting \eqref{e:Tint-zeta*} into \eqref{e:Muint},
the amount of the jump is estimated as
\begin{align}
\Delta\mu
&=
\frac{\hat q}{2}\frac{\Xi}{\Tc(\pex) }
\frac{(uv-1)\zeta(1-\zeta)}{[u\zeta+(1-\zeta)][\zeta+v(1-\zeta)]}+\oet,
\label{e:dMuint}
\end{align}
Another expression of $\Delta \mu$ is also obtained by 
substituting  \eqref{e:Tint} into \eqref{e:Muint}. The result is
\begin{align}
\Delta\mu
&=
\frac{\zeta(1-\zeta)}{2}
\frac{\hat q}{\Tc(\pex) }
\left[
-JL_x\left(\frac{1}{\kappa^\subG_\subC}-\frac{1}{\kappa^\subL_\subC}\right)
+
\Xi\bar\phi\left(\rho^\subL_\subC-{\rho^\subG_\subC}\right)
\right]
+\oet.
\end{align}

\subsection{
Super-cooled gas 
in the  liquid-gas coexistence
}\label{s:local meta-stable}

\begin{figure}[tb]
\centering
\includegraphics[scale=0.6]{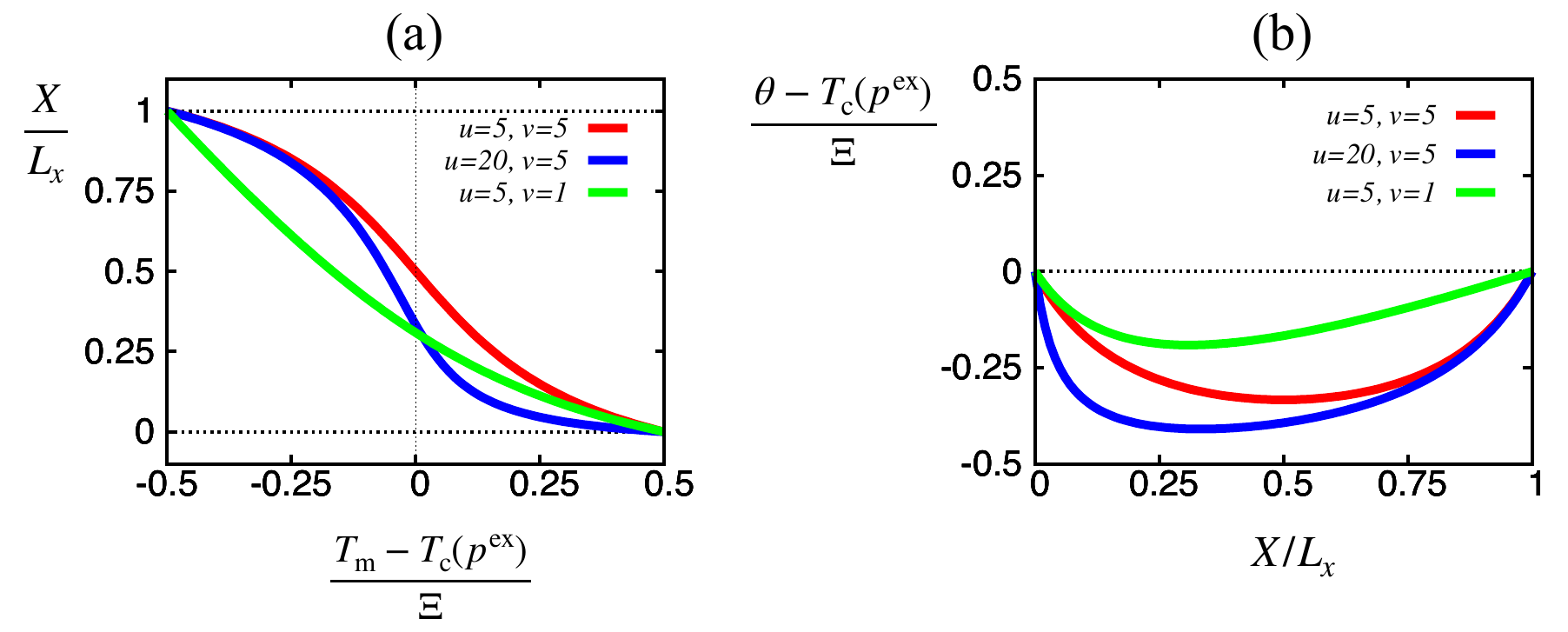}
\caption{Theoretical  results for the parameter values
  $(u,v)=(5,5)$, $(20,5)$ and $(5,1)$.
(a) The position of the liquid-gas interface in  \eqref{e:zeta*}
when $\mT$ deviates from $\Tc(\pex)$.
(b) The deviation of  the interface temperature $\Tint$ in  \eqref{e:Tint-zeta*}. All the three cases show $\Tint<\Tc(\pex)$, which indicates
the stabilization of the  super-cooled gas near the interface in
the heat conduction.}
\label{fig:Fig6}
\end{figure}

When $T_1<\Tc(\pex)<T_2$, we find the position $\xc$ at which
the local temperature satisfies $T(\xc;T_1,T_2,\pex)=\Tc(\pex)$.
Suppose $\Tint <\Tc(\pex)$. Then, the  position $X$ of the interface 
satisfies 
\begin{align}
0<\xint<\xc <L_x.
\end{align}
We observe the liquid in the region $0<x<\xint$ and the gas
in $\xint<x<L_x$.
Then the local temperature of the gas in $\xint<x<\xc$
is less than $\Tc(\pex)$. This means that a super-cooled gas
is observed in $\xint<x<\xc $, which is not stable in equilibrium.

We plot \eqref{e:zeta*} and \eqref{e:Tint-zeta*} in Fig.~\ref{fig:Fig6}
for three sets of $(u, v)$. It is clearly seen that the interface
temperature $\Tint$ deviates from the equilibrium transition
temperature $\Tc(\pex)$. 
The deviation of the global temperature $\bT$ from $\mT$ shows
the same figure with Fig.~\ref{fig:Fig6}(b)
due to the temperature relation \eqref{e:tempRelation}.
The jump of the chemical potential at the interface also
exhibits the same dependence on $\zeta=X/L_x$ 
since it is proportional to $\Tint-\Tc(\pex)$ as shown in \eqref{e:dMuint}.
In the presented  three sets of $(u,v)$, a super-cooled gas region appears
near the right side of the interface,
as shown schematically in Fig.~\ref{fig:Fig-meta}.

According to Fig.~ \ref{fig:Fig6}(a), the stable position of
the interface is shifted continuously as we increase the
temperature of both heat baths. This feature is different
from the numerical solution of the variational equation reported
in Ref. \cite{NS}, where the interface discontinuously appears.
A more accurate numerical calculation was necessary for the cases
with high  $\rho^\subL_\subC/\rho^\subG_\subC$.

\subsection{Examples}
\label{s:vdW}

In this subsection, we illustrate some examples of the liquid-gas
coexistence in heat conduction quantitatively,
where the particle density is measured in mol.

\begin{figure}[tb]
\centering
\includegraphics[scale=0.6]{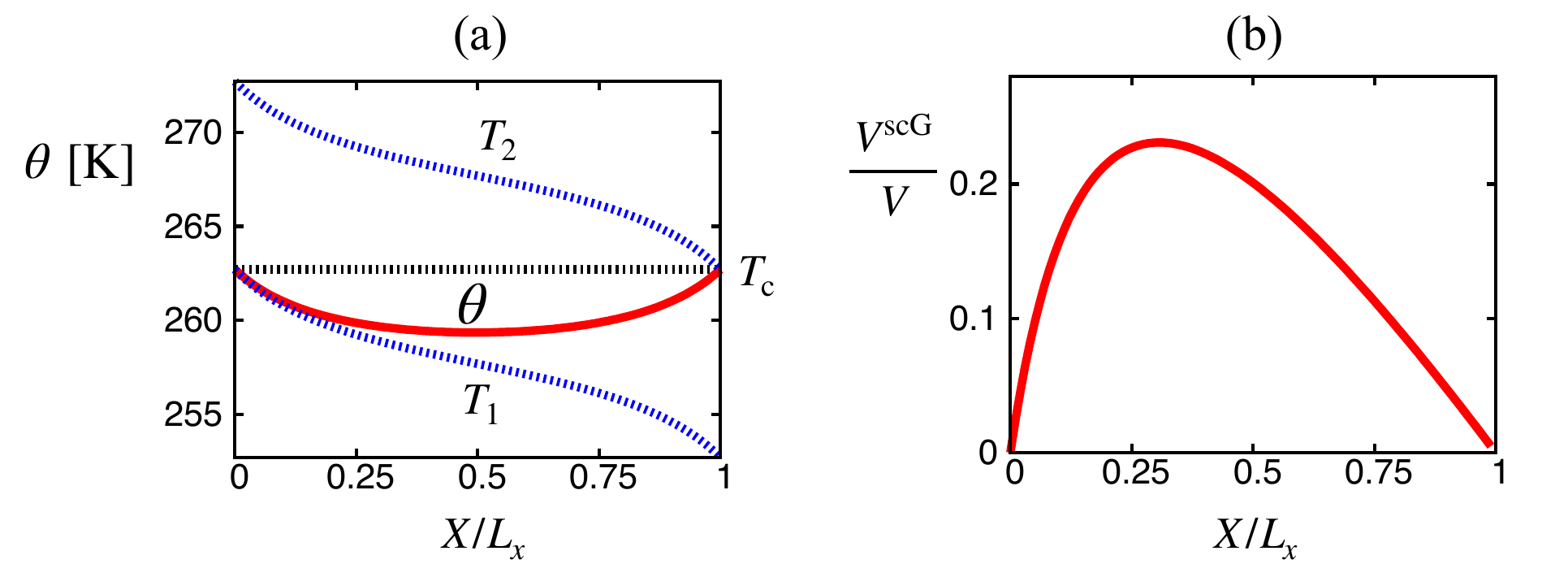}
\caption{(a) The interface temperature $\theta$
  and (b) volume fraction of the super-cooled gas
  for ${\rm CO}_2$ under $\pex=4\times 10^6$ Pa.
  $\Xi=10$ K.
  See the main text for parameter values. In (a),
  $\Tint$ (red solid line), $\Tc$ (black dotted line),
  $T_1$ and $T_2$ (blue dotted lines) are depicted
  simultaneously as a function  of the interface position.}
\label{fig:Fig-vdW}
\end{figure}

\begin{figure}[tb]
\centering
\includegraphics[scale=0.6]{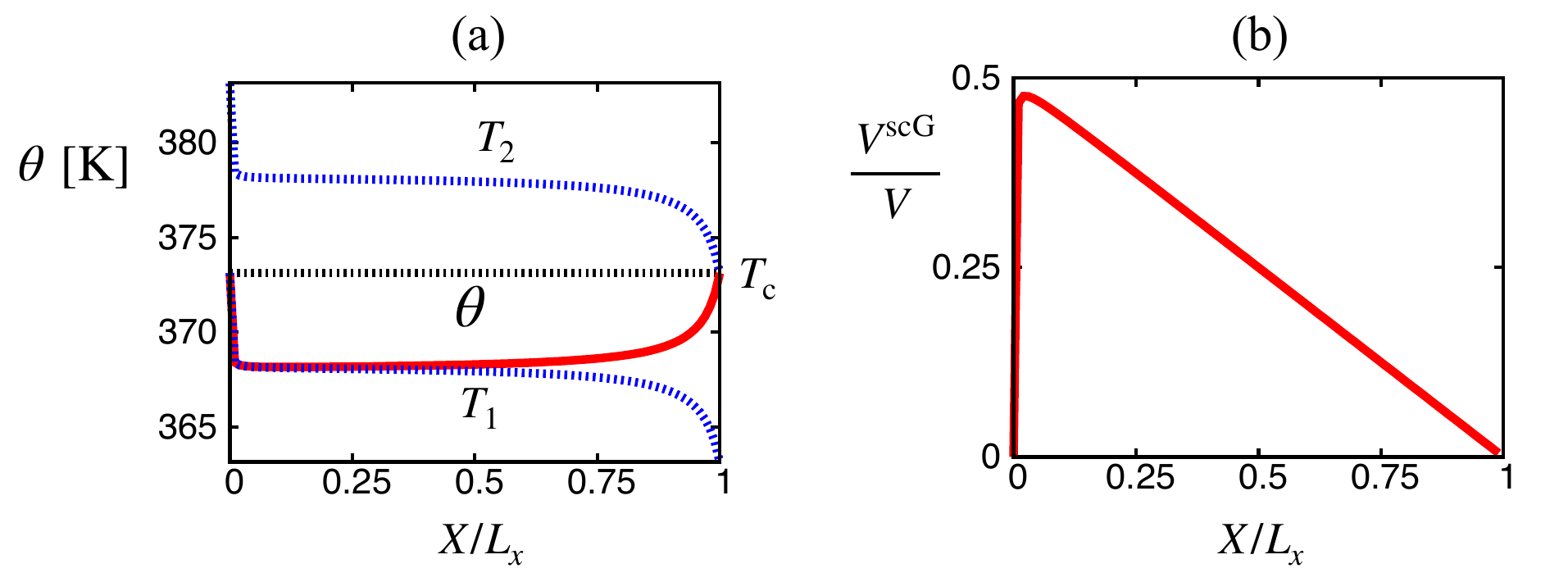}
\caption{(a) The interface temperature $\theta$
  and (b) volume fraction of the super-cooled gas
  for ${\rm H}_2{\rm O}$ under $\pex=1.013\times 10^5$ Pa.
  $\Xi=10$ K.
  See the main text for parameter values. In (a),
  $\Tint$ (red solid line), $\Tc$ (black dotted line),
  $T_1$ and $T_2$ (blue dotted lines) are depicted
  simultaneously as a function  of the interface position.}
\label{fig:Fig-H2O}
\end{figure}

First, we consider  the van der Waals equation of state
\begin{equation}
  p=\frac{RT \rho}{1- b \rho}-a \rho^2
\label{e:vdW}
\end{equation}  
of  $a=0.365$ Pa$\cdot$ m$^6$/mol$^2$ and $b=4.28\times 10^{-5}$ m$^3$/mol
for ${\rm CO}_2$ at a high pressure $\pex=4\times 10^6$ Pa \cite{CO2}.
The gas constant is $R=8.31$ J/K$\cdot$mol.  Substituting these parameters
into \eqref{e:vdW}, we obtain the transition temperature as $\Tc=262.7$ K,
and the mol  density at $\Tc$ as $\rho^\subL_\subC=1.38\times 10^4$ mol/m$^3$
and $\rho^\subG_\subC=2.70\times 10^3$ mol/m$^3$, which results in $u=5.1$.
From the database \cite{NIST}, we set the heat conductivity of the liquid
as $\kappa^\subL=0.1$ W/m$\cdot$ K and that of the gas as
$\kappa^\subG=0.02$ W/m$\cdot$ K, which yields $v=5$. The result is close
to the result $(u,v)=(5.5)$ which was already shown in Fig.~ \ref{fig:Fig6}. 
The temperature profile becomes linear both in the liquid and gas
regions in this example due to the constant heat conductivity.
Then the  volume fraction of the super-cooled gas is simply obtained as
\begin{align}
\frac{V^\mathrm{scG}}{V}
=\frac{L_x-X}{L_x}\frac{\Tc-\Tint(X)}{T_2(X)-\Tint(X)},
\end{align}
where $V^\mathrm{scG}$ is the volume of the super-cooled gas,
i.e., $V^{\rm scG}=(X-\xc)/L_x$. The interface temperature
$\Tint(X)$ is given by \eqref{e:Tint-zeta*} 
and $T_2$ results from \eqref{e:implicit-zeta*} as
\begin{align}
T_2(X)=\Tc(\pex)+\frac{\Xi}{2}-\frac{\Xi}{2}
  \frac{u \zeta^2-v(1-\zeta)^2}{[u\zeta +(1-\zeta)][\zeta+v(1-\zeta)]}+\oet,
\end{align}
where $\zeta=X/L_x$.
In Fig. \ref{fig:Fig-vdW}, we show the
interface temperature $\theta$ and the volume fraction $V^\mathrm{scG}/V$
as a function of the position of the interface when $\Xi=10$ K. We notice that the
interface temperature $\theta$ may deviate about $3.3$ K from
$\Tc$, and that the volume fraction of the super-cooled gas may exceed $1/3$.

Second, more importantly, we do not need the equation of state,
but have only to know the value of $u$, $v$, $\Tc(\pex)$, when we
obtain the phase diagram for a given material. For example, consider
pure water at $1.013 \times 10^5$ Pa, where $\Tc(\pex)=373.1$ K. 
From the database \cite{NIST}, we have $u=1604$ and $v=27.06$.  
From these, we can predict the interface temperature $\Tint$
and the volume fraction $V^\mathrm{scG}/V$ as a function of the
position of the interface as shown in Fig.~\ref{fig:Fig-H2O}.
As an example, when $T_1=368.0$ K and $T_2=378.0$ K, we obtain $\Tint=368.3$ K
and $X/L_x=0.4086$.

\section{Global thermodynamics for systems with a liquid-gas interface} 
\label{s:liquid-gas phase constant P}

Hereafter, all the quantities are defined in the steady state.
When a global quantity $\bA$ is determined for $(\bT, p, N,\Xi)$,
we express this relation as
\begin{align}
&\bA=A(\bT,p,N,\Xi).
\end{align}
Recall that $A(\bT,p,N,\Xi)=A(\bT,p,N)+O(\ep^2)$
for single-phase systems,
where $A(\bT,P,N)$ is the equilibrium function,
as discussed in \S\ref{s:single phase}.
For the systems with a liquid-gas interface, $\Xi$ dependence
appears even in the linear response regime.
In this section, we extend the global thermodynamics
introduced in \S\ref{s:single phase}
so as to describe the liquid-gas coexistence determined
in the previous sections.  In \S\ref{s:dAL+dAG},  we derive
formulas as preliminaries for later sections.  In \S\ref{s:der-dG},
we formally derive a fundamental relation of thermodynamics 
for systems with a liquid-gas interface.
However, since the derivative of $G$ is not defined at
$(\Tc(p),p)$, the formal expression is not properly defined. Then, in
\S\ref{s:eptozero}, we perform  a careful analysis of the limit
$\ep \to 0$. In \S\ref{s:entropy-LG-P}, we show the form of the entropy
and the volume in the appropriate limit.  In \S\ref{s:heat capacity},
we present formulas for  constant pressure heat capacity and  compressibility.

\subsection{Preliminaries} \label{s:dAL+dAG}

We first study  global extensive quantities 
\begin{align}
\bA=L_y L_z\intx  a(T(x),p).
\end{align}
It is obvious that they possess the additivity
\begin{equation}
  \bA=\bA^\subL+\bA^\subG,
  \label{e:A-add}
\end{equation}  
where $\bA^\subL$ and $\bA^\subG$ are defined as
global thermodynamic quantities in the liquid region and
the gas region, respectively.
By using $\bT^\subL$ and $\bT^\subG$, we  write
\begin{align}
  &\bA^\subL=  N^\subL\hat a^\subL(\bT^\subL,p)+O(\ep^2),  \label{e:AL}\\
    &\bA^\subG =N^\subG\hat a^\subG(\bT^\subG,p)+O(\ep^2).  \label{e:AG}
\end{align}

Now, we consider an infinitely small change
$(T_1, T_2, p, N)\rightarrow (T_1+\delta T_1, T_2+\delta T_2,p+\delta
p, N+\delta N)$,
which results in
\begin{equation}
 (\bT, p, N,\Xi)
 \to (\bT+\delta \bT, p+\delta p, N+\delta N,
 \Xi+\delta \Xi).
\end{equation}
Accordingly, any quantity $\phi$ for the steady state $(\bT, p, N,\Xi)$
changes as
\begin{equation}
\delta \phi =
\phi(\bT+\delta \bT, p+\delta p,
N+\delta N,\Xi+\delta \Xi)-\phi(\bT, p, N,\Xi).
\end{equation}
Then, \eqref{e:AL} and \eqref{e:AG} bring the following relations
among the infinitely small changes:
\begin{align}
\delta \bA^\subL
&=
N^\subL \left(\frac{\partial \hat a^\subL}
{\partial \bT^\subL}\right)_{p} \delta \bT^\subL
+
N^\subL \left(\frac{\partial \hat a^\subL}
{\partial p}\right)_{\bT^\subL} \delta p
+\hat a^\subL \delta N^\subL+O(\ep^2),
\label{e:dAL}
\\
\delta \bA^\subG
&=
N^\subG
\left(\frac{\partial \hat a^\subG}
{\partial \bT^\subG}\right)_{p} \delta \bT^\subG
+
N^\subG
\left(\frac{\partial \hat a^\subG}
    {\partial P}\right)_{\bT^\subG} \delta p
+\hat a^\subG \delta N^\subG+O(\ep^2).
\label{e:dAR}
\end{align}
We sum up the above two relations \eqref{e:dAL} and \eqref{e:dAR} and
substitute \eqref{e:bTL-bT} and \eqref{e:bTR-bT} into them.
Then, the sum becomes 
\begin{align}
\delta \bA &=
\left(
N^\subL \left(\frac{\partial \hat a^\subL}{\partial \bT^\subL}\right)_{p} 
+
N^\subG \left(\frac{\partial \hat a^\subG}{\partial \bT^\subG}\right)_{p} 
\right)
\left(\delta \bT+\frac{\Xi}{2}\delta\left(\frac{N^\subL}{N}\right)\right)
\nonumber\\
&+
\left(
N^\subL \left(\frac{\partial \hat a^\subL}{\partial p}\right)_{\bT^\subL} 
+
N^\subG
\left(\frac{\partial \hat a^\subG}{\partial p}\right)_{\bT^\subG} 
\right)\delta p
\nonumber\\
&-\frac{N^\subL N^\subG}{2N}
\left(\left(\frac{\partial \hat a^\subL}{\partial \bT^\subL}\right)_{p} 
-\left(\frac{\partial \hat a^\subG}{\partial \bT^\subG}\right)_{p} \right)\delta\Xi
\nonumber\\
&+\hat a^\subL\delta N^\subL +\hat a^\subG\delta N^\subG+\oet.
\label{e:add-dsum}
\end{align}
The fourth line is transformed into
\begin{align}
\hat a^\subL\delta N^\subL +\hat a^\subG\delta N^\subG
=
\left(\frac{N^\subL}{N}\hat a^\subL+\frac{N^\subG}{N}\hat a^\subG\right)
\delta N
+
(\hat a^\subL-\hat a^\subG) N\delta\left(\frac{N^\subL}{N}\right).
\label{e:add-dsum2}
\end{align}
By summarizing the terms proportional to $\delta(N^\subL/N)$ in
\eqref{e:add-dsum} and \eqref{e:add-dsum2}, the coefficient
of $N\delta(N^\subL/N)$ becomes 
\begin{align}
&\hat a^\subL
+
\frac{\Xi}{2}\frac{N^\subL}{N} \left(\frac{\partial \hat a^\subL}{\partial \bT^\subL}\right)_{p} 
-\hat a^\subG
+
\frac{\Xi}{2}\frac{N^\subG}{N} \left(\frac{\partial \hat a^\subG}{\partial \bT^\subG}\right)_{p} 
\nonumber\\
&~~~~
=\hat a\left(\bT+\frac{\Xi}{2}\frac{N^\subL-N^\subG}{N},p\right)
-\hat a\left(\bT+\frac{\Xi}{2}\frac{N^\subL-N^\subG}{N},p\right)+\oet.
\label{e:add-dsum3}
\end{align}
Here, we note that the sum of  
the two relations for $\bT^\subL$ and $\bT^\subG$
in \eqref{e:bT-bTL-bTR} results in
\begin{align}
\bT+\frac{\Xi}{2}\frac{N^\subL-N^\subG}{N}
&=\bT^\subL+\bT^\subG-\bT
\nonumber\\
&=\mT-\bT+\Tint+\oet ,
\label{e:T(X)}
\end{align}
where the transformation from the first line to the second
line is found by noting
$\bT^\subL=(T_1+\Tint)/2+\oet$ and $\bT^\subG=(T_2+\Tint)/2+\oet$
from \eqref{e:mT-bT}. Now, using the temperature relation
\eqref{e:tempRelation}, we find that
the right-hand side of (\ref{e:T(X)}) is $\Tc(p)$. We thus obtain
\begin{align}
\delta \bA&=
\left(
N^\subL \left(\frac{\partial \hat a^\subL}{\partial \bT^\subL}\right)_{p} 
+
N^\subG \left(\frac{\partial \hat a^\subG}{\partial \bT^\subG}\right)_{p} 
+\oet
\right)
\delta \bT
\nonumber\\
&+
\left(
N^\subL \left(\frac{\partial \hat a^\subL}{\partial p}\right)_{\bT^\subL} 
+
N^\subG
\left(\frac{\partial \hat a^\subG}{\partial p}\right)_{\bT^\subG} 
+\oet
\right)\delta p
\nonumber\\
&+\left(\frac{N^\subL}{N}\hat a^\subL+\frac{N^\subG}{N}\hat a^\subG+\oet\right)
\delta N
\nonumber\\
&-\frac{N^\subL N^\subG}{2N}
\left(\left(\frac{\partial \hat a^\subL}{\partial \bT^\subL}\right)_{p} 
-\left(\frac{\partial \hat a^\subG}{\partial \bT^\subG}\right)_{p}
+O(\ep)\right)\delta\Xi
\nonumber\\
&+
\left(
\hat a^\subL(\Tc(p),p)-\hat a^\subG(\Tc(p),p)
+\oet
\right)N\delta\left(\frac{N^\subL}{N}\right).
\label{e:add-A-LG}
\end{align}

Here, the last term should be expressed as a linear combination
of $\delta \bT$, $\delta p$, $\delta N$ and $\delta \Xi$.
Indeed. from \eqref{e:NL*}  which gives $N^\subL$ as
a function of $(\bT,p,\Xi)$, we derive 
\begin{align}
N\delta\left(\frac{N^\subL}{N}\right)=\left(\frac{N}{\Xi}+O(\ep^0)\right)
\left(
-\delta\bT+\frac{d\Tc}{dp}\delta p+\frac{N^\subG-N^\subL}{2N}\delta\Xi
\right).
\label{e:dNL-all}
\end{align}
We leave the formula \eqref{e:add-A-LG} and \eqref{e:dNL-all} as they
are for later convenience. 

\subsection{Formal derivation of fundamental relation} \label{s:der-dG}

We set 
\begin{align}
\bA=G,\quad \hat a=\mu
\end{align}
in the formulas in the previous  subsection.
Then, we recall the thermodynamic relations
\begin{align}
  \left(\frac{\partial \mu^\subLG}{\partial\bT^\subLG}\right)_{p}
  =-\hat s^\subLG,\quad
  \left(\frac{\partial \mu^\subLG}{\partial p}\right)_{\bT^\subLG}
  =\hat \phi^\subLG,
\end{align}
where $\hat\phi$ is the specific volume. We also define 
\begin{align}
\Psi\equiv\frac{\hat q}{\Tc}\frac{N^\subL N^\subG}{2N},
\label{e:Psi-LG}
\end{align}
where $\hat q$ is the latent heat defined in \eqref{e:latent-heat}.
By using them, we rewrite \eqref{e:add-A-LG} as
\begin{align}
\delta \bG
=&
-(\bS^\subL+\bS^\subG+\oet)\delta\bT+(V+\oet)\delta p +(\bmu+\oet)\delta N
\nonumber\\
&-\left(\Psi+O(\ep)\right)\delta\Xi
+(\mu^\subL(\Tc(p),p)-\mu^\subG(\Tc(p),p)+\oet)
N\delta\left(\frac{N^\subL}{N}\right).
\label{e:globalGibbs-LG-pre}
\end{align}
Here, $\bmu$ is the global chemical potential defined by 
\eqref{e:globalMu}. It should be noted that 
\begin{align}
\bmu=\frac{N^\subL}{N}\bmu^\subL+\frac{N^\subG}{N}\bmu^\subG,
\end{align}
where $\bmu^\subLG$ is the global chemical potential in  each region
such that $\bmu^\subL=\mu^\subL(\bT^\subL,p)+O(\ep^2)$ and
$\bmu^\subG=\mu^\subG(\bT^\subG,p)+O(\ep^2)$.

Now, noting  the relations
\begin{align}
\mu^\subL(\Tc(p),p)=\mu^\subG(\Tc(p),p),
\end{align}
and \eqref{e:dNL-all}, we obtain
\begin{align}
\delta G
&=
-(S+\oet)\delta\bT+(V+\oet)\delta p +(\bmu+\oet)\delta N
-\left(\Psi
+O(\ep)\right)\delta\Xi
\nonumber\\
&\qquad+(O(\ep))\left(
-\delta\bT+\frac{d\Tc}{dp}\delta p
   +\frac{N^\subG-N^\subL}{2N}\delta\Xi
\right) .
\label{e:globalGibbs-LG}
\end{align}
By formally considering the infinitely small change $\delta   $,
we obtain 
\begin{align}
d \bG
&=
-(\bS+O(\ep))d\bT+(V+O(\ep))d p +(\bmu+O(\ep^2))d N
-\left(\Psi+O(\ep)\right)d\Xi ,
\label{e:globalGibbs-LG2}
\end{align}
which corresponds to the fundamental relation of thermodynamics.

At first sight, one may be afraid that the
relation \eqref{e:globalGibbs-LG2} does not provide a non-equilibrium
extension because the relation involves the error of $O(\ep)$. 
However, it includes non-trivial information. The point is that
some equilibrium thermodynamic quantities are singular
at $T=\Tc(p)$. For example, $S$ and $V$ are not uniquely defined
at $(T,p)=(\Tc(p),p)$ in equilibrium. Nevertheless, \eqref{e:globalGibbs-LG2}
provides a definition of $S$ at $T=\Tc(p)$ as the limit $\ep \to 0$
of the derivative of $G$ with respect to $\tilde T$
while fixing $(p,\Xi,N)$. We study \eqref{e:globalGibbs-LG2}
more carefully by going back to \eqref{e:globalGibbs-LG}.

\subsection{Careful analysis of $\ep \to 0$} \label{s:eptozero}

We consider the limit $\ep \to 0$ with keeping the phase coexistence.
The parameter $r$ defined in \eqref{e:r} identifies the line
terminating to the equilibrium state as exemplified in Fig.~\ref{fig:Fig-r},
 and  undoes the
degeneration of the equilibrium state.
Since \eqref{e:r} is rewritten as
\begin{align}
  \bT =\Tc(p)+\Xi r,
  \label{e:bT-r}
\end{align}  
a global quantity $A$ is considered as a function of $r$
through
\begin{align}
  A(r,p,N,\Xi)=A(\bT(r,p,N,\Xi),p,N,\Xi).
\end{align}  
Then, let us define the equilibrium limit of entropy
and volume by 
\begin{align}
  S_{\rm eq}(r) & \equiv \lim_{\ep \to 0+} S(r,p,N,\Xi) , \\
  V_{\rm eq}(r) & \equiv \lim_{\ep \to 0+} V(r,p,N,\Xi) ,
\end{align}
where we explicitly write only the $r$ dependence for $S_\eq$ and $V_\eq$.

Substituting \eqref{e:bT-r} into \eqref{e:globalGibbs-LG},
we obtain
\begin{align}
\delta G
&=
-(S+\oet)\Xi \delta r+\left(V-S\der{\Tc}{p}+\oet \right)\delta p
+(\bmu+\oet)\delta N 
\nonumber\\
&\qquad
-\left(\Psi+S r+ O(\ep)\right)\delta\Xi
+(O(\ep))\left(
-\Xi \delta r \right) 
\label{e:globalGibbs-LG-3}
\end{align}
where we have used \eqref{N-r}. 
This relation leads to
\begin{equation}
\lim_{\ep \to 0} \frac{1}{\Xi}\pderf{G}{r}{p,\Xi,N} =-S_{\rm eq}(r).
\label{res-1}
\end{equation}
Since we find from \eqref{e:bT-r} that  the left-hand side
is equal to
\begin{align}
\pderf{G}{\bT}{p,N,\Xi},
\end{align}
we obtain
\begin{equation}
\lim_{\ep \to 0}  \pderf{G}{\bT}{p,N,\Xi} =-S_{\rm eq}(r).
\label{res-2}
\end{equation}
Similarly, \eqref{e:globalGibbs-LG-3} leads to
\begin{equation}
\lim_{\ep \to 0}\pderf{G}{p}{r,\Xi,N} =V_{\rm eq}(r)-S_\eq(r)\frac{d\Tc}{dp}.
\label{res-p}
\end{equation}
Here, from 
\begin{align}
  G(r,p,N,\Xi)=G(\bT(r,p,N,\Xi),p,N,\Xi)
\end{align}
with \eqref{e:bT-r}, we have  the identity
\begin{equation}
  \pderf{G}{p}{r,N,\Xi}=   \pderf{G}{\bT}{p,N,\Xi}\der{\Tc}{p}
  + \pderf{G}{p}{\bT,N,\Xi}.
\end{equation}  
Substituting this into \eqref{res-p}, we obtain
\begin{equation}
\lim_{\ep \to 0}  \pderf{G}{p}{\bT,N,\Xi} =V_{\rm eq}(r).
\label{res-3}
\end{equation}
Combining (\ref{res-2}) and (\ref{res-3}), we may write
\begin{align}
d \bG
=
&-(\bS_{\rm eq}(r)+O(\ep))d\bT+(V_{\rm eq}(r)+O(\ep))d p +(\bmu+O(\ep^2))d N\nonumber\\
&\qquad-\left(\Psi+O(\ep)\right)d\Xi ,
\label{e:globalGibbs-LG2-r}
\end{align}
which is the more precise expression of (\ref{e:globalGibbs-LG2})
and convey non-trivial information of thermodynamics.

\subsection{Explicit forms of $S_{\rm eq}$ and $V_{\rm eq}$ }
\label{s:entropy-LG-P}

By using \eqref{N-r}, we express any extensive quantity $\bA$  as
\begin{align}
\bA&=N^\subL\hat a^\subL(\bTL,p)+N^\subG\hat a^\subL(\bTG,p)+\oet \nonumber\\
&=N\left[\frac{\hat a^\subL(\bTL,p)+\hat a^\subG(\bTG,p)}{2}
+r (\hat a^\subG(\bTL,p)-\hat a^\subL(\bTL,p))\right]+\oet.
\end{align}
Letting $\hat a^\subLG_\subC(p)\equiv\hat a^\subLG(\Tc(p),p)$, we have 
\begin{align}
\lim_{\ep\rightarrow 0}\bA(r,p,N,\Xi)=N\left[
\frac{\hat a^\subL_\subC(p)+\hat a^\subG_\subC(p)}{2}
+r(\hat a^\subG_\subC(p)-\hat a^\subL_\subC(p))
\right].
\end{align}
This leads to
\begin{align}
\bS_\eq(r)
&=
N\left[
\frac{\hat s^\subL_\subC(p)+\hat s^\subG_\subC(p)}{2}
+r(\hat s^\subG_\subC(p)-\hat s^\subL_\subC(p))
\right]\nonumber\\
&=
N\left[
\frac{\hat s^\subL_\subC(p)+\hat s^\subG_\subC(p)}{2}
+r\frac{\hat q}{\Tc},
\right]
\end{align}
where we have used
the formula \eqref{e:latent-heat} for the latent heat $\hat q$.
Similarly, we obtain
\begin{align}
V_\eq(r)
&=
N\left[
\frac{\hat \phi^\subL_\subC(p)+\hat \phi^\subG_\subC(p)}{2}
+r(\hat \phi^\subG_\subC(p)-\hat \phi^\subL_\subC(p))
\right]\nonumber\\
&=
N\left[
\frac{\hat \phi^\subL_\subC(p)+\hat \phi^\subG_\subC(p)}{2}
+r\frac{\hat q}{\Tc}\frac{d\Tc}{dp}
\right],
\label{e:Veq-P}
\end{align}
where we have used the Clausius-Clapeyron relation
\begin{align}
(\hat \phi^{\subG}_{\subC}-\hat \phi^{\subL}_{\subC})
=\frac{d\Tc}{dP}\frac{\hat q}{\Tc}.
\end{align}
Note that $V_\eq(r)$ in \eqref{e:Veq-P} is consistent
with $V(\bT,p,\Xi)$ in \eqref{e:V*} at the limit $\ep\rightarrow 0$.

\subsection{Heat capacity and compressibility} \label{s:heat capacity}

Let $H(\tilde T, p, N, \Xi)$ be the enthalpy for the heat
conduction system with the phase coexistence.
Since $H$ is given by the spatial integral of the local enthalpy,
it satisfies 
\begin{align}
&H(\bT,p,N,\Xi)=U(\bT,p,N,\Xi)+pV(\bT,p,N,\Xi).
\end{align}
This leads to 
\begin{align}
\delta H=\delta U+p\delta V
\end{align}
in an infinitely small quasi-static process at constant
pressure. We interpret this $\delta H$ as the quasi-static heat
$d'Q$ in this process.  We thus define  heat capacity at constant pressure as
\begin{align}
\bC_p\equiv \pderf{\bH}{\bT}{p,N,\Xi} .
\label{e:globalCp-def}
\end{align}
Substituting $A=H$ into \eqref{e:add-A-LG} with $(p, N, \Xi)$ fixed,
we have
\begin{align}
\pderf{\bH}{\bT}{p,N,\Xi}=N^\subL \hat c_p^\subL+N^\subG \hat c_p^\subG
+(\hat h^\subL(\Tc)-\hat h^\subG(\Tc)+O(\ep^2)) \left(\frac{\partial N^\subL}{\partial \bT}\right)_{p,N,\Xi},
\label{e:CP0}
\end{align}
where the specific heat for the liquid (gas) is defined as
\begin{align}
\hat c_p^\subLG= \left(\frac{\partial \hat h^\subLG}{\partial \bT^\subLG}\right)_{p}.
\end{align}
By using \eqref{e:NL*} and \eqref{e:globalCp-def}, we rewrite
\eqref{e:CP0} as 
\begin{align}
\bC_p=N^\subL \hat c_p^\subL+N^\subG \hat c_p^\subG+\frac{N\hat q }{\Xi}+O(\ep)
\label{e:CP}
\end{align}
with the latent heat $\hat q=\hat h^\subG_\subC-\hat h^\subL_\subC$.
Note that the third term mainly contributes to $C_p$ in \eqref{e:CP}
because it diverges as $\Xi\rightarrow 0$.
This is consistent with the singularity  at $\Tc(p)$
for equilibrium cases, where $C_p$ is given by the derivative
of a discontinuous function (enthalpy) at $T=\Tc(p)$.
The first and the second terms correspond to the heat capacity
of the liquid region and the gas region, respectively.
Thus, the third term may be interpreted as the interface contribution.
The expression \eqref{e:CP} also clarifies the violation of
additivity for $C_p$.

Similarly to the heat capacity, we can also study the additivity
of other response functions. As one example, we show the singular
nature of the compressibility from the viewpoint of the violation
of the additivity. The compressibility
in heat conduction is defined as 
\begin{align}
\alpha_{\bT}
=
-\frac{1}{V}\left(
\frac{\partial V}{\partial p}
\right)_{\bT,N,\Xi}.
\end{align}
By taking $A=V$ in the formula \eqref{e:add-A-LG}, we obtain
\begin{align}
-V\alpha_{\bT}
&=
N^\subL \left(\frac{\partial \hat \phi^\subL}{\partial p}\right)_{\bT^\subL} 
+
N^\subG
\left(\frac{\partial \hat \phi^\subG}{\partial p}\right)_{\bT^\subG} 
-(\hat \phi^\subG(\Tc)-\hat\phi^\subL(\Tc))\left(\frac{\partial N^\subL}{\partial p}\right)_{\bT,\Xi}
\nonumber\\
&=
N^\subL \left(\frac{\partial \hat \phi^\subL}{\partial p}\right)_{\bT^\subL} 
+
N^\subG
\left(\frac{\partial \hat \phi^\subG}{\partial p}\right)_{\bT^\subG} 
-(\hat \phi^\subG(\Tc)-\hat\phi^\subL(\Tc))
\frac{d\Tc}{dp}\frac{N}{\Xi}.
\end{align}
Similarly to $C_p$, the main contribution in $\Xi\rightarrow 0$
is the non-additive term, which diverges in the equilibrium limit.

\section{Thermodynamic relations in $(\bT, V, N, \Xi)$ } 
\label{s:liquid-gas phase constant V}

\begin{figure}[tb]
\centering
\includegraphics[scale=0.55]{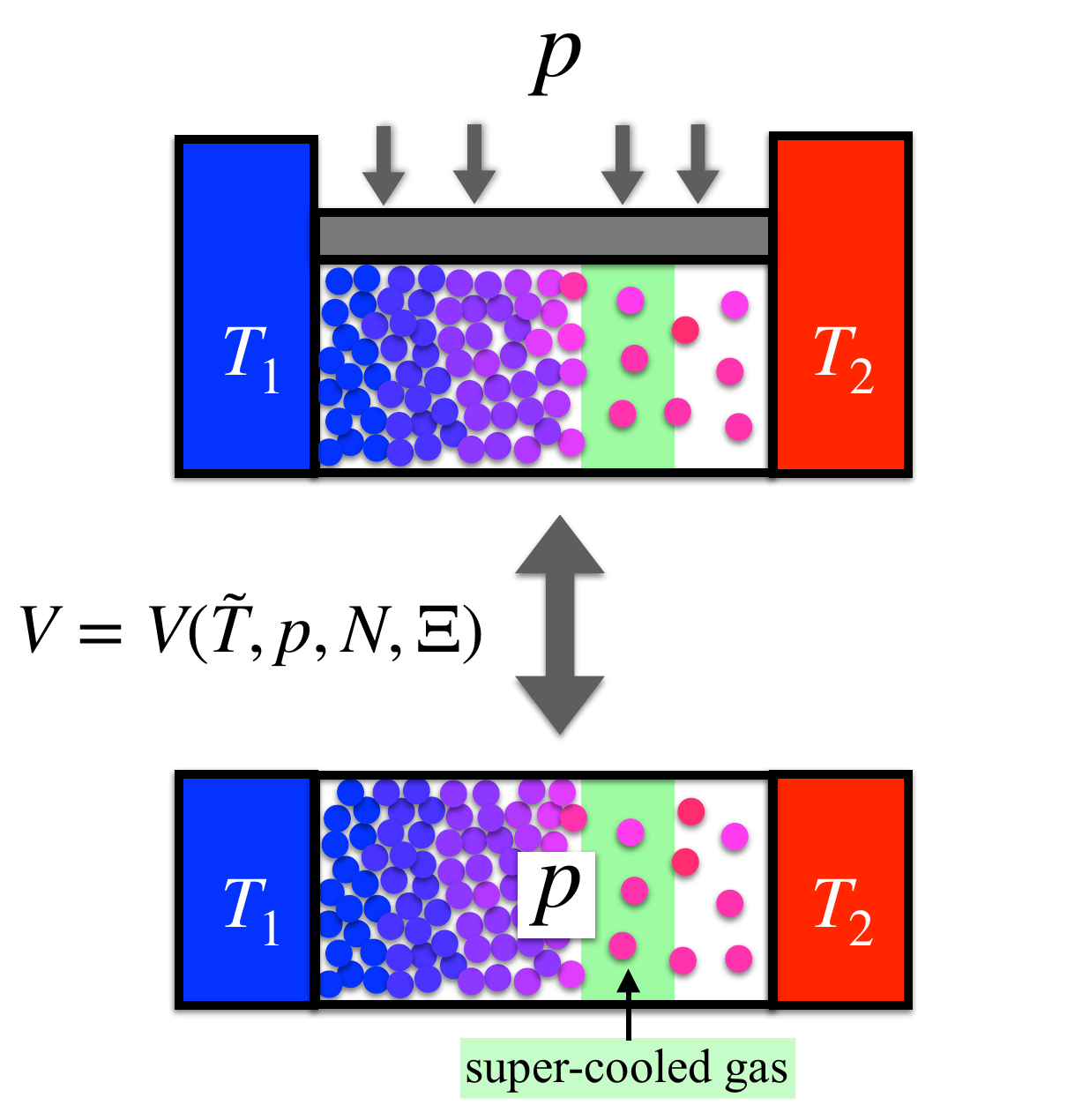}
\caption{Equivalence of the steady states for  different
  environments. The top figure shows the steady state in the
  system at the constant pressure $p$, whose volume is $V(\bT,p,N,\Xi)$.
  The bottom figure shows the steady state in the constant volume system
  where the volume $V$ is chosen as $V=V(\bT,p,N,\Xi)$. The resulting
  pressure $p(\bT,V,N,\Xi)$ is equal to $p$.
 }
\label{fig:FigV}
\end{figure}

The fundamental relation \eqref{e:globalGibbs-LG2}
derived in the previous section 
is expressed in terms of the
state variable $(\bT, p, N, \Xi)$. As is familiar with
thermodynamics, one may consider other fundamental relations. The most familiar one to physicists
may be the form using the state variable $(\bT, V, N, \Xi)$,
because this form is directly related to experimental
configurations with  $(\bT, V, N, \Xi)$ fixed, instead of
$(\bT, p, N, \Xi)$ fixed in the previous sections.
Since the same steady state  can be realized 
when the fixing condition is replaced from  $p={\rm const}$
to $V={\rm const}$,  
there is one-to-one correspondence between
$(\bT, V, N, \Xi)$ and $(\bT, p, N, \Xi)$.
See Fig.~\ref{fig:FigV}. 
In this section, we show thermodynamic relations with volume $V$.
In \S\ref{s:thermo-V}, we derive the fundamental relation in
$(\bT, V, N, \Xi)$. In \S\ref{s:F-TV}, we express the free energy
$F$ by using the saturated pressure $\Ps(T)$ for equilibrium
systems. 
Since the value of $\bT$ is not easily controlled in experiments,
in \S\ref{s:control T1-T2}, we formulate $T_1$ and $T_2$
as functions of $\bT$ and $V$.
Below, we fix $N$ without loss of generality and
sometimes abbreviate $A(\bT, N)$ as $A(\bT)$.

\subsection{Fundamental relation}\label{s:thermo-V}

Since $\bF$ and $\bG$ are given as the spatial integral of
the local free energies, the uniformity of the pressure leads to
\begin{align}
\bG=\bF+pV .
\label{e:GtoF}
\end{align}
Substituting this  into \eqref{e:globalGibbs-LG}, we have
\begin{align}
\delta F
&=
-(S+\oet)\delta\bT-(p+\oet)\delta V +(\bmu+\oet)\delta N
\nonumber\\
&\qquad\qquad
-\left(\Psi+O(\ep)\right)\delta\Xi
+(
\oet)N\delta\left(\frac{N^\subL}{N}\right),
\label{e:globalGibbsF-LG}
\end{align}
where the error of $\oet \delta p$ in \eqref{e:globalGibbs-LG} 
is regarded as $\oet \delta p(\bT, V, N, \Xi)$ which results
in the error of $\oet \delta V$ besides $\oet \delta \bT$ and so on.
The most important thing here is the order estimate of $N\delta(N^\subL/N)$,
which has been estimated as $O(\ep^{-1})$ when $N^\subL$ is a function of $(\bT,p,N,\Xi)$.
See \eqref{e:dNL-all}.
As a function of $(\bT, V,N,\Xi)$, $N\delta(N^\subL/N)$ becomes of $O(\ep^0)$
because the liquid-gas coexistence stably appears at equilibrium ($\ep=0$)
in  a certain range of $(\bT,V,N)$ and, therefore,  the term is not singular in the limit $\ep\to 0$.
Indeed, the second line is estimated as
\begin{align}
\left(\frac{\partial N^\subL}{\partial \bT}\right)_{V,\Xi}
&=
\left(\frac{\partial \NL}{\partial \bT}\right)_{p,\Xi}
+\left(\frac{\partial \NL}{\partial p}\right)_{p,\Xi}
\left(\frac{\partial p}{\partial \bT}\right)_{V,\Xi}
\nonumber\\
&=
-\frac{N}{\Xi}\frac{d\Tc}{dp}
\left(
\frac{d\Ps}{d\bT}
-
\left(\frac{\partial p}{\partial \bT}\right)_{V,\Xi}
\right)
=O(\ep^0),
\end{align}
where we have used
\begin{equation}
  \left(\frac{\partial p}{\partial \bT}\right)_{V,\Xi}
  =\frac{d\Ps}{d\bT}+O(\ep).
\end{equation}  
Thus, the last term in \eqref{e:globalGibbsF-LG} remains to be negligible
as  $\oet$ term, and we conclude
\begin{align}
d \bF
&=
-(\bS+\oet)d\bT-(p+\oet)d V +(\bmu+\oet)d N
-\left(\Psi+O(\ep)\right)d\Xi.
\label{e:globalGibbsF-LG2}
\end{align}
It should be noted that there are the errors of $O(\ep^2)$
in contrast to \eqref{e:globalGibbs-LG2}.

\subsection{Free energy for the coexistence phase}\label{s:F-TV}

\begin{figure}[tb]
\centering
\includegraphics[scale=0.55]{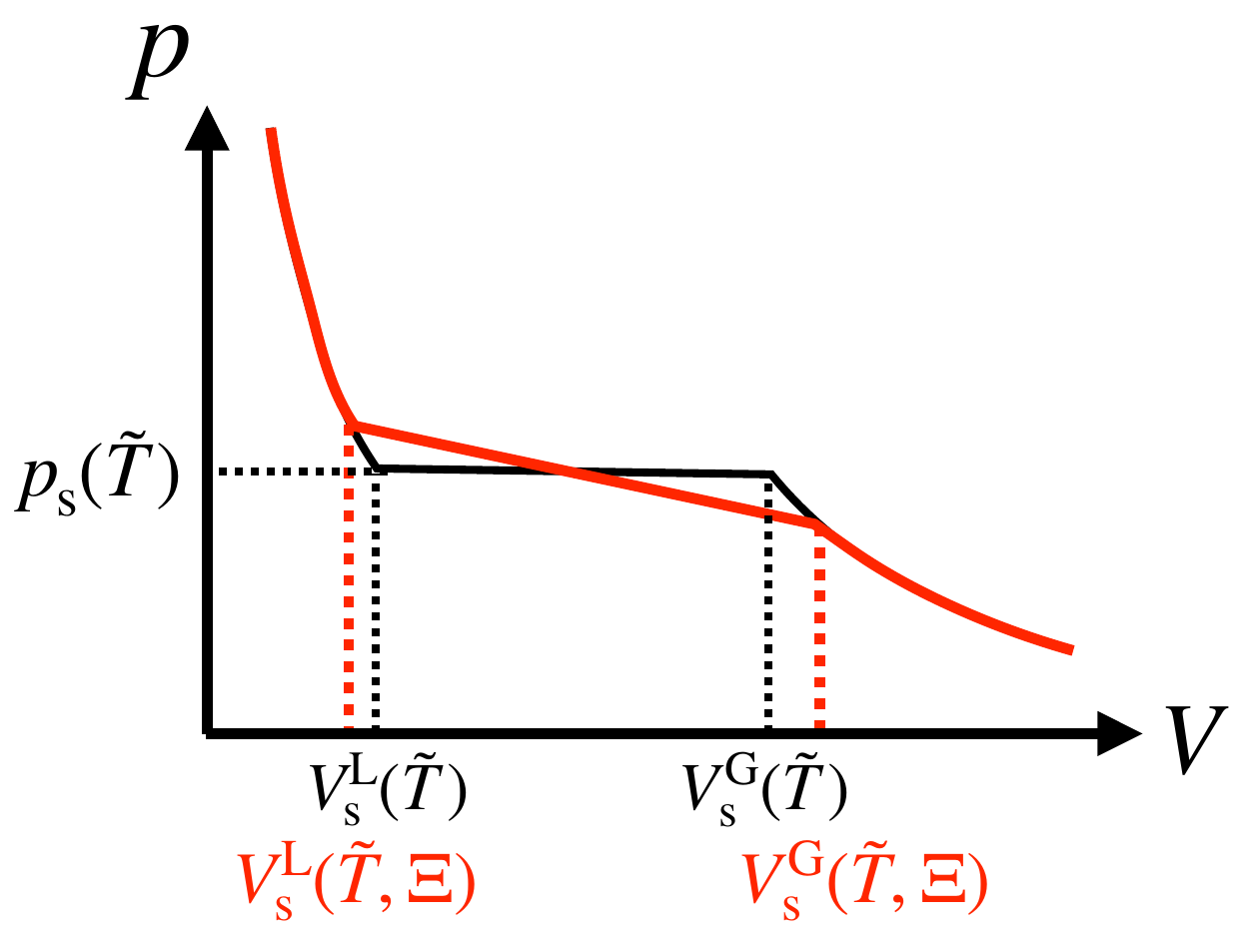}
\caption{$p$-$V$ curves with $\bT$ fixed for $\Xi=0$ (black line)
  and for $\Xi \not = 0$ (red line). For the equilibrium case $(\Xi=0)$,
  the pressure is kept constant to be a  saturated pressure $\Ps(\bT)$
  in the region $V_{\rm s}^\subL(\bT)<V<V_{\rm s}^\subG(\bT)$, where the
  liquid and the gas coexist.}
\label{fig:FigPV}
\end{figure}

In the equilibrium coexistence phase, as shown in Fig.~\ref{fig:FigPV},
the saturated volumes, $\Vs^\subL(T)$ and $\Vs^\subG(T)$, and the saturated
pressure $\Ps(T)$ are defined for a given temperature $T$. 
Even for the  coexistence phase in heat conduction, we define
the saturated volumes, $\Vs^\subL(\tilde T,\Xi)$ and $\Vs^\subG(\tilde T,\Xi)$,
at which the liquid and the gas start to coexist.
Note that the pressure is not kept constant in
$\Vs^\subL(\bT,\Xi)<V<\Vs^\subG(\bT,\Xi)$, differently from the equilibrium case.

\begin{figure}[tb]
\centering
\includegraphics[scale=0.5]{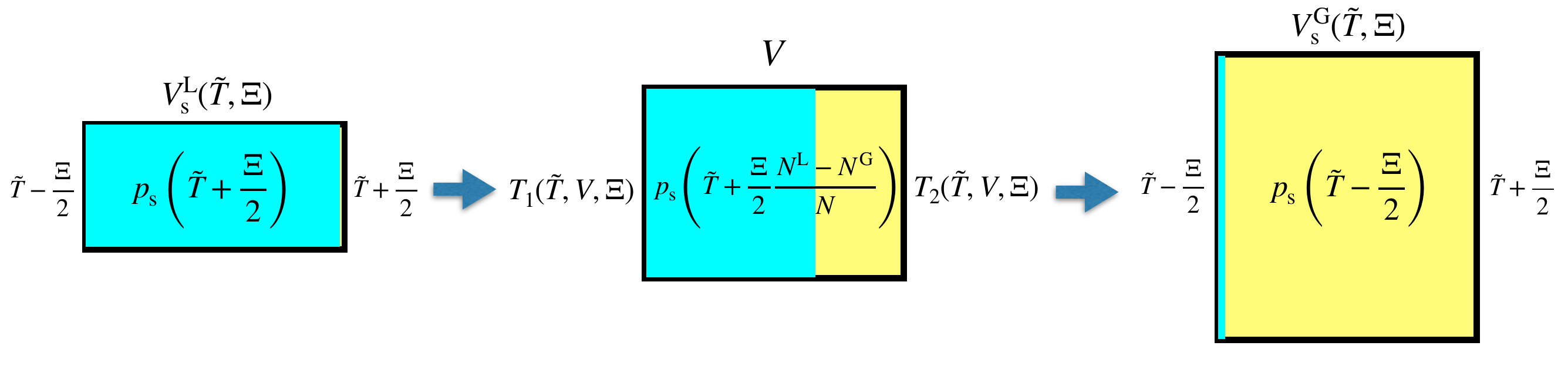}
\caption{Dependence of the steady state on $V$ for a given $\bT$. 
  The liquid region and the gas region in the container are
  painted by light blue  and yellow, respectively.
  The pressure changes with increasing $V$ according to \eqref{e:P-V-Xi}.
  The temperatures of the two heat baths, $T_1$ and $T_2$, are
  controlled to realize fixing $(\bT,V,\Xi)$ as
  formulated in \eqref{e:T1-V} and \eqref{e:T2-V}. }
\label{fig:FigClapeyron}
\end{figure}

We first express $\Vs^\subL(\bT,\Xi)$, $\Vs^\subG(\bT,\Xi) $
and $p(\tilde T, V,\Xi)$  in terms of  $\Vs^\subL(\bT)$,  $\Vs^\subG(\bT)$
and $\Ps(\bT)$. Since the steady state satisfies \eqref{e:NL*}, we have
\begin{align}
\Tc(p)=\bT+\frac{\Xi}{2}\frac{N^\subL-N^\subG}{N}+\oet.
\end{align}
Solving this in $p$, we obtain
\begin{align}
  p(\tilde T, V,\Xi)
  =\Ps\left(\bT+\frac{\Xi}{2}\frac{N^\subL-N^\subG}{N}\right)+\oet.
\label{e:P-NL-Xi}
\end{align}
For equilibrium cases, the relations $N^\subL+N^\subG=N$
and $N^\subL \Vs^\subL+N^\subG\Vs^\subG=NV$ lead to 
\begin{align}
N^\subL=N\frac{\Vs^\subG(\bT)-V}{\Vs^\subG(\bT)-\Vs^\subL(\bT)}, 
\quad
N^\subG=N\frac{ V-\Vs^\subL(\bT)}{\Vs^\subG(\bT)-\Vs^\subL(\bT)}.
\label{287}
\end{align}
By using these expressions, we rewrite \eqref{e:P-NL-Xi} as 
\begin{align}
  p(\bT,V,\Xi)
  =\Ps\left(\bT-\Xi
  \frac{V-V_\mathrm{m}(\bT)}{\Vs^\subG(\bT)-\Vs^\subL(\bT)}\right)+\oet,
\label{e:P-V-Xi}
\end{align}
where $V_\mathrm{m}(\bT)=(\Vs^\subL(\bT)+\Vs^\subG(\bT))/2$. 
Figure \ref{fig:FigPV} shows a $p$-$V$ curve described by
\eqref{e:P-V-Xi}, in which
$p(\bT,V,\Xi)$ is linearly decreasing with $V$ in the range
$\Vs^\subL(\bT,\Xi)<V<\Vs^\subG(\bT,\Xi)$.
Configurations in the phase coexistence  are
exemplified in Fig.~\ref{fig:FigClapeyron}.
Then, 
the saturated volumes under heat conduction are expressed as
\begin{align}
  \Vs^\subL(\bT, \Xi)=V^\subL
  \left(\bT, \Ps\left(\bT+\frac{\Xi}{2}\right)\right)
  ,\quad
  \Vs^\subG(\bT,  \Xi)=
  V^\subG\left( \bT,
  \Ps\left(\bT-\frac{\Xi}{2}\right)\right), 
\end{align}
where 
$V=V^\subL(T,p)$ and $V=V^\subG(T,p)$ are the equilibrium equation
of state for the liquid and  the gas, respectively.

Next, we study the free energy $F(\tilde T,V,\Xi)$.
We first consider the free energy difference
\begin{equation}
  \Delta^{\rm LG} F\equiv
  \bF(\bT,\Vs^\subG(\bT,\Xi),\Xi)-\bF(\bT,\Vs^\subL(\bT,\Xi),\Xi).
\end{equation}  
Since the saturated state is in a single phase, we have
\begin{equation}
  \bF(\bT,\Vs^\subLG(\bT,\Xi),\Xi)=F^\subLG(\bT,\Vs^\subLG(\bT,\Xi))+\oet.
\end{equation}  
We then transform
\begin{align}
\Delta^{\rm LG} F
=& 
F^\subG(\bT,\Vs^\subG(\bT,\Xi)) -F^\subL(\bT,\Vs^\subL(\bT,\Xi)) +\oet \nonumber \\
=&
F^\subG(\bT,\Vs^\subG(\bT)) -F^\subL(\bT,\Vs^\subL(\bT)) \nonumber \\
& \quad  -\Ps(\tilde T)\Xi\left( 
                 \pderf{\Vs^\subG(\bT,\Xi)}{\Xi}{\bT}- 
                 \pderf{\Vs^\subL(\bT,\Xi)}{\Xi}{\bT}
                 \right)+ \oet  \nonumber \\
=&  -\Ps(\bT)(\Vs^\subG(\bT,\Xi)-\Vs^\subL(\bT,\Xi))+\oet ,           
\label{e:FG-FL}
\end{align}
where we have used the
Taylor expansion in  $\Xi$ to obtain the second line,
and substituted the equilibrium relation
\begin{align}
  F^\subG(\bT,\Vs^\subG(\bT)) -F^\subL(\bT,\Vs^\subL(\bT))
  =-\Ps(\bT)(\Vs^\subG(\bT)-\Vs^\subL(\bT))
\end{align}
to obtain the last line. 

\begin{figure}[tb]
\centering
\includegraphics[scale=0.43]{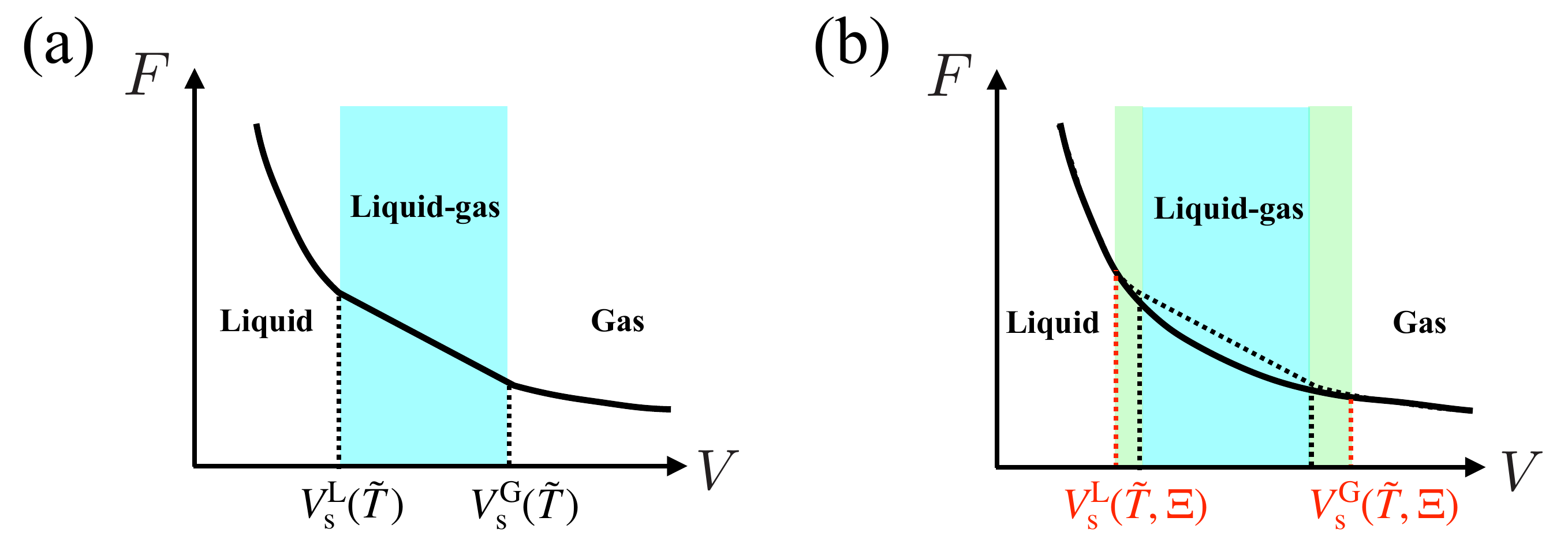}
\caption{Schematic figure of the global Helmholtz
  free energy $F$ with $\bT$ fixed when $\bT$ is less than
  the critical temperature.
  (a) Equilibrium cases.  $F$ is a linear function in the light
  blue area, $\Vs^\subL(\bT)<V<\Vs^\subG(\bT)$, where the system
  shows the liquid-gas coexistence.
  (b) Heat conduction cases with  $\Xi>0$.
  The solid line becomes a quadratic curve
  in $\Vs^\subL(\bT,\Xi)<V<\Vs^\subG(\bT,\Xi)$
  painted by light blue and green,
  where the system shows the liquid-gas coexistence.
 $F$ is smooth at $V=\Vs^\subLG(\bT,\Xi)$.  
  The dotted line represents $F$ at equilibrium in (a). 
  The solid and dotted lines are the same
  in $V<\Vs^\subL(\bT,\Xi)$ and $V>\Vs^\subG(\bT,\Xi)$. }
\label{fig:Fig-globalF}
\end{figure}

Second, from \eqref{e:globalGibbsF-LG2}, we have
\begin{align}
  \bF(\bT,V,\Xi)
  = F^\subL(\bT,\Vs^\subL(\bT,\Xi))-\int_{\Vs^\subL(\bT,\Xi)}^V P(\bT,V',\Xi) ~dV'.
\label{e:globalF-LG-op}
\end{align}
Performing the integration with \eqref{e:P-V-Xi}, we obtain
\begin{align}
\bF(\bT,V,\Xi)
&=
F^\subL(\bT,\Vs^\subL(\bT,\Xi))
-\Ps(\bT)(V-\Vs^\subL(\bT,\Xi))- B\Xi
\label{e:globalF-LG-Vform}
\end{align}
with 
\begin{align}
B=\frac{d\Ps(\bT)}{d\bT}\frac{\Vs^\subG(\bT)-\Vs^\subL(\bT)}{2}
\left[\frac{1}{4}-\left(\frac{V-V_\mathrm{m}(\bT)}{\Vs^\subG(\bT)
    -\Vs^\subL(\bT)}\right)^2\right]+O(\ep) .
\label{e:Psi-LG-V}
\end{align}
Here, by using \eqref{287}, we can confirm 
\begin{align}
  \frac{N^\subL N^\subG}{N^2}
=
\frac{1}{4}-
\left(\frac{V-V_\mathrm{m}(\bT)}
     {\Vs^\subG(\bT)-\Vs^\subL(\bT)}\right)^2 .
\end{align}
Combining it with the Clausius-Clapeyron relation
\begin{align}
\frac{d\Ps(\bT)}{d\bT}({\Vs^\subG(\bT)-\Vs^\subL(\bT)})=N\frac{\hat q}{\bT},
\end{align}
we derive
\begin{align}
  B=\Psi ,
\label{B-psi}
\end{align}
where $\Psi$ was introduced in \eqref{e:Psi-LG}.
Summarizing these, we schematically draw the graph $\bF$
in  Fig.~\ref{fig:Fig-globalF}.
For equilibrium cases, the free energy of the liquid-gas coexistence
phase is expressed as  a common tangent at $\Vs^\subL(\bT)$ and $\Vs^\subG(\bT)$
as shown in Fig.~\ref{fig:Fig-globalF}(a).
For heat conduction cases, the free energy of the coexistence phase
is expressed as a common tangential quadratic curve at $\Vs^\subL(\bT,\Xi)$
and $\Vs^\subG(\bT,\Xi)$ as shown in Fig~\ref{fig:Fig-globalF}(b),
and therefore $F$ keeps the convexity on $V$.

\begin{figure}[tb]
\centering
\includegraphics[scale=0.43]{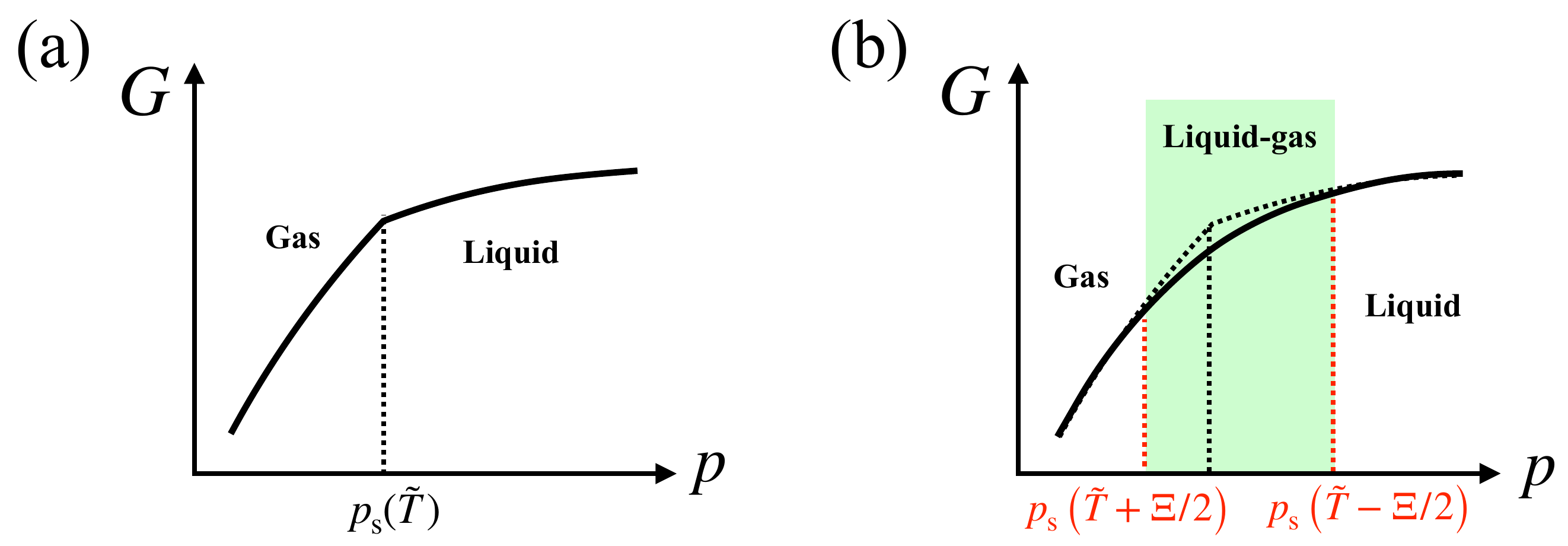}
\caption{Schematic figure of the global Gibbs free energy $G$ with
  $\bT$ fixed, where $\bT$ is less than the critical temperature.
  (a) Equilibrium cases. $G$ exhibits the first order transition
  at $p=\Ps(\bT)$, at which $G$ is continuous but not smooth. 
  (b) Heat conduction cases with $\Xi>0$.  $G$ contains a
  quadratic curve in $\Ps(\bT+\Xi/2)<p<\Ps(\bT-\Xi/2)$,
  in which the system shows the liquid-gas coexistence (solid line).
  $G$ is smooth at $p=\Ps(\bT\pm\Xi/2)$.
  The dotted line represents $G$ at equilibrium in (a).
  The solid and dotted lines are the same in $p<\Ps^\subL(\bT+\Xi/2)$
  and $p>\Ps^\subG(\bT-\Xi/2)$.}
\label{fig:Fig-globalG}
\end{figure}

Since  the convexity of $F$ is concluded, the Gibbs free energy $G$
is expressed as the Legendre transform of $F$:
\begin{align}
G(\bT,p,\Xi)=\min_{V} [F(\bT,V,\Xi)+pV].
\end{align}
The graph
$F$ in Fig.~\ref{fig:Fig-globalF}(b) yields the graph $G$
in Fig.~\ref{fig:Fig-globalG}(b),
while it contains an error of $O(\ep^2)$ as well as $F$
formulated explicitly in \eqref{e:globalF-LG-Vform}.
The liquid-gas coexistence states are
expressed as a common tangential quadratic curve
to the graph $G$ at $\Ps(\bT+\Xi/2)$ and $\Ps(\bT-\Xi/2)$.

\subsection{$T_1$, $T_2$ and $\theta$ as functions of  $(\bT,V,N,\Xi)$ }
\label{s:control T1-T2}

We have studied the dependence of the steady states on
$V$ for $(\bT, \Xi)$ fixed. 
However, when we fix  values of $T_1$ and $T_2$, 
the change of $V$ yields the change of $\bT$.
Here we propose a protocol to control $T_1$ and $T_2$,
by which $\bT$ is fixed when changing $V$. In other words,
we derive $T_1$ and $T_2$ as functions of $(\bT, V, N, \Xi)$.
We also give the interface temperature $\theta$ and its deviation from $\Tc$ as functions of $(\bT, V, N, \Xi)$.

In \eqref{e:implicit-zeta*}, $\mT$ is expressed in terms
of $\Tc(p)$, $\Xi$, $u$, $v$, and $\zeta$, where the value
of $u$ and $v$ are determined for each $p$. For the coexistence phase
in heat conduction, the pressure is determined as \eqref{e:P-V-Xi}
and can be replaced by $\Ps(\bT)$ for the argument of $u$ and $v$.
We thus represent $u=u(\Ps(\bT))$ and $v=v(\Ps(\bT))$, and simply
write $u(\bT)$ and $v(\bT)$. Then, we  rewrite \eqref{e:implicit-zeta*} as 
\begin{align}
  \mT=\Tc(p(\bT,V,\Xi))-\frac{\Xi}{2}
  \frac{u \zeta^2-v(1-\zeta)^2}{[u\zeta+(1-\zeta)][\zeta+v(1-\zeta)]}+\oet
  \label{e:mT-1}
\end{align}
with 
\begin{align}
u(\bT)=\frac{\Vs^\subG(\bT)}{\Vs^\subL(\bT)},
\qquad
v(\bT)=\frac{\kappa_\subC^\subL(\bT)}{\kappa_\subC^\subG(\bT)}.
\end{align}
Furthermore, 
from
\begin{align}
  \zeta=\frac{N^\subL}{N} \frac{\Vs^\subL}{V},
\end{align}
and $N^\subL \Vs^\subL+N^\subG \Vs^\subG=NV$ at equilibrium,
we obtain 
\begin{align}
\zeta
&=\frac{(\Vs^\subG(\bT)-V)\Vs^\subL(\bT)}{(\Vs^\subG(\bT)-\Vs^\subL(\bT))V}+O(\ep).
\end{align}
These expressions of $u$, $v$ and $\zeta$ lead to
\begin{align}
  &\frac{1}{u\zeta+(1-\zeta)}
=\frac{V}{\Vs^\subG},\\
  &\frac{1}{\zeta+v(1-\zeta)}
=\frac{V(\Vs^\subG-\Vs^\subL)}{\Vs^\subL(\Vs^\subG-V)+v\Vs^\subG(V-\Vs^\subL)},\\
&
u\zeta^2-v(1-\zeta)^2
=\frac{\Vs^\subG[\Vs^\subL(\Vs^\subG-V)^2-v\Vs^\subG(V-\Vs^\subL)^2]}
      {V^2(\Vs^\subG-\Vs^\subL)^2 },
\end{align}
with which \eqref{e:mT-1} is rewritten as
\begin{align}
  \mT=\Tc(p)-\frac{\Xi}{2}
  \frac{1}{\Vs^\subG-\Vs^\subL}
  \frac{\Vs^\subL(\Vs^\subG-V)^2-v\Vs^\subG(V-\Vs^\subL)^2}
       {\Vs^\subL(\Vs^\subG-V)+v\Vs^\subG(V-\Vs^\subL)}.
  \label{e:mT-2}
\end{align}

Here, let us recall that $\Tc(p)=\bT-\Xi r$  defined in \eqref{e:r}
and that 
\begin{align}
r=\frac{V-V_\mathrm{m}(\bT)}{\Vs^\subG(\bT)-\Vs^\subL(\bT)}+O(\ep) 
\end{align}
as used in \eqref{e:P-V-Xi}.
Substituting this formula into \eqref{e:mT-2} and noting
\begin{align}
V-V_\mathrm{m}(\bT)=\frac{V-\Vs^\subG(\bT)}{2}+\frac{V-\Vs^\subL(\bT)}{2},
\end{align}
we finally arrive at
\begin{align}
\mT(\bT,V,\Xi)=
\bT-\frac{\Xi}{2}
\frac{(V-\Vs^\subL)(\Vs^\subG-V)(\Vs^\subL-v \Vs^\subG)}
  {(\Vs^\subG-\Vs^\subL)(\Vs^\subL(\Vs^\subG-V)+v\Vs^\subG(V-\Vs^\subL))}
+O(\ep^2).
\label{e:mT-V}
\end{align}
Since $\mT=(T_1+T_2)/2$ and $\Xi=T_2-T_1$, we obtain
\begin{align}
&T_1(\bT,V,\Xi)=
\bT-\frac{\Xi}{2}-\frac{\Xi}{2}
\frac{(V-\Vs^\subL)(\Vs^\subG-V)(\Vs^\subL-v \Vs^\subG)}{(\Vs^\subG-\Vs^\subL)(\Vs^\subL(\Vs^\subG-V)+v\Vs^\subG(V-\Vs^\subL))}
+O(\ep^2),
\label{e:T1-V}\\
&T_2(\bT,V,\Xi)=
\bT+\frac{\Xi}{2}-\frac{\Xi}{2}
\frac{(V-\Vs^\subL)(\Vs^\subG-V)(\Vs^\subL-v \Vs^\subG)}{(\Vs^\subG-\Vs^\subL)(\Vs^\subL(\Vs^\subG-V)+v\Vs^\subG(V-\Vs^\subL))}
+O(\ep^2).
\label{e:T2-V}
\end{align}
These formulas inform how to operate $T_1$ and $T_2$
in order to fix the global temperature $\bT$ when changing $V$ and $\Xi$.

Substituting the temperature relation \eqref{e:tempRelation}
into \eqref{e:mT-2}, we obtain the interface temperature as
\begin{align}
  \theta(\bT,V,\Xi)=\bT+\frac{\Xi}{2}
  \frac{1}{\Vs^\subG-\Vs^\subL}
  \frac{\Vs^\subL(\Vs^\subG-V)^2-v\Vs^\subG(V-\Vs^\subL)^2}
       {\Vs^\subL(\Vs^\subG-V)+v\Vs^\subG(V-\Vs^\subL)}.
\end{align}
Similarly,  \eqref{e:mT-V} and \eqref{e:tempRelation} lead to the
deviation from the transition temperature,
\begin{align}
&\theta(\bT,V,\Xi)-\Tc(p(\bT,V,\Xi))
\nonumber\\
&\qquad =\frac{\Xi}{2}
\frac{(V-\Vs^\subL)(\Vs^\subG-V)(\Vs^\subL-v \Vs^\subG)}
  {(\Vs^\subG-\Vs^\subL)(\Vs^\subL(\Vs^\subG-V)+v\Vs^\subG(V-\Vs^\subL))}
+O(\ep^2).
\end{align}

\section{Other assumptions
  leading to the results of the variational principle} 
\label{s:equivalence}

To this point, it has been assumed that the steady states are
determined by the variational principle proposed
in \S\ref{s:variational principle}. 
In this section, we  introduce other assumptions from which
the result of the variational principle is obtained.
First, in \S\ref{s:var-V}, we formulate the variational
principle for the constant volume systems and show that
its solution is equivalent to the previous one \eqref{e:NL*}.
In \S\ref{s:thermo to var}, we start with the fundamental
relation of thermodynamics  without assuming the variational principle. 
We  then derive the result of the variational principle.
In \S\ref{s:singularity}, we notice a
singularity relation which is a simple assumption for the
singularity in the limit $\ep \to 0$. We  confirm that this
is equivalent to the result of the variational principle.
Finally, in \S\ref{s:scaling}
we argue some scaling behavior of the system which
also leads to the result of the variational principle.

\subsection{Variational principle for constant volume systems}
\label{s:var-V}

In this subsection, we study constant volume systems,
where the steady state is characterized by $(\bT, V, \Xi, N)$.
As shown in Fig.~\ref{fig:FigV}, this steady state is equivalent to
the steady state of the system at the constant pressure $\pex$
when the value of $\pex$ is  chosen as the pressure $p(\bT,V,N,\Xi)$.
We assume the mechanical balance everywhere. That is, 
\begin{align}
p(x)=p(x')
\end{align}
for any $x$ and $x'$ in $[0,L_x]$, where $p(x)=p(T(x),\rho(x))$.
This condition determines $V^\subL$ and $V^\subG=V-V^\subL$
when $N^\subL$ and $N^\subG=N-N^\subL$ are determined. The problem
is then to determine $N^\subL$. After briefly reviewing the corresponding
variational principle for equilibrium cases, we propose the variational
principle for heat conduction systems and determine the value of $N^\subL$. 

\subsubsection{Equilibrium systems}

We define the variational function as
\begin{align}
{\cal F}({\cal N}^\subL; T, V, N)
&=F^\subL(T,V^\subL,{\cal N}^\subL)+F^\subG(T, V-V^\subL,N-{\cal N}^\subL),
\end{align}
in which
$F^\subLG(T,V,N)$ is the Helmholtz free energy of the liquid (gas).
$V^\subL$ and $V^\subG$ are determined by
\begin{align}
p(T,{\cal N}^\subL/V^\subL)=p(T,(N-{\cal N}^\subL)/V^\subG),
\end{align}
and $V^\subL+V^\subG=V$. 
Then, the variational principle for determining ${\cal N}^\subL$
is 
\begin{align}
  \left.\frac{\partial {\cal F}({\cal N}^\subL;T,V,N)}
       {\partial {\cal N}^\subL}\right|_{{\cal N}^\subL=N^\subL}=0,
\qquad 
\left.\frac{\partial^2 {\cal F}({\cal N}^\subL;T,V,N)}
     {\partial ({\cal N}^\subL)^2}\right|_{{\cal N}^\subL=N^\subL}>0,
\label{e:var-F-eq}
\end{align}
which is equivalent to  the equality of the chemical potential.

\subsubsection{Heat conduction systems}

Let us consider a variational principle for the heat conduction systems 
characterized by $(\bT,V,N,\Xi)$. We assume that the steady state
with a liquid-gas interface is determined by the variational function
\begin{align}
  {\cal F}({\cal N}^\subL; \bT, V,N, \Xi)
  &\equiv L_yL_z\intx~ f(T(x),\rho(x))\nonumber\\
  &=F^\subL(\bT^\subL,V^\subL,{\cal N}^\subL)
  +F^\subG(\bT^\subG,V-V^\subL,N-{\cal N}^\subL)+O(\ep^2),
\label{e:calF}
\end{align}
where we have used a similar formula
as \eqref{e:A-add} with \eqref{e:AL} and \eqref{e:AG}
to obtain the second line. As we formulated in \S\ref{s:LG-SS},
thermodynamic quantities are expressed as a function
of ${\cal N}^\subL$ such that $\bT^\subLG=\bT^\subLG({\cal N}^\subL;\bT,V,N,\Xi)$.
The variation of ${\cal F}$ is then written as
\begin{align}
\delta {\cal F}=\delta F^\subL+\delta F^\subG.
\label{e:dFLFG}
\end{align}
Remembering that the relation \eqref{e:globalGibbs-X} holds in each region,
we have
\begin{align}
  \delta F^\subLG=-S^\subLG\delta \bT^\subLG
  -p\delta V^\subLG+\bmu^\subLG\delta {\cal N}^\subLG.
  \label{e:varF-LG-0}
\end{align}
Substituting 
$\delta {\cal N}^\subG=-\delta {\cal N}^\subL$, $\delta V^\subL+\delta V^\subG=0$
and \eqref{e:dTLTG} into  \eqref{e:varF-LG-0}, we rewrite 
\eqref{e:dFLFG} as
\begin{align}
  \delta {\cal F}
  =(\mu^\subL(T^{\rm s},p)-\mu^\subG(T^{\rm s},p)+\oet)\delta {\cal N}^\subL,
\end{align}
where $T^{\rm s}$ is given in \eqref{e:Ts}.
Then, the variational principle
\begin{align}
  \left.\frac{\partial {\cal F}({\cal N}^\subL;\bT,V,N,\Xi)}
       {\partial {\cal N}^\subL}\right|_{{\cal N}^\subL=N^\subL}=0,
\qquad 
\left.\frac{\partial^2 {\cal F}({\cal N}^\subL;\bT,V,N,\Xi)}{\partial ({\cal N}^\subL)^2}\right|_{{\cal N}^\subL=N^\subL}>0
\label{e:var-F-neq}
\end{align}
determines ${\cal N}^\subL$ as that satisfying 
\begin{align}
\mu^\subL(T^{\rm s},p)=\mu^\subG(T^{\rm s},p)+\oet.
\label{e:sol-F}
\end{align}
This is equivalent to \eqref{e:sol1} so that
\eqref{e:sol-F} leads to \eqref{e:NL*}. Therefore,
the variational principle for the constant volume system
is equivalent to that for the system at constant pressure.

\subsection{Thermodynamic relation}\label{s:thermo to var}

In this subsection, we do not assume any variational principles,
but assume the fundamental relation of thermodynamics
\begin{align}
d F
=-S d \bT-p d V +\bmu d N -\Psi d\Xi
\label{ad-as}
\end{align}
for the liquid-gas coexistence phase in the linear response regime.

First of all, it should be noted that \eqref{e:add-dsum},  \eqref{e:add-dsum2},  \eqref{e:add-dsum3}  and \eqref{e:T(X)} hold regardless of the variational
principle.
Then, we have 
\begin{align}
\delta \bA=&
\left(
N^\subL \left(\frac{\partial \hat a^\subL}{\partial \bT^\subL}\right)_{p} 
+
N^\subG \left(\frac{\partial \hat a^\subG}{\partial \bT^\subG}\right)_{p} 
+\oet
\right)
\delta \bT
\nonumber\\
&+
\left(
N^\subL \left(\frac{\partial \hat a^\subL}{\partial p}\right)_{\bT^\subL} 
+
N^\subG
\left(\frac{\partial \hat a^\subG}{\partial p}\right)_{\bT^\subG} 
+\oet
\right)\delta p
\nonumber\\
&+\left(\frac{N^\subL}{N}\hat a^\subL+\frac{N^\subG}{N}\hat a^\subG+\oet\right)
\delta N
\nonumber\\
&-\frac{N^\subL N^\subG}{2N}
\left(\left(\frac{\partial \hat a^\subL}{\partial \bT^\subL}\right)_{p} 
-\left(\frac{\partial \hat a^\subG}{\partial \bT^\subG}\right)_{p} +O(\ep)\right)\delta\Xi
\nonumber\\
&+
\left(
\hat a^\subL(\mT-\bT-\Tint,p)-\hat a^\subG(\mT-\bT-\Tint,p)
+\oet
\right)N\delta\left(\frac{N^\subL}{N}\right)
\label{e:add-A-LG-noVar}
\end{align}
instead of \eqref{e:add-A-LG}.
Letting $A=G$ and $\hat a=\bmu$, and using $G=F+pV$, we obtain
\begin{align}
\delta F
&=
-(S+\oet)\delta\bT-(p+\oet)\delta V +(\bmu+\oet)\delta N
\nonumber\\
&\qquad\qquad
-\left(\Psi+O(\ep)\right)\delta\Xi
+(D+\oet)N\delta\left(\frac{N^\subL}{N}\right),
\label{e:re-globalGibbsF-LG}
\end{align}
where
\begin{align}
  D\equiv \mu^{\rm L}(\mT-\bT-\Tint,p)-
  \mu^{\rm G}(\mT-\bT-\Tint,p).
\end{align}
Since $N \delta (N^\subL/ {N} )=O(\ep^0)$, the assumption
(\ref{ad-as}) leads to
\begin{align}
  D=0.
\end{align}  
This means the temperature relation \eqref{e:tempRelation}.
As shown in \S\ref{s:Temp relation}, \eqref{e:tempRelation}
leads to \eqref{e:NL*}
so that  the thermodynamic relation is equivalent to the
variational principle. 

\begin{figure}[tb]
\centering
\includegraphics[scale=0.6]{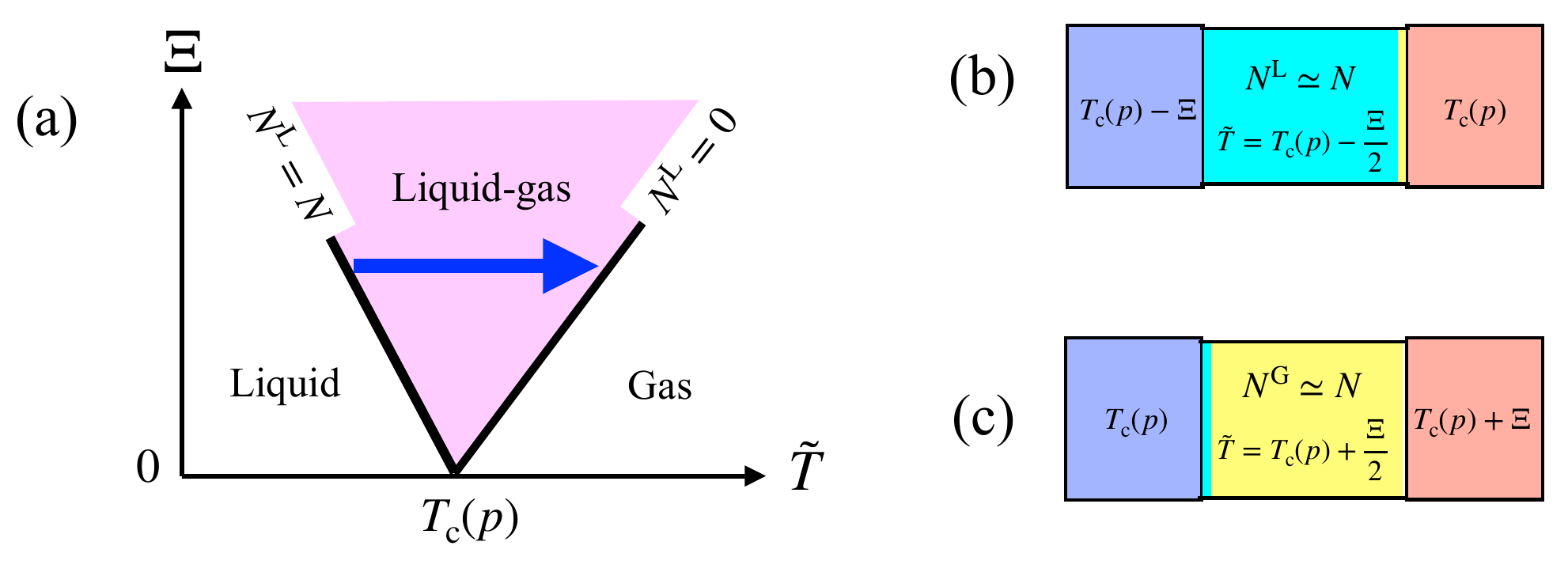}
\caption{(a) Phase diagram for a given $p$. The line depicted as
  $N^\subL=N$ corresponds to $\bT=\Tc(p)-{\Xi}/{2}$ from \eqref{e:bT-Tc},
  at which the gas region starts to appear as shown in (b).
  Similarly, the line as $N^\subL=0$, i.e., $N^\subG=N$, is
  $\bT=\Tc(p)-\frac{\Xi}{2}$, at which the liquid region
  disappears as shown in (c).}
\label{fig:diagram}
\end{figure}

\subsection{Singularity relation}\label{s:singularity}

We here consider a protocol to shift the liquid-gas interface
from $X=L_x$ to $X=0$ with fixing $\Xi$ and $p$.
This protocol is obtained by varying $(T_1, T_2)$ as 
\begin{align}
  \left(\Tc(p)-\Xi, \Tc(p)\right)
 \quad\rightarrow\quad 
 \left(\Tc(p), \Tc(p)+\Xi\right),
\end{align}
as shown in Figs.~\ref{fig:diagram}(b) and (c),
where $(\bT, N^\subL)$ is changed as
\begin{align}
  \left(\Tc(p)-\frac{\Xi}{2}, N\right)
 \quad\rightarrow\quad 
 \left(\Tc(p)+\frac{\Xi}{2}, 0\right).
\end{align}
On the phase diagram of $(\bT, \Xi)$ for a given $p$,
the change occurs along the line displayed by the arrow in Fig.~\ref{fig:diagram}(a)
when $\Xi$ is sufficiently small.
Therefore, it is natural to assume that
\begin{align}
\left(\frac{\partial\bT}{\partial \NL}\right)_{p,\Xi}
&= -\frac{\Xi}{N}+O(\ep^2)
\label{e:scale-4}
\end{align}
for sufficiently small $\Xi$.
This differential equation \eqref{e:scale-4} is written as
\begin{align}
\left(\frac{\partial\NL}{\partial \bT}\right)_{p,\Xi}=-\frac{N}{\Xi}+O(\ep^0),
\label{e:singularRelation}
\end{align}
which we call a {\it singularity relation}.
Solving \eqref{e:scale-4}
with the boundary conditions 
\begin{align}
\bT(N^\subL=N-0,p,\Xi)=\Tc(p)-\frac{\Xi}{2}+\oet
\label{e:limitL}
\end{align}
 in Fig.~\ref{fig:diagram}(b)
and 
\begin{align}
\bT(N^\subL=+0,p,\Xi)=\Tc(p)+\frac{\Xi}{2}+\oet
\label{e:limitG}
\end{align}
in Fig.~\ref{fig:diagram}(c),
we obtain
\begin{align}
\bT&=\Tc(p)+\frac{\Xi}{2}\frac{\Ng-\NL}{N}+\oet,
\label{e:bT-Tc}
\end{align}
which is the same form as  the result of
the variational principle \eqref{e:NL*}.

The relation \eqref{e:singularRelation} indicates
\begin{align}
\left|
\left(\frac{\partial \NL}{\partial \bT}\right)_{P,\Xi}
\right|
\xrightarrow{\Xi\rightarrow 0}
\infty,
\label{e:singularNL}
\end{align}
which corresponds to an expression of the singularity
associated with  the first-order transition for equilibrium cases.
This singularity is consistent with the discontinuous change of $N^\subL$
at $T=\Tc(p)$. It should be noted that \eqref{e:singularNL}
is connected to the singularity of constant pressure
heat capacity $C_p$ and compressibility $\alpha_T$ as shown
in \S\ref{s:heat capacity}.

\subsection{Scaling relation}
\label{s:scaling}


We characterize the coexistence phase by 
$(\pex,N^\subL,\Xi, N)$. As examples, we write 
\begin{align}
 \bT^\subL &=  \bT^\subL(\pex, N^\subL, \Xi, N), \\
 \bT^\subG &=  \bT^\subG(\pex, N^\subL, \Xi, N).
\end{align}
From the homogeneity in the direction perpendicular to $x$,
$ \bT^\subL$ and  $\bT^\subG$ are invariant for 
$(\pex,  N^\subL, \Xi, N) \to 
(\pex, \lambda N^\subL, \Xi,\lambda N)$.
We thus write 
\begin{eqnarray}
 \bT^\subL &=&  \bT^\subL(\pex, s, \Xi), \\
 \bT^\subG &=&  \bT^\subG(\pex, s, \Xi),
\end{eqnarray}
where
\begin{equation}
  s=\frac{N^\subL}{N}.
\end{equation}  
By noting $\bT^\subG-\bT^\subL=\Xi/2$,
we express  $\bT^\subL$ and $\bT^\subG$ as 
\begin{eqnarray}
  \bT^\subL
  &=& \Tc(\pex)-  \tau(\pex, s)\frac{\Xi}{2}+O(\ep^2),
  \label{TmL}\\
  \bT^\subG
  &=& \Tc(\pex)+ (1- \tau(\pex, s))\frac{\Xi}{2}+O(\ep^2).
  \label{TmG}
\end{eqnarray}


\begin{figure}[b]
\begin{center}
\includegraphics[scale=0.75]{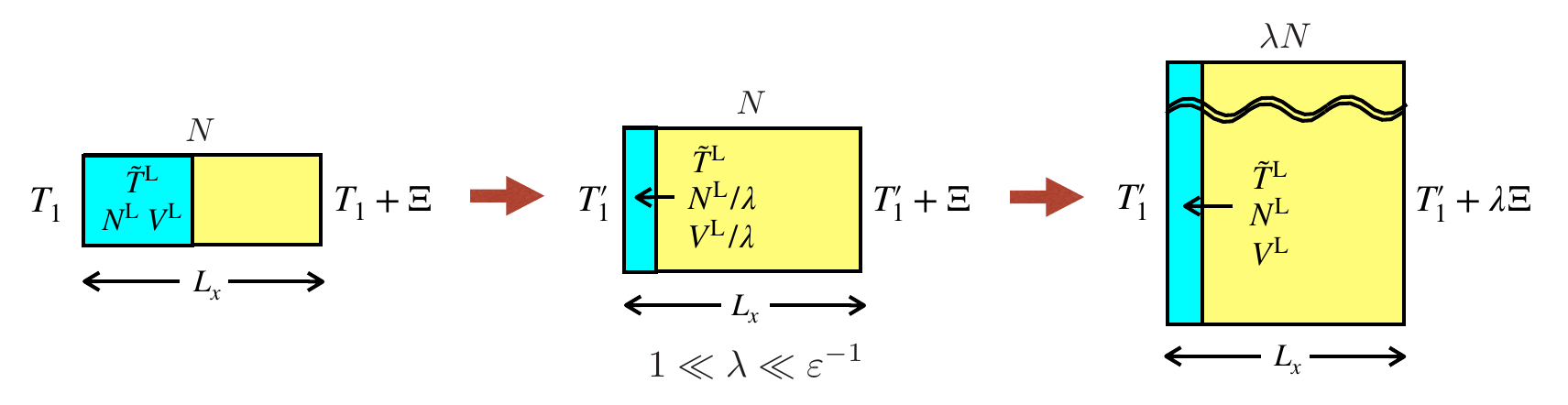}
\caption{Scaling to extend the gas region with keeping the state in the liquid region and the horizontal length $L_x$. In order to fix $\bT^\subL$, the temperature of the left heat bath is changed from $T_1$ to $T_1'$. }
\label{fig-s1}
\end{center}
\end{figure}

Now, we attempt to extend the gas region while keeping the thermodynamic
state in the liquid region.  See Fig. \ref{fig-s1}.
First, we increase $\Xi$ as
$\Xi \to \lambda \Xi$ with $\bT^\subL$ fixed,
where we take $ \lambda  \ll \ep^{-1}$ such that the system
is still in the linear response regime. 
We estimate the $\lambda$
dependence of quantities in this asymptotic regime. First, 
the heat flux is proportional to $\lambda$ as a common value
to the liquid and the gas region. 
Since $\bT^\subL$ is fixed, it is plausible that the temperature 
difference in the liquid region remains to be $O(\lambda^0)$,
and then  this leads that the horizontal length of the liquid 
region is proportional to $1/\lambda$. 
On the other hand, in the gas region, the temperature
difference is proportional to $\lambda$ and the horizontal length
is $L_x$ in the leading order estimate. Thus, the total volume
saturates at the finite value with which the whole system is occupied by
the gas with the temperature less than  $\Tc(\pex)+\lambda \Xi$. 
From these, we find that  the volume of the liquid region is
proportional to $1/\lambda$. Although its proportional coefficient
is not determined in the asymptotic region, we assume the scaling
relation that the volume of the liquid is  $V^\subL/\lambda$ as it is 
consistent with the case $\lambda=1$.  Since $\pex$ and $\bT^\subL$
is fixed in this operation, the density of the liquid
is also fixed. Thus, the particle number of the liquid becomes
$N^\subL/\lambda$. Next, we increase $N$ as $N \to \lambda N$.
Then, we have $N^\subL/\lambda  \to  N^\subL$ and
$V^\subL/\lambda  \to  V^\subL$. That is, for a series of processes
\begin{equation}
  (\pex, N^\subL, \Xi, N)
  \to
  (\pex, N^\subL, \lambda \Xi, \lambda N),
\end{equation}
we have 
\begin{equation}
  \bT^{\subL}  (\pex, N^\subL, \lambda \Xi,  \lambda N)
= \bT^{\subL}  (\pex,   N^\subL, \Xi,  N).
\label{inv}
\end{equation}
Since this condition cannot be concluded from 
the local equilibrium thermodynamics, we impose (\ref{inv}) 
as a requirement for the steady state.

By using (\ref{TmL}) for (\ref{inv}), we have
\begin{equation}
\tau\left(\pex, \frac{s}{\lambda } \right)\lambda
=
\tau(\pex, s)
\label{inv3}
\end{equation}
for $1 \ll \lambda  \ll \ep^{-1}$.
Expanding 
\begin{equation}
\tau(\pex, s) = a_0+a_1 s +a_2 s^2+O(s^3) , 
\label{asympt}
\end{equation}
and substituting it to (\ref{inv3}), we obtain
\begin{equation}
a_0(\lambda-1)  +a_2 s^2(\lambda^{-1}-1)+\cdots=0
\label{asy2}
\end{equation}
for $1 \ll \lambda  \ll \ep^{-1}$. This gives
$a_0=a_2=a_n=0$ for $n \ge 3$. Thus,
\begin{equation}
\tau\left(\pex, s \right)=a_1 s,
\label{result-1}
\end{equation}
which restricts the $s$ dependence of $\tau$.

Next, we  extend the liquid region while keeping the gas
region. In this case, we consider the  transformation 
$(\Xi, N) \to  (\lambda\Xi, \lambda N)$,
while fixing $(\bT^\subG,N^\subG)$. From (\ref{TmG}), we have
\begin{equation}
\left[
1-\tau\left(\pex, 1-\frac{N^\subG}{\lambda N} \right)
\right] \lambda
=
1-\tau\left(\pex, 1-\frac{N^\subG}{N} \right).
\label{inv-G}
\end{equation}
Expanding 
\begin{equation}
\tau(\pex, 1-u) = b_0+b_1 u +O(u^2) 
\label{asympt-gas}
\end{equation}
in the limit  $u \to 0$, we obtain
$b_0=1$ and $b_n=0$ for $n \ge 2$. 
This leads to 
\begin{equation}
b_1  \left( 1-s \right)
=
1-\tau\left(\pex, s \right).
\label{result-2}
\end{equation}
By combining  (\ref{result-1}) and (\ref{result-2}),
we obtain
\begin{equation}
b_1  ( 1-s)
=
1-a_1 s.
\label{result-3}
\end{equation}
Since this holds for any $s$, we obtain $b_1=a_1=1$.
By substituting (\ref{result-1}) with $a_1=1$ into (\ref{TmL}), 
 we have arrived at
\begin{align}
\bT^\subL=\Tc(\pex)-\frac{\Xi}{2}\frac{N^\subL}{N}+O(\ep^2).
\end{align}
Substituting \eqref{e:bT-bTL-bTR} into this relation,
we obtain the solution \eqref{e:NL*} of the variational principle.

\renewcommand{\intr}{\int_{{\cal D}} d^3{\bm r}~}

\section{Generalization for  single phase systems}
\label{s:general configuration}

In previous sections,  global thermodynamics has been developed for
systems in a rectangular container as shown in
Fig. \ref{fig:Fig-setup}. In this section, we consider
an arbitrarily shaped container ${\cal D}$ in contact with two heat
baths of $T_1$ and $T_2$. Here, we assume that the response
of a local quantity is not singular for the change of the system
parameters $(T_1, T_2, V)\rightarrow(T_1+\delta T_1,T_2+\delta T_2,V+\delta V)$, where $V$ is the volume of the container.
We also assume that local steady states inside the container
are well described by the local equilibrium thermodynamics of
local quantities $T(\bm r)$, $u(\bm r)$, $s(\bm r)$, and so on.
In \S\ref{s:general-A}, we define  global thermodynamic quantities
as they are consistent with the definitions given in \S\ref{s:single phase}.
We construct  global thermodynamics in the linear response regime
in \S\ref{s:arbitrary shaped}. Then, we extend the thermodynamic formulas
beyond the linear response regime in \S\ref{s:second order}. 
In what follows, we restrict to single-phase systems without
liquid-gas interfaces. 

\subsection{Global thermodynamic quantities in an arbitrarily
shaped container}\label{s:general-A}

We define all global thermodynamic quantities similarly to the case
of rectangular containers. 
Explicitly, the global temperature
and the global chemical potential are  defined as
\begin{align}
\bT\equiv\frac{\intr T(\bm r)\rho(\bm r)}{\intr \rho({\bm r})},
\label{e:globalT-r}
\end{align}
and
\begin{align}
\bmu\equiv\frac{\intr \mu(\bm r)\rho(\bm r)}{\intr \rho({\bm r})}.
\label{e:globalMu-r}
\end{align}
More generally, global extensive quantities $\bA$ are defined
by the spatial integration of local thermodynamic quantities $a(\bm r)$
per unit volume or $\hat a(\bm r)$ per one particle as
\begin{align}
\bA(\bT,V,N,\Xi)\equiv\intr a(\bm r)=\intr \hat a(\bm r)\rho(\bm r).
\label{e:globalEx-r}
\end{align}
It should be noted that all global quantities transform consistently
for the change of the reference state on entropy and internal energy as discussed in 
\S\ref{s:global quantities}.

Due to the mechanical balance,
the local pressure $p(\bm r)$ is homogeneous. That is,
\begin{align}
p(\bm r)=p.
\end{align}
We here note that  the pressure $p(\bm r)$ does not satisfy the
local equilibrium equation of state beyond the linear response
regime. Concretely, the contribution out of the local equilibrium
pressure was explicitly calculated in the kinetic regime by analyzing
the Boltzmann equation \cite{Kim-Hayakawa}, and also the long-range
correlation yields the additional pressure depending on the system
size \cite{Aoki,Wada,Sengers2}. 
Thus, we express 
\begin{align}
p(\bm r)=p^\LE(\bm r)+\oet,
\label{p-ple}
\end{align}
where $p^\LE(\bm r)$ is the pressure determined from 
the local equilibrium equation of state
such that
\begin{align}
p^\LE(\bm r)\equiv
-\pderf{f(\bm r)}{\hat\phi(\bm r)}{T(\bm r)}
\label{e:def-p^LE}
\end{align}
with $\hat\phi(\bm r)=1/\rho(\bm r)$.
Let us introduce a global version of the local equilibrium pressure as
\begin{align}
\bP^\LE\equiv \frac{1}{V}{\intr p^\LE(\bm r)}
\label{e:globalPLE-r}
\end{align}
This may deviate from the global pressure as
\begin{align}
\bP^\LE=\bP+\oet.
\label{ple-p}
\end{align}

\subsection{Equivalence of non-equilibrium global quantities with
  equilibrium quantities}\label{s:arbitrary shaped}

We  show that the results in \S\ref{s:single phase} hold
in the container with an arbitrary shape in the linear response regime. 
As we did in \S\ref{s:globalGibbs-linear},
we introduce a function $\eta(\bm r)=O(\ep)$ as
\begin{align}
\eta(\bm r)\equiv T(\bm r)-\bT.
\end{align}
From the definition of $\bT$, we have $\intr \eta(\bm r)=0$.
Any local quantity $\hat a(\bm r)=\hat a(T(\bm r), p(\bm r))$ per one particle
is expanded in $\eta(\bm r)$ as
\begin{align}
\hat a(T(\bm r), p(\bm r))=
\hat a(\bT,p) +\left(\frac{\partial\hat a(\bT,p)}
     {\partial \bT}\right)_p\eta(\bm r)
+\oet.
\end{align}
We then derive
\begin{align}
\bA=\intr \hat a(T(\bm r),p(\bm r))\rho(\bm r)=
\hat a(\bT,p) N +\oet.
\label{e:coincidence}
\end{align}
This indicates that all global thermodynamic quantities are equivalent
to those in equilibrium by adopting the global temperature $\bT$
in \eqref{e:globalT-r}, and that  global thermodynamics for
heat conduction systems are generally mapped to equilibrium
thermodynamics, regardless of the shape of the container.
It should be noted that we cannot apply the trapezoidal rule
argued in \S\ref{s:trapezoidal} to  such general configurations.
Thus, in contrast to the case of  rectangular 
containers, we have 
\begin{align}
\bT\neq \mT,
\end{align}
even in the linear response regime.

\subsection{Fundamental relation beyond the linear response regime}
\label{s:second order}

Since global thermodynamics relies on  local equilibrium
thermodynamics, its validity depends on the extent  how the local
states are close to the local equilibrium.
For instance, the local equilibrium thermodynamic relations may
be valid up to  $O(\ep)$ and a non-equilibrium driving may bring
a slight deviation of $\oet$ from the local equilibrium. 
We here expect that the deviation from the local equilibrium
is small even beyond the linear response regime, and attempt
to extend the global thermodynamics to the second order
of $\ep$. Indeed, in \S\ref{s:globalGibbs-2nd}
and \S\ref{s:globalLegendre-2nd},
we prove
\begin{align}
  &\delta \bF= -\bS\delta\bT-\bP^\LE\delta V +\bmu\delta N
  -\Psi\delta\Xi+O(\ep^3\delta\nu,\ep^2\delta\Xi),
  \label{e:globalGibbs-second}\\
&\bF=\bU-\bT \bS-\Psi\Xi+\oep,\label{e:globalLeGendre-second}	
\end{align}
where $\delta\nu$ is either $\delta\bT$, $\delta V$ or $\delta N$.
$\Psi$ is the conjugate variable to $\Xi$, which is  identified as
\begin{align}
\Psi=\alpha\frac{N\hat c_p}{\bT}\Xi.
\label{e:Psi}
\end{align}
This is a quantity of $O(\ep)$ with constant pressure
specific heat $\hat c_p$. $\alpha$ is a geometrical factor
independent of $\bT$ and $p$ as examined in \S\ref{s:alpha}.
That is, we can extend the thermodynamic framework  with an error of
$O(\ep^3)$ beyond the linear response regime. 

\subsubsection{Derivation of
  \eqref{e:globalGibbs-second}}\label{s:globalGibbs-2nd}

Any local density $\hat a(\bm r)=\hat a(T(\bm r), p^\LE(\bm r))$
per one particle is expanded as
\begin{align}
&\hat a(T(\bm r), p^\LE(\bm r)) \nonumber \\
&\qquad  =\hat a(\bT,\bP^\LE) 
+\left(\frac{\partial\hat a(\bT,\bP^\LE)}{\partial \bT}\right)_p \eta(\bm r)
+\frac{1}{2}\left(\frac{\partial^2\hat a(\bT,\bP^\LE)}{\partial \bT^2}\right)_p
\eta(\bm r)^2 \nonumber \\
& \qquad\qquad+\left(\frac{\partial\hat a(\bT,\bP^\LE)}{\partial p}\right)_T
(p^\LE(\bm r)-p^\LE)+\oep. 
\end{align}
Then, we have
\begin{align}
&\intr \hat a(T(\bm r),p^\LE(\bm r))\rho(\bm r) \nonumber \\
&\qquad=\hat a(\bT,\bP^\LE) N
+\frac{1}{2}\left(\frac{\partial^2\hat a(\bT,\bP^\LE)}{\partial \bT^2}
\right)_p\intr \eta(\bm r)^2\rho(\bm r)\nonumber\\
&\qquad \qquad
+\pderf{\hat a(\bT,p^{\rm LE})}{p}{\bT}\intr p^\LE(\bm r)\left(\rho(\bm r)-\frac{N}{V}\right)
+\oep
\label{e:10.16}
\end{align}
from the definition of $\eta(\bm r)$ and $p^\LE$.
Since $p^\LE(\bm r)=p+O(\ep^2)$ and $\rho(\bm r)=N/V+O(\ep)$, 
the integral of the third term in the right-hand side of \eqref{e:10.16} is expressed as
\begin{align}
p\intr\left(\rho(\bm r)-\frac{N}{V}\right)+O(\ep^3).
\label{e:10.17}
\end{align}
As the integral of \eqref{e:10.17} vanishes, the third term is estimated as $O(\ep^3)$.
Next, by applying $\rho(\bm r)=N/V+O(\ep)$ to the second term of \eqref{e:10.16},
we obtain
\begin{align}
\intr \hat a(\bm r)\rho(\bm r)=
\left[
\hat a(\bT,\bP^\LE) 
+\frac{\alpha}{2}\left(\frac{\partial^2\hat a(\bT,\bP^\LE)}
{\partial \bT^2}\right)_p\Xi^2\right]N+\oep,
\end{align}
where
\begin{align}
\alpha\equiv \frac{1}{V}\intr\left(\frac{T(\bm r)-\bT}{T_2-T_1}\right)^2.
\label{e:alpha}
\end{align}
$\alpha$  is a geometrical factor depending on the configuration
of the container as will be examined in \S\ref{s:alpha}.
Because $\alpha=O(\ep^0)$, we conclude that
global quantities are extended to those including $O(\ep^2)$
contribution:
\begin{align}
\bA(\bT,\bP^\LE,N,\Xi)=
\hat a(\bT,\bP^\LE) N
+\frac{\alpha N}{2}
\left(\frac{\partial^2\hat a(\bT,\bP^\LE)}{\partial \bT^2}\right)_p\Xi^2+\oep.
\label{e:global-a-r}
\end{align}
With this formula, we write  the global thermodynamic quantities as
\begin{align}
&\frac{\bS}{N}=
\hat s(\bT,\bP^\LE)+\frac{\alpha }{2}\left(\frac{\partial^2\hat s}{\partial \bT^2}\right)_p\Xi^2+\oep,\\
&\frac{V}{N}=
\hat \phi(\bT,\bP^\LE)+\frac{\alpha }{2}\left(\frac{\partial^2\hat \phi}{\partial \bT^2}\right)_p\Xi^2+\oep,\\
&\bmu(\bT,\bP^\LE,\Xi)=
\mu(\bT,\bP^\LE) 
+\frac{\alpha}{2}\left(\frac{\partial^2\mu}{\partial \bT^2}\right)_p\Xi^2+\oep.
\label{e:globalMu-apprx}
\end{align}

Performing the spatial integration of the local relation
$f(\bm r)=\mu(\bm r)\rho(\bm r)-p^\LE(\bm r)$, 
we have the global relation
\begin{align}
\bF=\bmu N-\bP^\LE V.
\label{e:globalF-r}
\end{align}
For the later purpose, we define a new quantity $\Psi$ as
\begin{align}
\Psi\equiv -N\left(\frac{\partial\bmu}{\partial\Xi}\right)_{T,p}.
\end{align}
Using \eqref{e:globalMu-apprx}, we obtain
\begin{align}
\Psi=-\alpha N \left(\frac{\partial^2\mu}{\partial \bT^2}\right)_p\Xi=\alpha\frac{N\hat c_p}{\bT}\Xi,
\end{align}
which is the form given in \eqref{e:Psi}.

Now, consider a parameter change
$V\rightarrow V+\delta V$, $T_1\rightarrow T_1+\delta T_1$
and $T_2\rightarrow T_2+\delta T_2$.
Accordingly, the local quantities are changed as
\begin{align}
T(\bm r)\rightarrow T(\bm r)+\delta T(\bm r),
\quad
\rho(\bm r)\rightarrow \rho(\bm r)+\delta \rho(\bm r),
\end{align}
for $\bm r\in {\cal D}$, and the global quantities as
\begin{align}
\bT\rightarrow \bT+\delta\bT,
\quad
\bP^\LE\rightarrow \bP^\LE+\delta \bP^\LE,
\quad
\Xi\rightarrow \Xi+\delta\Xi.
\end{align}
This leads to 
\begin{align}
\delta \bF&=N\delta\bmu-\bP^\LE\delta V-V\delta \bP^\LE,
\label{e:global-dF-r0}
\end{align}
where we have used \eqref{e:globalF-r}.
From \eqref{e:globalMu-apprx}, we derive 
\begin{align}
  \delta \bmu&=\left[\pderf{\mu}{\bT}{p}
    +\frac{\alpha}{2}\left(\frac{\partial^3\mu}{\partial \bT^3}\right)_p
    \Xi^2+O(\ep^3)\right]\delta\bT\nonumber\\
 &\quad +\left[\pderf{\mu}{p}{\bT}+\frac{\alpha}{2}
    \left(\frac{\partial^2}{\partial \bT^2}
    \left(\frac{\partial\mu}{\partial P}\right)_T\right)_p\Xi^2+O(\ep^3)\right]
  \delta \bP^\LE\nonumber \\
&\quad+\left[\alpha\left(\frac{\partial^2\mu}{\partial \bT^2}\right)_p\Xi+O(\ep^2)\right]\delta\Xi,\nonumber\\
&=-\left[\hat s+\frac{\alpha}{2}\left(\frac{\partial^2\hat s}{\partial\bT^2}\right)_p\Xi^2+O(\ep^3)\right]\delta\bT
+\left[\hat \phi+\frac{\alpha}{2}\left(\frac{\partial^2\hat\phi}{\partial \bT^2}\right)\Xi^2+O(\ep^3)\right]\delta \bP^\LE\nonumber\\
&\quad+\left[\alpha\left(\frac{\partial^2\mu}{\partial \bT^2}\right)_p\Xi+O(\ep^2)\right]\delta\Xi,\nonumber\\
&=-\left(\frac{\bS}{N}+O(\ep^3)\right)\delta\bT+\left(\frac{V}{N}+O(\ep^3)\right)\delta \bP^\LE-\left(\frac{\Psi}{N}+O(\ep^2)\right)\delta\Xi.
\label{338}
\end{align}
Substituting this into \eqref{e:global-dF-r0}, we obtain
\begin{align}
  \delta F= -\bS\delta\bT-\bP^\LE\delta V -\Psi\delta\Xi
  +O(\ep^3\delta\bT, \ep^3\delta V,\ep^2\delta\Xi),
\label{339}  
\end{align}
where $\delta p^\LE$ is rewritten by the linear combination of $\delta \bT$, $\delta V$ and $\delta\Xi$.
Moreover, for the change $N\rightarrow N+\delta N$ with $(\bT,V,\Xi)$ fixed,
\eqref{e:globalF-r} leads to
\begin{align}
\delta \bF= N\delta \bmu+ \bmu \delta N- V\delta \bP^\LE.
\label{e:globalF-r-d}
\end{align}
Substituting  (\ref{338}) into this, we obtain
\begin{align}
\delta \bF&= \bmu \delta N+O(\ep^3\delta p^\LE)\nonumber\\
&=\bmu\delta N+O(\ep^3\delta N).
\label{e:globalF-r-d-2}
\end{align}
By combining (\ref{339}) and (\ref{e:globalF-r-d-2}), 
we  have arrived at the formula \eqref{e:globalGibbs-second}. 

\subsubsection{Derivation of \eqref{e:globalLeGendre-second}}
\label{s:globalLegendre-2nd}

Since $u(x)=f(x)+T(x)s(x)$, we have
\begin{align}
\bU=\bF+\bT \bS +\intr~\hat s(\bm r) \eta(\bm r)\rho(\bm r).
\label{e:U-F0}
\end{align}
Substituting the expansion
\begin{align}
\hat s(\bm r)
=
\hat s(T(\bm r),p^\LE(\bm r))
&=\hat s(\bT,\bP^\LE)+\eta(\bm r)\left.\left(\frac{\partial \hat s}{\partial \bT}\right)_p\right|_{T=\bT}+\oet
\nonumber\\
&=\hat s(\bT,\bP^\LE)+\eta(\bm r)\frac{c_p(\bT,p)}{\bT}
+\oet,
\label{e:s-approx}
\end{align}
and $\rho(\bm r)=N/V+O(\ep)$ into \eqref{e:U-F0}, we obtain 
\begin{align}
&\intr~\hat s(\bm r) \eta(\bm r)\rho(\bm r)\nonumber\\
&\qquad=\hat s(\bT,p)\intr~\eta(\bm r)\rho(\bm r)+\frac{c_p(\bT,p)}{\bT}\frac{N}{V}\intr~\eta(\bm r)^2+\oep\nonumber\\
&\qquad=\alpha\frac{N\hat c_p}{\bT}\Xi^2+\oep ,
\end{align}
where the first term in the right-hand side of the first line turns out
to be zero from the definition of $\bT$.
Thus, \eqref{e:U-F0} is rewritten as
\begin{align}
\bU=\bF+\bT \bS+\Psi\Xi+\oep,
\end{align}
which corresponds to \eqref{e:globalLeGendre-second}.

\subsubsection{Factor $\alpha$
  determined by the geometry of containers }\label{s:alpha}

\begin{figure}[tb]
\centering
\includegraphics[scale=0.4]{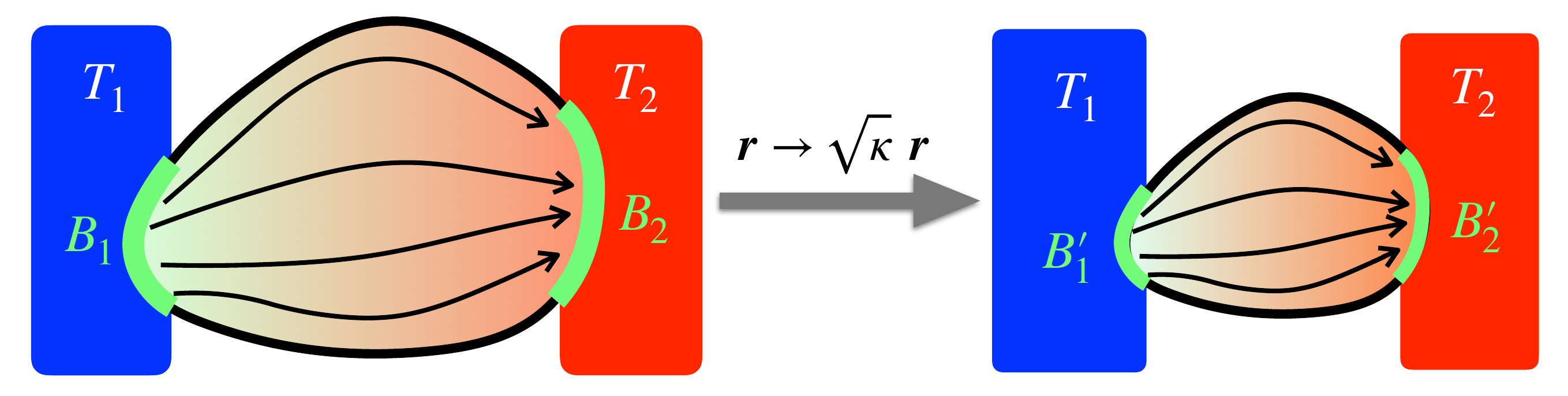}
\caption{An arbitrarily shaped container in contact with two heat baths.
  The value of $\bT$ is unchanged with respect to the scale transformation
   of the container with fixing $T_1$ and $T_2$.}
\label{fig:Fig-general}
\end{figure}

Hereafter, we show that $\alpha$ is a function of the geometry
of the container and independent of $\bT$ and $p$.
We consider a container as shown in Fig.~\ref{fig:Fig-general},
which is in contact with  the first heat bath of $T_1$ in a
region of the surface $\bm B_1$
and with the second bath of $T_2$ in $\bm B_2$. 
The rest of the surface of the container is thermally insulating.
The steady profiles of temperature, density, and heat flux are
$T(\bm r)$, $\rho(\bm r)$, and $\bm j(\bm r)=-\kappa(T(\bm r), p)\nabla T$,
respectively.

The profile $T(\bm r)$ is determined by the condition
$\nabla \cdot \bm j=0$ for the steady state, which is written as
\begin{align}
\left\{
\begin{array}{ll}
&\nabla\cdot \left( \kappa(\bT, p) \nabla T(\bm r) \right)=0, \\
&T(\bm r \in \bm B_1)=T_1, \quad T(\bm r \in \bm B_2)=T_2, 
\end{array}
\right.
\label{e:T(r)-eqn0}
\end{align}
with an error of $\oet$. 
By introducing  a scale transformation $\bm r\rightarrow \bm r'$ 
\begin{align}
\bm r=\sqrt{\kappa(\bT,p)}~\bm r',
\end{align}
as shown  in Fig.~\ref{fig:Fig-general}, we express
\eqref{e:T(r)-eqn0} as 
\begin{align}
\left\{
\begin{array}{ll}
&\Delta't(\bm r')=0,\\
&t(\bm r' \in \bm B_1')=T_1, \quad t(\bm r' \in \bm B_2')=T_2,
\end{array}
\right.
\label{e:T(r)-eqn}
\end{align}
where $t(\bm r')=T\left(\sqrt{\kappa(\bT,p)}~\bm r'\right)$.
Letting the solution of \eqref{e:T(r)-eqn} be $t_*(\bm r')$,
the global temperature is written as
\begin{align}
\bT=\frac{\intrd\rho(t_*(\bm r'), p) t_*(\bm r')}{\intrd\rho(t_*(\bm r'), p)}.
\label{e:globalT-t(r)}
\end{align}
We here expand $t_*(\bm r')$ in $\Xi/\mT$:
\begin{align}
&t_*(\bm r')=\mT\left(1+\frac{\Xi}{\mT}\Phi(\bm r')+\oet\right),\label{e:t(r)-expand}\\
&T_1=\mT-\frac{\Xi}{2},\quad T_2=\mT+\frac{\Xi}{2}.
\end{align}
With these expansions, \eqref{e:T(r)-eqn} is written as
\begin{align}
\begin{array}{ll}
&\Delta'\Phi(\bm r')=0,\\
&\Phi(\bm r' \in \bm B_1')=-\displaystyle\frac{1}{2}, \quad \Phi(\bm r' \in \bm B_2')=\frac{1}{2}.
\label{e:Phi(r)-eqn}
\end{array}
\end{align}
It is obvious that the solution of \eqref{e:Phi(r)-eqn} depends
only on the geometry of the surface but is independent of the scale of
the container and  all physical parameters. More explicitly, we confirm
that $\Phi(\bm r')$ is independent of $\bT$ and $p$.
\eqref{e:globalT-t(r)} yields 
\begin{align}
\bT-\mT&=\Xi\frac{\intrd\rho(t_*(\bm r'), p) \Phi(\bm r')+O(\ep)}{\intrd\rho(t_*(\bm r'), p)}\nonumber \\
&=\Xi\intrd \Phi(\bm r')+\oet,
\label{e:bT-expand}
\end{align}
which indicates  $\bT-\mT$ is independent of $\bT$ and $p$.

At last, we examine the parameter $\alpha$ which is rewritten as
\begin{align}
\alpha=\frac{1}{V'}\intrd \left(\frac{t_*(\bm r')-\bT}{\Xi}\right)^2.
\end{align}
Using \eqref{e:t(r)-expand} and \eqref{e:bT-expand}, we have
\begin{align}
\alpha=\frac{1}{V'}\intrd\left[\Phi(\bm r')-\intrd\Phi(\bm r')\right]^2,
\end{align}
which is determined by the function $\Phi$ given by \eqref{e:Phi(r)-eqn}.
For instance, calculating $\alpha$ for rectangular containers, we obtain
\begin{align}
\alpha=\frac{1}{12}.
\end{align}

\section{Concluding Remarks}


In this long paper, we have developed a thermodynamic framework
for heat conduction states, which we call global thermodynamics.
The key concept of the framework is the
global temperature defined by \eqref{e:globalT}. We describe spatially
inhomogeneous systems by a set of global thermodynamic variables.
Although the method is rather different from a standard continuum
description,  global thermodynamics is built on local equilibrium
thermodynamics and the transportation equation. Thus, it is not
contradictory with  established theories. Nevertheless, our
framework provides a new prediction on the thermodynamic properties
of a liquid-gas interface in heat conduction.
As we explained in \S\ref{s:problem-LG}, the rule of connecting
two phases at the interface 
is out of the standard hydrodynamics with local equilibrium
thermodynamics. Even for this case, global thermodynamics can determine
the way how to connect the two phases by the  variational principle
that is a natural extension of the Maxwell construction for
equilibrium cases. 


Although global thermodynamics is self-consistent, self-contained, and
natural, it does not ensure the validity of the framework. Similarly
to other universal theories such as thermodynamics, hydrodynamics, and
statistical mechanics,  global thermodynamics involves a fundamental
assumption on the determination of the steady state. In other words,
if our quantitative prediction on the phase coexistence is denied by
experiments,  global thermodynamics could not be valid. In this case,
we will attempt to understand the experimental results carefully so
that we have a correct description of phenomena out of equilibrium.
This is of course quite stimulating, and we sincerely wait for
experiments. Numerical simulations of microscopic dynamics such as
molecular dynamics also provide useful materials for further studies.
To the present, no deviation of the interface temperature from the transition
temperature was observed \cite{Bedeaux00,Ogushi}.
We conjecture that this is due to the insufficiency of the
separation of scales, because the deviation should be observed
in systems with enough separation of scales, as discussed in
Ref. \cite{SNIN}. 


Putting aside the ultimate validity of  global thermodynamics,
we emphasize that the framework itself is quite fruitful.
This paper presents only the minimum backbone, but exhibits
non-trivial simple relations among several quantities.
These somewhat miraculous statements may suggest 
a deeper structure behind  global thermodynamics. Moreover,
importantly, the story does not end. 
We did not discuss the Clausius equality for the cases with a liquid-gas interface, 
while we confirmed it for  single-phase systems in \S\ref{Clausius}.
The equality connects ``heat" to the entropy change, and the ``heat''
should be defined as the renormalized one, called ``excess heat'',
after subtracting the persistent heat as we reviewed in {\it Introduction}.
It is quite natural to study the liquid-gas coexistence by considering an extended form of the Clausius relation. Since we find that this provides another
source of further development of the framework, we have decided
to argue it in a separate paper. 


From a theoretical viewpoint in a broad context, we  expect
that we may derive  global thermodynamics based on a more microscopic
description. Since we know the basic framework of non-equilibrium
statistical mechanics, e.g. the linear response theory, we obtain
a linear response formula of the temperature of the interface.
It is quite formal, and it is not evident that the result satisfies
the temperature relation \eqref{e:tempRelation}. Related to this problem,
recently, we have calculated the temperature of the interface for 
a stochastic order parameter equation  describing the phase
coexistence in  heat conduction \cite{SNIN}. The result shows that
the interface temperature deviates from the equilibrium transition
temperature. It should be noted that global thermodynamics can be
formulated for systems that exhibits an order-disorder transition
in heat conduction. By finding a connection between the two theories,
we will understand a microscopic mechanism of the interesting
macroscopic phenomenon. 


Our theory may be naturally generalized to that for heat conduction systems
where a liquid  flows in and a gas flows out. When a steady state is 
observed, the system  exhibits  perpetual evaporation at the interface.
The latent heat is persistently  generated at the interface and as a result
additional heat flux  occurs  from the interface to the liquid region and
gas region.  This phenomenon was experimentally studied  for pure
water \cite{Fang-Ward}, and  a macroscopic  temperature gap 
as much as about $8$K was found at the liquid-gas interface.
As a related phenomenon, the temperature gradient in the inverse
direction to the imposed temperature difference was observed for
the liquid-gas coexistence of a heat conducting Helium far from
equilibrium \cite{Urban}. These phenomena may be closely related to
our prediction on the phase coexistence in heat conduction. We
will attempt to explain such phenomena in global thermodynamics.
Furthermore, we also expect that the description of global quantities
may be useful for more complicated systems. Even for phenomena
far from equilibrium such as motility-induced phase separation \cite{mips},
there may be an appropriate set of global variables on the basis
of local variables whose space-time evolution may be described
by another evolution rule. We will accumulate such examples. 


We hope that this paper is a good starting point of many studies. 


\section*{Acknowledgment}
The authors thank  Masato Itami, Michikazu Kobayashi, Christian Maes,
Yohei Nakayama, Kazuya Saito, Hal Tasaki, and Satoshi Yukawa
for their useful comments. 
One of the authors (N.N.) also thanks Yoshiyuki Chiba, Daiki Takei, Yuya Kai, Kyosuke Tachi and Akira Yoshida for their efforts on numerical experiments of subjects related to \S \ref{s:single phase}.
The present study was supported by KAKENHI (Nos. JP17H01148, JP19H05496, JP19H05795 and  JP19K03647).




\begin{thebibliography}{99}


  
\bibitem{Landau-Lifshitz-Fluid}
Landau, L. D., Lifshitz, E. M. :
 Fluid Mechanics.
Pergamon Press, Oxford (1959).

\bibitem{Zubarev} 
 Zubarev,  D. N. :
Nonequilibrium Statistical Thermodynamics.
Consultants Bureau, New York (1974).

\bibitem{Mclennan}
Mclennan, J. A. :
Phys. Fluids {\bf 3}, 493-502 (1960);
Introduction to Non-equilibrium Statistical Mechanics. Prentice-Hall (1988).



\bibitem{Onsager1931}
Onsager, L. :
 Reciprocal relations in irreversible processes. I. Phys. Rev. {\bf 37}, 405-426 (1931);
 Reciprocal relations in irreversible processes. II. Phys. Rev. {\bf 38}, 2265-2279 (1931).
 
 

\bibitem{Groot-Mazur}
de Groot, S.  and Mazur, P. :
 Non-equilibrium Thermodynamics.  North-Holland, Amsterdam (1962)




\bibitem{Schmitz}
Schmitz,  R. :
Fluctuations in nonequilibrium fluids,
Physics Reports {\bf 171}, 1-58 (1988).



\bibitem{Kubo}
Kubo, R.,  Toda, M., Hashitsume, N. :
Statistical Physics II.  Springer, (1985).

\bibitem{Nakano}
Nakano, H. :
Linear response theory - historical perspective. 
Int. J. Mod. Phys. B 7:2397-2467 (1993).

\bibitem{Zwanzig}
Zwanzig, R. :
  Memory Effects in Irreversible Thermodynamics,
  Phys. Rev. {\bf 124}, 983-992 (1961).

\bibitem{Mori}
 Mori,  H. :
Transport, Collective Motion, and Brownian Motion, Prog. Theor. Phys. {\bf 33}, 423-455 (1965)


\bibitem{Kawasaki-Gunton}
Kawasaki, K., Gunton, J. D. :
Theory of nonlinear transport processes: nonlinear shear viscosity and normal stress effects,
 Phys. Rev. {\bf A} 8, 2048-2064 (1973).



\bibitem{Ashkin}
Ashkin, A.,  Sch\"utze, K.,  Dziedzic, J. M.,  Euteneuer, U., Schliwa,  M. :
Force generation of organelle transport measured in vivo by an infrared laser trap, 
Nature {\bf 348}, 346-348 (1990).

\bibitem{Bustamante}
Smith, S. B., Finzi, L.  Bustamante, C. :
Direct mechanical measurements of the elasticity of single DNA molecules by using magnetic beads, 
Science {\bf 258}, 1122-1126 (1992).

\bibitem{Svoboda}
Svoboda, K.,  Schmidt C. F.,  Schnapp B. J.,  Block, S. M. :
Direct observation of kinesin stepping by optical trapping interferometry, 
Nature {\bf 365}, 721-727 (1993).

\bibitem{Chu}
Perkins, T. T.,  Quake, S. R., Smith, D. E.,  Chu, S. :
Relaxation of a single DNA molecule observed by optical microscopy, 
Science {\bf 264}, 822-826 (1994)

\bibitem{Noji} 
 Noji, H.,  Yasuda, R.,  Yoshida, M.,  Kinosita Jr, K. :
  Direct observation of the rotation of F\(_1\)-ATPase. 
Nature {\bf 386}, 299-302 (1997).


\bibitem{SeifertRPP} 
Seifert, U. :
  Stochastic thermodynamics, fluctuation theorems
  and molecular machines, Rep. Prog. Phys. {\bf 75}, 126001 (2012).

\bibitem{Sekimoto-book}
 Sekimoto, K. :
 Stochastic Energetics, Lect. Notes Phys. 799.
Springer-Verlag, Berlin (2010).

\bibitem{Kalges-Just-Jarzynski}
Klages, R.,  Just, W., Jarzynski, C. (eds.) :
Nonequilibrium Statistical Physics of Small Systems.
Wiley-VCH, Weinheim (2013) 




\bibitem{Evans-Cohen-Morriss}
 Evans, D. J.,  Cohen, E. G. D.,  Morriss, G. P. :
  Probability of second law violations in shearing
  steady states,  Phys. Rev. Lett. {\bf 71}, 2401-2404 (1993).



\bibitem{JarzynskiPRL}
Jarzynski, C. :
Nonequilibrium equality for free energy differences,
Phys. Rev. Lett. {\bf 78}, 2690-2693 (1997).


\bibitem{Gallavotti}
 Gallavotti, G.,  Cohen, E. G. D. :
Dynamical Ensembles in Nonequilibrium Statistical Mechanics,
{ Phys. Rev. Lett.}
\textbf{74},  2694-2697 (1995).

\bibitem{Kurchan}
Kurchan,  J. :
Fluctuation theorem for stochastic dynamics,
{J. Phys. A: Math. Gen.}
\textbf{31},  3719-3729 (1998).

\bibitem{LS}
 Lebowitz, J. L.,  Spohn, H. :
A Gallavotti--Cohen-type symmetry in the large
deviation functional for stochastic dynamics,
J. Stat. Phys. \textbf{95}, 333-365 (1999). 

\bibitem{Maes}
 Maes, C. :
The fluctuation theorem as a Gibbs property,
J. Stat. Phys. \textbf{95},  367-392 (1999).

\bibitem{Crooks}
Crooks, G. E. :
Entropy production fluctuation theorem and the nonequilibrium
work relation for free energy differences,
{Phys. Rev. E} \textbf{60}, 2721-2726 (1999).
 
\bibitem{JarzynskiDFT}
Jarzynski, C. :
J. Stat. Phys. {\bf 98}, 77-102 (2000)
 
 

 
 \bibitem{CrooksNRL}
Crooks, G. :
 Path ensembles averages in systems driven far-from-equilibrium,
 {Phys. Rev. E} \textbf{61}, 2361-2366 (1999).
 

\bibitem{KN}
 Komatsu, T. S.,  Nakagawa, N. :
Expression for the stationary distribution in nonequilibrium steady states,
Phys. Rev. Lett. {\bf 100}, 030601 (2008).

\bibitem{KNST-rep}
Komatsu, T. S.,  Nakagawa, N.,  Sasa, S.-i.,  Tasaki, H. :
Representation of nonequilibrium steady states in large mechanical systems,
J. Stat. Phys. {\bf 134}, 401-423 (2009).


\bibitem{Maes-rep}
 Maes, C.,  Neto\v{c}n\'{y}, K. :
Rigorous meaning of McLennan ensembles,
J. Math. Phys. {\bf 51}, 015219 (2010).


\bibitem{SasaFluid}
 Sasa, S.-i. :
Derivation of Hydrodynamics from the Hamiltonian Description of Particle Systems,
Phys. Rev. Lett. {\bf 112}, 100602 (2014).






\bibitem{Bodineau-Derrida}
Bodineau, T.,  Derrida, D. :
 Current fluctuations in nonequilibrium diffusive systems:
 an additivity principle,  
 Phys. Rev. Lett. {\bf 92}, 180601 (2004).

\bibitem{BertiniETAL} 
Bertini, L.,  De Sole, A.,  Gabrielli, D.,  Jona-Lasinio, G.,  Landim, C. :
Current fluctuations in stochastic lattice gases', 
Phys. Rev. Lett. {\bf 94}, 030601 (2005).

\bibitem{Maes-LD}
 Maes, C., Neto\v{c}n\'{y}, K. :
Minimum entropy production principle from a dynamical fluctuation law,
J.  Math.  Phys. {\bf 48}, 053306 (2007).


\bibitem{Nemoto}
Nemoto, T.,  Sasa, S.-i. :
Thermodynamic formula for the cumulant generating function of
time-averaged current, 
Phys. Rev. E {\bf 84}, 061113 (2011).


\bibitem{Bertini-rev}
Bertini, L.,  De Sole, A.,  Gabrielli, D.,  Jona-Lasinio, G.,  Landim, C. :
Macroscopic fluctuation theory, 
Rev. Mod. Phys. {\bf 87}, 593-636 (2015).


\bibitem{Information}
Parrondo, J. M. R.,  Horowitz, J. M.,  Sagawa, T. :
Thermodynamics of information,
Nature Physics {\bf 11}, 131-139 (2015).


\bibitem{SNIN}
 Sasa, S.-i.,  Nakagawa, N.,  Itami, M.,  Nakayama, N. :
Stochastic order parameter dynamics for phase coexistence in heat conduction,
in preparation.



\bibitem{min-ent}
 Jaynes, E. T. : 
  The minimum entropy production principle,
  Ann. Rev. Phys. Chem. {\bf 31}, 579- 601 (1980).

\bibitem{Klein}
 Klein, M. J.,  Meijer, P. H. E. :
  Principle of minimum entropy production,
  Phys. Rev. {\bf 96}, 250-255 (1954).


\bibitem{Callen} 
Callen, H. B., 
 Thermodynamics and an Introduction to Thermostatistics, 2nd ed.
Wiley, New York (1985) 

\bibitem{Prigogine-Kondepudi}
Prigogine, I.,  Kondepudi, D. :
Modern Thermodynamics : From Heat Engines to Dissipative Structures, Second Edition.
Wiley,  Chichester (1998)

\bibitem{Einstein} 
   Einstein, A. :
  Theorie der Opaleszenz von homogenen Fl\"usigkeiten
  und Fl\"ussigkeitsgemischen in der N\"ahe des kritischen Zustandes,
  (The Theory of the Opalescence of Homogeneous Fluids and Liquid
  Mixtures near the Critical State), Annalen der Physik {\bf 33},
  1275 - 1298, (1910). 

\bibitem{Oono} 
  Oono, Y. :
  Perspectives on Statistical Thermodynamics.
  Cambridge University Press (2017)
  
  

\bibitem{Keizer}  
Keizer, J. :
Statistical thermodynamics of nonequilibrium processes. Springer, (1987)



\bibitem{Eu}
Eu, B. C. :
Kinetic Theory and Irreversible Thermodynamics. Wiley, New York (1992)

\bibitem{Jou-book}
Jou, D.,  Casas-Vazquez, J.,  Lebon, G. :
Extended Irreversible Thermodynamics 3rd revised and enlarged edn.
Springer, Berlin (2001)


\bibitem{Bedeauz86}
  Bedeaux,  D. :
  Nonequilibrium thermodynamics and  statistical physics of surfaces, 
  Adv. Chem. Phys. {\bf 64} 47-109 (1986).



\bibitem{Jou}
Casas-Vazquez, J.,  Jou, D. :
Temperature in non-equilibrium states: a review of open problems and current proposals,
Rep. Prog. Phys. {\bf 66}, 1937-2023 (2003).


\bibitem{Sasa-Tasaki} 
Sasa, S.-i.,  Tasaki, H. :
Steady state thermodynamics, 
J. Stat. Phys. {\bf 125}, 125-224 (2006).



\bibitem{Hayashi-Sasa}
Hayashi, K.,  Sasa, S.-i. :
Thermodynamic relations in a driven lattice gas: numerical experiments.
Phy. Rev. {\bf E} 68, 035104(R) (2003).

\bibitem{Bertin}
Bertin, E.,  Martens, K.,   Dauchot, O., Droz, M. :
Intensive thermodynamic parameters in nonequilibrium systems,
Phys. Rev. E {\bf 75}, 031120 (2007).

\bibitem{Seifert-contact}
Pradhan, P.,  Ramsperger, R.,  Seifert, U. :
Approximate thermodynamic structure for driven lattice gases in contact,
Phys. Rev. E {\bf 84}, 041104 (2011).

\bibitem{Dickman}
Dickman, R. :
Failure of steady-state thermodynamics in nonuniform driven lattice gases,
Phys. Rev. E {\bf 90}, 062123 (2014).



\bibitem{Landauer}
Landauer, R. :
$dQ=TdS$ far from equilibrium,
Phys. Rev. {\bf A} 18, 255-266 (1978)

\bibitem{Oono-Paniconi}
Oono, Y., Paniconi, M. :
Steady state thermodynamics,
Prog. Theor. Phys. Suppl. {\bf 130}, 29-44  (1998).

  
\bibitem{Hatano-Sasa} 
Hatano, T.,  Sasa, S.-i. :
  Steady-state thermodynamics of Langevin systems,
  Phys. Rev. Lett. {\bf 86}, 3463-3466 (2001).

\bibitem{Ruelle}
Ruelle,  D. :
Proc. Natl. Acad. Sci. U.S.A. {\bf 100}, 3054 (2003).

\bibitem{KNST}
Komatsu, T. S.,  Nakagawa, N.,  Sasa, S.-i.,  Tasaki, H. :
Steady-state thermodynamics for heat conduction: Microscopic derivation,
Phys. Rev. Lett. {\bf 100}, 230602 (2008).


\bibitem{NN}
Nakagawa, N. :
Work relation and the second law of thermodynamics
in nonequilibrium steady states,
Phys. Rev. E {\bf 85}, 051115 (2012).
  
  
\bibitem{Jona-thermo}
Bertini, L.,  De Sole, A.,  Gabrielli, D.,  Jona-Lasinio, G.,  Landim, C. :
Clausius inequality and optimality of
quasistatic transformations for nonequilibrium stationary states,
Phys. Rev. Lett. {\bf 110}, 020601  (2013).

  
\bibitem{Maes-thermo}
Maes, C.,  Neto\v{c}n\'{y}, K. :
A nonequilibrium extension of the Clausius heat theorem,
J. Stat. Phys. {\bf 154}, 188-203 (2014).

\bibitem{Spinney-Ford}
Spinney, R. E.,  Ford, I. J. :
Nonequilibrium Thermodynamics of Stochastic Systems with Odd and Even Variables,
Phys. Rev. Lett. {\bf 108}, 170603  (2012).

\bibitem{Chiba-Nakagawa}
Chiba, Y.,  Nakagawa, N. :
Numerical determination of entropy associated with excess heat in steady-state thermodynamics,
Phys. Rev. E {\bf 94}, 022115 (2016).


 
 \bibitem{NS}
Nakagawa, N., Sasa, S.-i. :
Liquid-Gas Transitions in Steady Heat Conduction,
Phys. Rev. Lett. {\bf 119}, 260602 (2017)



\bibitem{Bedeaux00} 
R{\o}sjorde,  A., Fossmo, W. D.,  Bedeaux, D.,  Kjelstrup, S.,  Hafskjold, B. :
 Nonequilibrium molecular dynamics simulations of steady-state heat and mass transport in condensation: I. Local equilibrium,
 J. Coll. Int. Sci. {\bf 232}, 178-185 (2000).

 
\bibitem{Ogushi}
 Ogushi, F.,  Yukawa, S., Ito, N. :
  Asymmetric structure of gas-liquid interface,
  J. Phys. Soc. Jpn. {\bf 75}, 073001 (2006).
 


  
\bibitem{Bedeaux03} 
 Bedeaux, D.,  Johannessen, E.,  R{\o}sjorde, A. :
  The nonequilibrium van der Waals square
  gradient model.(I). The model and its numerical solution,
  Physica A {\bf 330}, 329-353 (2003).
  
\bibitem{Onuki}
 Onuki, A., 
  Dynamic van der Waals theory,
  Phys. Rev. E {\bf 75}, 036304 (2007).


\bibitem{CO2}
Iso, N., Uematsu, K., Mashimo, K. :
Kiso Butsuri Kagaku. Tokyo Kyogakusha, Tokyo (1997)

\bibitem{NIST}
NIST Chemistry WebBook, SRD69.
https://webbook.nist.gov/chemistry/


\bibitem{Kim-Hayakawa} 
Kim, H. D.,  Hayakawa, H. :
Kinetic theory of a dilute gas system under steady heat conduction.
J. Phys. Soc. Jpn. {\bf 72}, 1904-1916 (2003); 73, 1609 (2003).


\bibitem{Aoki}
Aoki, K.  Kusnezov, D. :
On the violations of local equilibrium and linear response,
arXiv:nlin/0105063.

\bibitem{Wada}
Wada, H.,  Sasa,  S.-i. :
Anomalous pressure in fluctuating shear flow,
Phys. Rev. E {\bf 67}, 065302(R) (2003).

\bibitem{Sengers2}
Kirkpatrick, T. R.,  Ortiz de Z\'arate, J. M.,  Sengers, J. V. :
Fluctuation-induced pressures in fluids in thermal nonequilibrium
steady states, Phys. Rev. E {\bf 89}, 022145 (2014).

\bibitem{Fang-Ward}
Fang, G.,  Ward, C. A. :
Temperature measured close to the interface of an evaporating liquid,
Phys. Rev. E {\bf 59}, 417-428 (1999)

\bibitem{Urban}
Urbana, P., Schmoranzerb, D.,  Hanzelkaa, P.,  Sreenivasanc, K. R.,  Skrbekb, L. :
Anomalous heat transport and condensation in convection of cryogenic helium,
Proc. Nat. Acad. Sci. {\bf 110},  8036-8039 (2013).

\bibitem{mips}
Cates, M. E.,  Tailleur, J. :
Motility-Induced Phase Separation,
Annual Review of Condensed Matter Physics {\bf 6}, 219-244 (2015).



\end{thebibliography}
\end{document}